\documentclass{article}

\usepackage{arxiv}
\usepackage{subcaption}
\usepackage{amsmath}
\usepackage[utf8]{inputenc} 
\usepackage[T1]{fontenc}    
\usepackage{hyperref}       
\usepackage{url}            
\usepackage{booktabs}       
\usepackage{amsfonts}       
\usepackage{nicefrac}       
\usepackage{microtype}      
\usepackage{lipsum}
\usepackage{graphicx}
\graphicspath{ {./images/} }
\usepackage{natbib}

\title{Scale by scale analysis of magnetoconvection with uniform wall-normal and wall-parallel magnetic fields at low magnetic Reynolds number}

\author{
 Jake Ineson \\
  Department of Mechanical and Aerospace Engineering\\
  University of Manchester\\
  Oxford Road, Manchester, M13 9PL \\
  jake.ineson@manchester.ac.uk \\
   \And
 Aleksander Dubas \\
  UKAEA (United Kingdom Atomic Energy Authority)\\
  Culham Campus, Abingdon, Oxfordshire, OX14 3DB \\
  \And
 Alex Skillen \\
  Department of Mechanical and Aerospace Engineering\\
  University of Manchester\\
  Oxford Road, Manchester, M13 9PL \\
  alex.skillen@manchester.ac.uk \\
}

\begin{document}
\maketitle
\begin{abstract}
\noindent Rayleigh-Bénard convection under an imposed inductionless magnetic field is analysed statistically from the perspective of single-point and multi-scale energy budgets. The data is obtained from direct numerical simulations with a Rayleigh number of $10^6$, a Prandtl number of $1$ and Hartmann numbers of $0$, $20$, $40$ and $80$. Wall-parallel and wall-normal magnetic fields are considered as two separate cases. The initial analysis focuses qualitatively on the influence of the magnetic field upon the coherent structures. A central contribution of this work is the interpretation of these structural modifications through magnetohydrodynamically modified turbulent kinetic energy budgets. For example, in the wall-normal case, the thinning of the thermal plumes can be attributed to the damping of the pressure-diffusion mechanisms due to the Lorentz dissipation. In the wall-parallel configuration, Joule dissipation induces a pressure–strain redistribution mechanism that preferentially transfers kinetic energy from the wall-normal velocity component to the field-perpendicular, wall-parallel velocity component but less so to the field-parallel velocity component. This description is then extended to scale-space by considering budgets relating second- and third-order structure functions. Here, the anisotropy is accounted for by analysing directional structure functions. Despite the anisotropy, the Lorentz force appears as an isotropic sink damping intermediate and large scales of motion. The result of this is a lack of transfer between scales of motion and hence a flow with suppressed small-scale turbulence. These results establish a link between qualitative observations and long-term energy balances, providing new insight into magnetoconvective turbulence and informing future modelling and theoretical approaches to such flows. 
\end{abstract}

\section{Introduction}
\label{sec:introduction}

Magnetoconvection (MC) occurs when electrically conducting fluids are subjected to both temperature gradients and an applied magnetic field. The temperature gradients cause thermal expansion, leading to density gradients within the fluid. The resulting buoyancy forces then drive thermal convection. As the fluid moves, its flow intersects the applied magnetic field lines, inducing electric currents within the flow. The applied field then acts to distort and suppress the flow through the action of the Lorentz force. This alteration of the velocity through the Lorentz force leads to different flow phenomena and heat transfer behaviour when compared with non-magnetohydrodynamic flows. 

The magnetoconvective phenomena discussed above govern astrophysical \citep{Weiss_2003}, geophysical \citep{Takahashi_Shimizu_2012} and many industrially relevant systems, such as fusion breeder blankets \citep{smolentsev2010mhd} and semiconductor crystal growth \citep{VIZMAN2002545}. Despite the complexity of these systems, such flows are governed by a small set of dimensionless groups \citep{grossmann2000scaling} for a given set of boundary and initial conditions. Of particular relevance to this study are, 

\refstepcounter{equation}
$$
  Ra \equiv  \frac{g \beta \Delta \theta H^3}{\nu \alpha}, \quad
  Pr \equiv \frac{\nu}{\alpha}, \quad
  Ha \equiv B_0 H \sqrt{\frac{\sigma}{\nu \rho}}, \quad
  Re_m \equiv \frac{UH}{\eta},
  \eqno{(\theequation{\mathit{a},\mathit{b}},\mathit{c},\mathit{d})}
$$

\noindent where $g$ is the acceleration due to gravity, $\beta$ is the coefficient of thermal expansion, $\Delta\theta$ is the characteristic temperature difference, $H$ is a characteristic length scale, $\nu$ is the kinematic viscosity, $\alpha$ is the thermal diffusivity, $B_0$ is the magnitude of the applied magnetic field, $\sigma$ is the electrical conductivity, $\rho$ is the mass density and $\eta$ is the magnetic diffusivity. The Rayleigh number $Ra$ and the Hartmann number $Ha$ quantify the relative strength of the buoyancy and Lorentz forces, respectively, whilst the Prandtl number $Pr$ is the ratio of momentum to thermal diffusivity, corresponding to the working fluid. A key distinction between directions of study within this field is the value of the magnetic Reynolds number $Re_m$. For example, astrophysical flows typically involve high $Re_m$, leading to flows dominated by induced fields and magnetic reconnection \citep{doi:10.1126/sciadv.abn7627}. Conversely, industrially relevant liquid metal flows, such as those in fusion breeder blankets, typically have low $Re_m$, leading to negligible induced fields but still significant damping of the flow variables \citep{Pothérat_Dymkou_2010}. This work focuses on the latter `quasi-static' regime, with $Re_m \ll 1$, where the Lorentz force acts mainly as an additional dissipation mechanism. A key distinction between viscous and Joule dissipation, however, is that Joule dissipation acts anisotropically, favourably damping field-parallel gradients of flow variables. This leads to the presence of thin Hartmann layers, the damping of small-scale structures and the elongation of flow structures along the field direction. For large enough $Ha$, the flow becomes quasi-two-dimensional (Q2D) where there is notably less variation in flow variables along the field direction \citep{sommeria1982and}.


There are numerous current directions of study in MC at low $Re_m$. For example, much work is dedicated to closing global scaling relations for the Nusselt number, $Nu$, and the Reynolds number, $Re$ as functions of the governing dimensionless groups, $Nu=Nu(Ra,Pr,Ha)$ and $Re=Re(Ra,Pr,Ha)$. On this topic, \citet{Teimurazov_McCormack_Linkmann_Shishkina_2024} provides a unifying heat transfer model for Rayleigh-Bénard convection under the influence of a wall-normal field, while \citet{Chen_Yang_Ni_2024} experimentally studies scaling in the uniform wall-parallel field case. The general decline in $Nu$ with increasing $Ha$ is also supported by numerous DNS studies \citep{yan2019heat, liu2018wall}. More work is needed to quantify the influence of parameters such as flow geometry, wall-conductivity and magnetic field temporal and spatial dependence on scaling in MC, as well as exploring more of the $(Ra,Pr,Ha)$ parameter space. Despite the overall decline in $Nu$, the presence of large magnetoconvective fluctuations (MCFs) has been reported in forced convection simulations \citep{BELYAEV2021106773}, suggesting that, counter-intuitively, the Lorentz force can lead to intermittency, appearing in the form of a greater temperature variance within the flow. Other work focuses on turbulence modelling in the quasi-static regime \citep{wilson2015application, Fan_Chen_Ni_2025}. A challenge in this area is the failure of many traditional models to account for the anisotropy of the flow. Additionally, challenges exist in the closure of magnetohydrodynamic (MHD)-specific terms such as the electric field-velocity covariance, where no consistent generalised model exists. The lack of MHD DNS data also remains a challenge in MC, making it difficult to validate models. Other approaches aim to understand MC from the perspective of instability \citep{McCormack_Teimurazov_Shishkina_Linkmann_2023}. There are also fusion blanket relevant studies \citep{FRANKLIN2025115128}. 

We note that most MC studies so far have focused on global properties such as $Nu$, but the literature is missing studies providing a detailed statistical description of magneto-convective phenomena. When studying MC, much inspiration can be taken from previous influential studies of the non-MHD natural convection. An example of a prior study is that of \citet{togni2015physical}, which provided a detailed physical- and scale-space description of the classic Rayleigh-Bénard convection (RBC). To the authors' knowledge, such a detailed statistical description is missing from the MC literature despite the fact that this has the potential to provide significant physical insight.

The aim of this paper is to provide a physical- and scale-space description of magneto-convection in the RBC setup, extending the approach of \citet{togni2015physical} by including the effects of MHD into the analysis. This involves analysing the behaviour of thermal plumes from the perspective of single-point statistics, such as turbulent kinetic energy and temperature variance, and then extending this approach to the scale-space by studying exact equations relating the structure functions of velocity and temperature \citep{hill2002exact}. This approach gives insight into the regenerative nature of buoyancy-driven turbulence and how the presence of the Lorentz force influences this behaviour. This analysis has been conducted for MC with both wall-normal and wall-parallel fields, leading to drastically different phenomena. The aim here is to describe qualitatively observed MC phenomena in terms of long-term energy budgets. This has the potential to offer a more complete theoretical understanding of MC; such insights are expected to prove useful to future modelling efforts.

The structure of this paper is as follows. Section \ref{sec:theory} introduces the governing equations, numerical methods, and simulation parameters. Section \ref{sec:topology} then provides instantaneous snapshots of the flow fields, highlighting qualitatively the influence of the magnetic field on the thermal coherent structures for both field orientations. In section \ref{sec:physbudgets} the long-term statistical behaviour of the flow is analysed from the perspective of turbulent kinetic energy and temperature variance budgets with respect to the inhomogeneous wall-normal coordinate. Then, in section \ref{sec:hill}, this analysis is extended to the space of scales from the perspective of budgets relating second and third order structure functions of velocity and temperature \citep{hill2002exact}. A key feature of this section is studying the field-normal and field-parallel structure functions separately, highlighting and quantifying anisotropy present within the flow. Finally, in section \ref{sec:conclusion}, conclusions, final remarks and opportunities for future work are highlighted.

\section{Governing equations and numerical method}\label{sec:theory}

Inductionless magnetoconvection in the RBC system is governed by transport equations describing conservation of momentum, mass and energy,

\begin{subequations}
\label{eq:NS}
\begin{equation}
    \label{eq:momentum}
    \frac{\partial u_i}{\partial t}
    + u_j \frac{\partial u_i}{\partial x_j}
    = -\frac{\partial p}{\partial x_i}
    + \sqrt{\frac{Pr}{Ra}} \left(\frac{\partial^2u_i}{\partial x_j \partial x_j}
    + Ha^2 \varepsilon_{ijk}J_jB_k \right)
    + \theta g_i,
\end{equation}

\begin{equation}
    \label{eq:continuity}
    \frac{\partial u_i}{\partial x_i} = 0,
\end{equation}

\begin{equation}
    \label{eq:energy}
    \frac{\partial \theta}{\partial t}
    + u_j \frac{\partial \theta}{\partial x_j} 
    = \frac{1}{\sqrt{PrRa}} \frac{\partial^2 \theta}{\partial x_j \partial x_j},
\end{equation}
\end{subequations}
where the governing variables are the velocity $u_i$, the pressure $p$, the temperature $\theta$, the current density $J_i$ and the magnetic field $B_i$. The gravitational field vector is $g_i=\delta_{i2}$. The displacement and time are $x_i$ and $t$ respectively and $i,j,k = 1,2,3$ with $\delta_{ij}$ the Kronecker delta and $\varepsilon_{ijk}$ the Levi-Cevita symbol. Throughout this paper, the following notation will be used: $u_1, u_2, u_3 = u, v, w$; $x_1, x_2, x_3 = x, y, z$; $J_1, J_2, J_3 = J_x, J_y, J_z$ and $B_1, B_2, B_3 = B_x, B_y, B_z$. As written, equations (\ref{eq:momentum}-\ref{eq:energy}) are in non-dimensional form. The relevant units are the free-fall velocity, $u_f = \sqrt{g \beta  \Delta \theta H}$, the wall separation distance $H$, the wall temperature difference $\Delta \theta$, the free-fall time $\tau_f = H/u_f$, the magnitude of the applied magnetic field $B_0$ and the free-fall current $J_f=\sigma u_fB_0$. 

In order to close equations (\ref{eq:momentum}-\ref{eq:energy}), a relationship is required for $J_i$. It is assumed that the flow is inductionless, implying that the Lorentz force influences the flow but the magnetic field induced by the flow is negligible. Therefore, the applied field, $B_i$, remains constant. The regime of validity of this inductionless approximation is low $Re_m$, $Re_m = u_f H / \eta \ll 1$. The closure for $J_i$ comes in the form of Ohm's law,

\begin{equation}
    \label{eq:ohms}
    J_i = -\frac{\partial \phi}{\partial x_i}
    + \varepsilon_{ijk}u_jB_k,
\end{equation}

\noindent where $\phi$ is the electric potential which has been non-dimensionalised by $\phi_f= \sigma u_f B_0 H$. Asserting conservation of current within the system, by requiring $\partial J_i / \partial x_i =0$, yields a Poisson equation for the electric potential,

\begin{equation}
    \label{eq:poisson}
    \frac{\partial^2 \phi}{\partial x_i \partial x_i}
    = \frac{\partial}{\partial x_i}
    (\varepsilon_{ijk}u_jB_k),
\end{equation}

\noindent closing the system of equations.

\begin{figure}
    \centering
    \includegraphics[width= 0.9\linewidth]{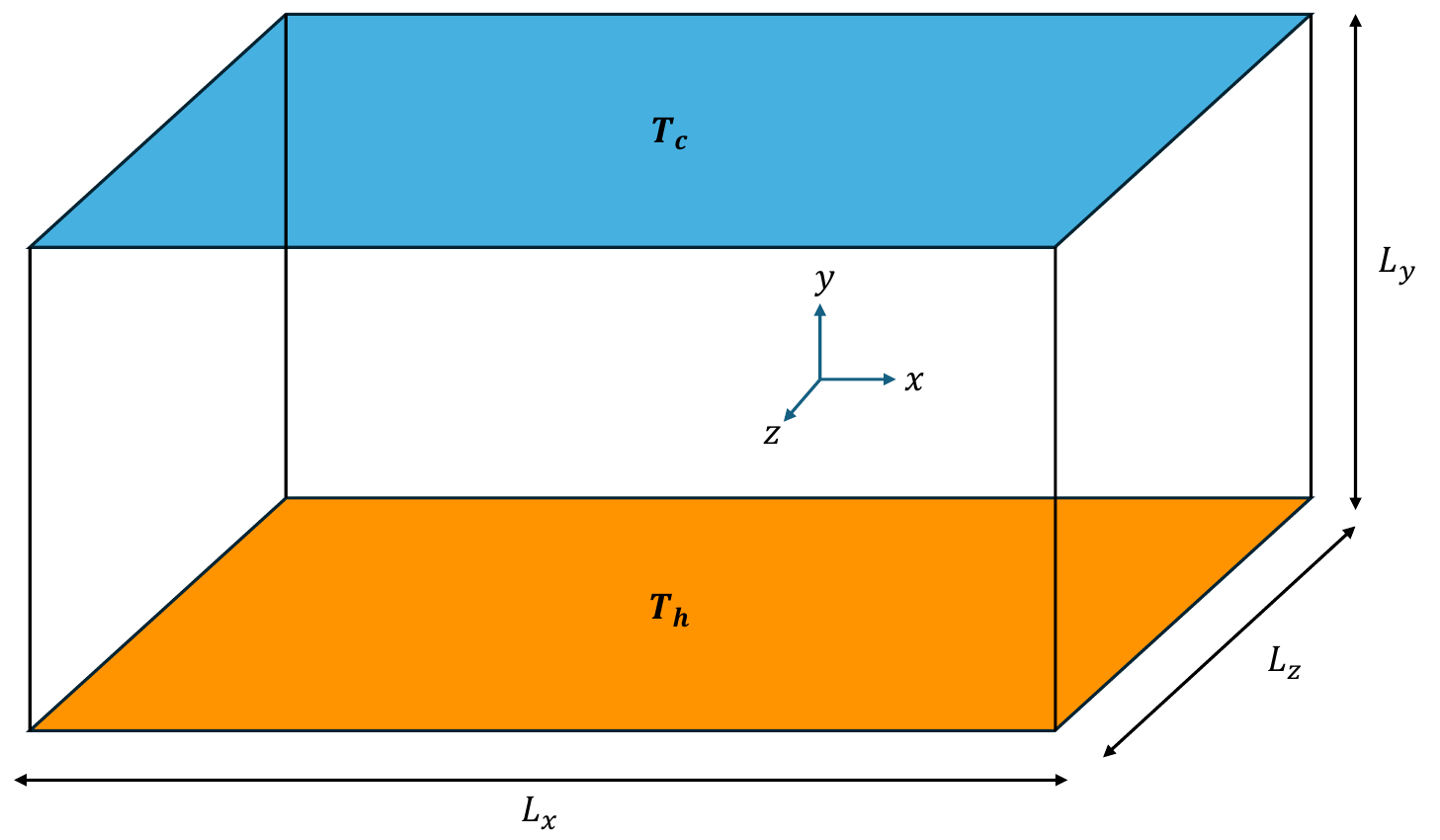}
    \caption{Simulation domain setup where the walls are held at $T_h=0.5$ and $T_c = -0.5$. Both fields along the $y$ and $x$ axis are considered in this setup.}
    \label{fig:case_setup}
\end{figure}

The system of equations (\ref{eq:momentum}-\ref{eq:energy}) is solved using Incompact3d, an open-source high-fidelity DNS solver \citep{bartholomew_2020}. The domain is discretised by a 6th order compact finite difference scheme  and time advancement is handled using a fractional step method, where incompressibility is enforced intermediately by solving a pressure Poisson equation in Fourier space through a modified wave number approach \citep{laizet2009high}. The mesh is Cartesian with stretching in a single direction as also described in \citet{laizet2009high}. The parallelisation strategy is a 2D pencil decomposition \citep{Li20102DECOMPFFTA, laizet2011incompact3d} and the simulations were conducted using the ARCHER2 UK National Super Computing Service. The code has been validated rigorously for flows involving buoyancy forces \citep{guo2024turbulence, fregni2022direct, VICENTECRUZ2024109640} and more recently, MHD has been implemented into the software, and validated \citep{FANG2025109400}.

\begin{table}
  \begin{center}
\def~{\hphantom{0}}
  \begin{tabular}{lccccccc}
      Case   & Field Direction & $Ha$ & $L_x \times L_y \times L_z$ &  $N_x \times Ny \times Nz$    & $\Delta t$ & $T$ & $\Delta \tau$ \\[3pt]
       S     & None            &  ~0  & $8 \times 1 \times 8$       &  $256 \times 151 \times 256$  & $10^{-4}$  & 400 & 2             \\
       A     & $x$             &  20  & $12 \times 1 \times 12$     &  $384 \times 151 \times 384$  & $10^{-4}$  & 400 & 2             \\
       B     & $x$             &  40  & $12 \times 1 \times 12$     &  $384 \times 151 \times 384$  & $10^{-4}$  & 400 & 2             \\
       C     & $x$             &  80  & $12 \times 1 \times 12$     &  $384 \times 151 \times 384$  & $10^{-4}$  & 400 & 2             \\
       D     & $y$             &  20  & $8  \times 1 \times 8$      &  $256 \times 151 \times 256$  & $10^{-4}$  & 400 & 2             \\           
       E     & $y$             &  40  & $8  \times 1 \times 8$      &  $256 \times 151 \times 256$  & $10^{-4}$  & 400 & 2             \\           
       F     & $y$             &  80  & $8  \times 1 \times 8$      &  $256 \times 151 \times 256$  & $10^{-4}$  & 400 & 2             \\
  \end{tabular}
  \caption{A summary of the cases simulated and their respective parameters. Case S is a non-MHD case in the same setup, for the purpose of comparison. A mesh convergence, with doubled $N_y$, was carried out for case E.}
  \label{tab:sim_params}
  \end{center}
\end{table}

\begin{table}
  \begin{center}
\def~{\hphantom{0}}
  \begin{tabular}{lcccc}
      Case & $\langle \partial \theta/ \partial y \rangle_{y=0,H}$ & $1 + \sqrt{RaPr} \langle v'\theta'\rangle_V$ & $1 + \sqrt{RaPr}\langle \varepsilon + \varepsilon_J \rangle_V$ & $\sqrt{RaPr}\langle \chi \rangle_V$\\[3pt] 
       S    & 8.31                & 8.31                                         & 8.29                                           & 8.28    \\
       A    & 7.39                & 7.39                                         & 7.39                                           & 7.39    \\
       B    & 6.87                & 6.88                                         & 6.88                                           & 6.87    \\  
       C    & 7.33                & 7.33                                         & 7.35                                           & 7.32    \\  
       D    & 8.12                & 8.11                                         & 8.10                                           & 8.10    \\      
       E    & 7.36                & 7.36                                         & 7.36                                           & 7.35    \\     
       F    & 5.92                & 5.91                                         & 5.92                                           & 5.91    \\   
  \end{tabular}
  \caption{A table containing details on verification (equations \ref{eq:Nu_1}-\ref{eq:Nu_2}) and the Nusselt number values, for all simulations.}
  \label{tab:nu_values}
  \end{center}
\end{table}

Cases with wall-normal and wall-parallel magnetic fields are considered separately and this study is limited to cases with electrically insulating walls. This case setup is visualised in figure \ref{fig:case_setup}. The physical domain size is $L_x \times L_y \times L_z$ which is discretised by a Cartesian mesh of size $N_x \times N_y \times N_z$. Mesh refinement is applied close to the walls, with $\beta=0.29$, where $\beta$ is the stretching parameter described in \citep{laizet2009high}. The time-step is $\Delta t$. The flow is initially stationary and a linear temperature gradient is imposed between the two walls. The flow then undergoes an initial transient period for a time $\tau_i$, after which it reaches a statistically stationary state. The flow statistics in the subsequent sections are all gathered once this initial transient has vanished. Statistics are gathered for a time period of $T$ every $\Delta \tau$ units of time. All details of the simulations parameters can be found in table \ref{tab:sim_params}. The larger domain sizes for cases A-C was motivated by \textit{a posteriori} based upon much spatially larger observed thermal plumes. The choice of $T$ was selected to achieve convergence given the long time-scales associated with MC. Naturally, the focus of this paper will be on cases A-F. The non-MHD case, case S, was conducted for the purpose of benchmarking and comparison, useful for further highlighting the influence of the magnetic field upon the flow.

To ensure the validity, accuracy and convergence of the results a number of measures are taken. First of all, the following consistency relations for $Nu$ are used to verify the simulations, 

\begin{subequations}
\label{eq:nu_relations}
\begin{equation}
      Nu = \left\langle \frac{\partial \theta}{\partial y} \right \rangle_{y=0,H}
     \label{eq:Nu_1}
\end{equation}
\begin{equation}
     Nu = 1 + \sqrt{Ra Pr} \langle v' \theta' \rangle_V
     \label{eq:Nu_2}
\end{equation}

\begin{equation}
     Nu  = 1 +\sqrt{RaPr} \langle \varepsilon +\varepsilon_J \rangle_V
     \label{eq:Nu_3}
\end{equation}

\begin{equation}
     Nu = \sqrt{PrRa}\langle \chi \rangle_V
    \label{eq:Nu_4}
\end{equation}
\end{subequations}

\noindent details of which can be found in table \ref{tab:nu_values}. Here, $\langle \cdot \rangle_V$ is the volume average, $\varepsilon=\partial u_i / \partial x_j\partial u_i / \partial x_j$ is the pseudo-dissipation of turbulent kinetic energy, $\chi = \partial T / \partial x_j\partial T / \partial x_j$ is the dissipation of temperature variance and $\varepsilon_J=Ha^2 \sqrt{Pr/Ra} \varepsilon_{ijk} u_iJ_jB_k$ is the work done by the Lorentz force. When accounting for MHD, equations \ref{eq:Nu_1}-\ref{eq:Nu_4} are the same as initially proposed in \citet{VERZICCO_CAMUSSI_2003}. That is with the exception of equation \ref{eq:Nu_3} where $\varepsilon_J$ must be added to account for the extra dissipation due to the presence of the magnetic field. Along with these identities, the conservation of current, $\partial J_i/\partial x_i=0$, is also enforced to machine accuracy through the modified wave-number approach, as detailed in \citet{FANG2025109400}.

Secondly, the temporal and spatial resolution of the flow is checked \textit{a posteriori} by calculating the Kolmogorov length- and time-scales, $\eta_k$ and $\tau_k$ respectively, and comparing it to the mesh spacing and time-step. Details of these calculations can be found in table \ref{tab:kolg_values}. With regards to these scales, the strictest resolution requirements are for case S. However, with the presence of a magnetic field, the smallest length scales are likely associated with the velocity boundary layer, which thins as $Ha$ increases. A mesh-convergence study, with doubled wall-normal resolution, is conducted for case E. This shows acceptable convergence on the Nusselt number and the energy dissipation values. With regards to timescales, MHD introduces the Joule damping time, $\tau_J = \sqrt{Ra / Pr} Ha^{-2}$. For the worst case scenario with $Ha=80$, $\tau_J \simeq 10^{-1}$, showing that similar to the Kolmogorov timescales, the MHD timescales are also greatly over-resolved.

Finally, the statistical convergence of the data can be observed in section \ref{sec:physbudgets}, where all of the budget equations studied balance to a high degree of accuracy. 

\begin{table}
  \begin{center}
\def~{\hphantom{0}}
  \begin{tabular}{lccc}
      Case   &  $\max(\Delta y /\eta_k)$ & $\max(\Delta x / \eta_k)$ & $\max(\Delta t/\tau_k)$\\[3pt]
       S     & 0.56                      & 2.61                      & 0.0006                 \\
       A     & 0.47                      & 2.40                      & 0.0005                 \\
       B     & 0.41                      & 2.45                      & 0.0006                 \\
       C     & 0.39                      & 2.82                      & 0.0008                 \\
       D     & 0.49                      & 2.32                      & 0.0005                 \\           
       E    & 0.40                      & 2.06                      & 0.0004                 \\           
       F     & 0.33                      & 1.83                      & 0.0003                 \\
  \end{tabular}
  \caption{A comparison between the mesh spacings and timesteps, and their respective Kolmogorov microscales, $\eta_k$ and $\tau_k$, for each simulation.}
  \label{tab:kolg_values}
  \end{center}
\end{table}

\section{Flow snapshots and coherent structures}\label{sec:topology}



\begin{figure}[htbp]
  \centering
  \includegraphics[width=\linewidth]{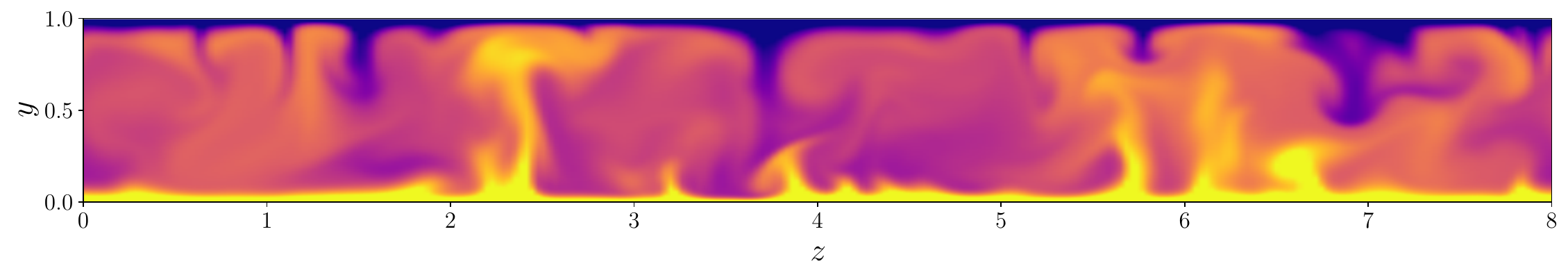}
  \includegraphics[width=\linewidth]{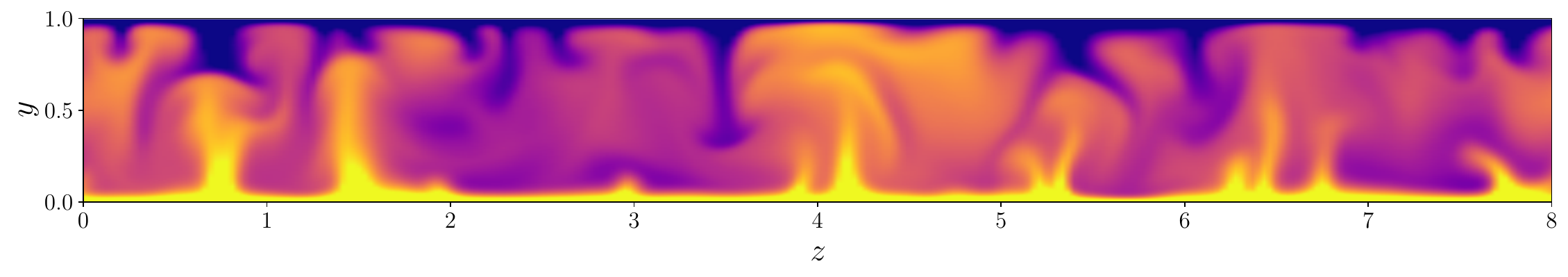}
  \includegraphics[width=\linewidth]{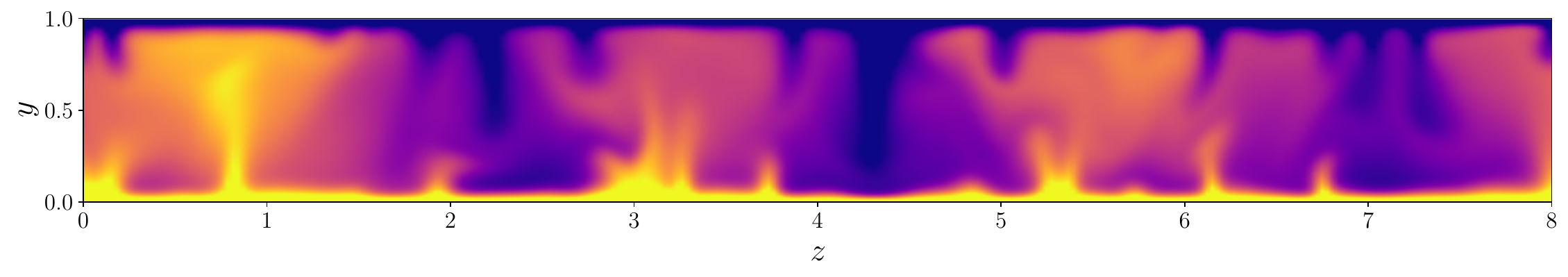}  \includegraphics[width=\linewidth]{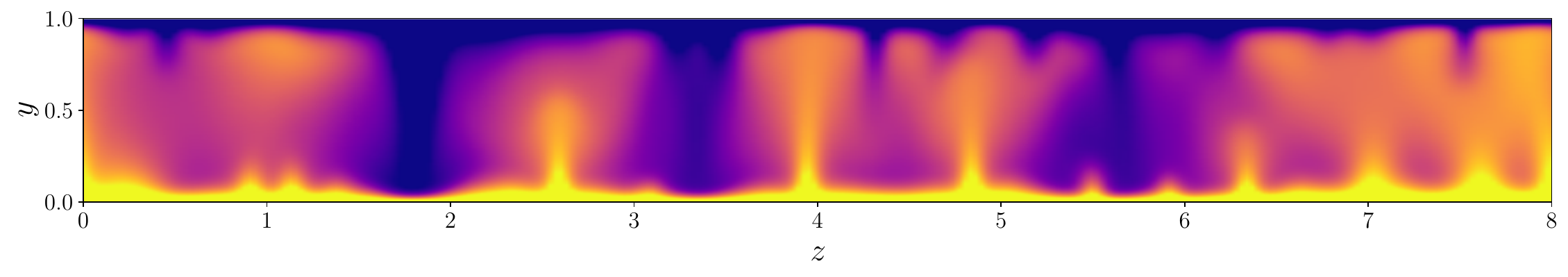}
  \includegraphics[width=0.7\linewidth]{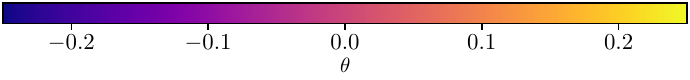}
  \caption{Instantaneous temperature plots in the $y-z$ plane with a wall-normal magnetic field. (Top to bottom) cases S, D, E and F respectively. $\textbf{g} \downarrow \textbf{B} \uparrow$.}
  \label{fig:yz_temperature_plots_hay}
\end{figure}
\begin{figure}[htbp]
  \centering
  \includegraphics[width=\linewidth]{figures/temperature_yz_ha0.pdf}
  \includegraphics[width=\linewidth]{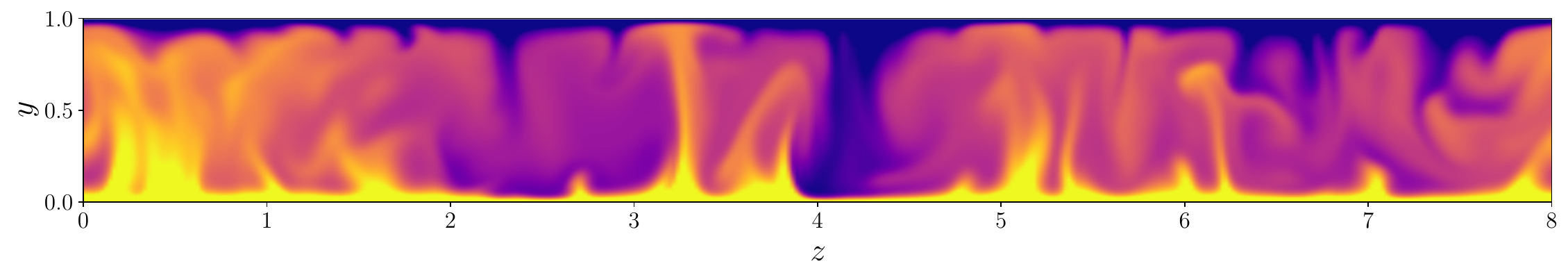}
  \includegraphics[width=\linewidth]{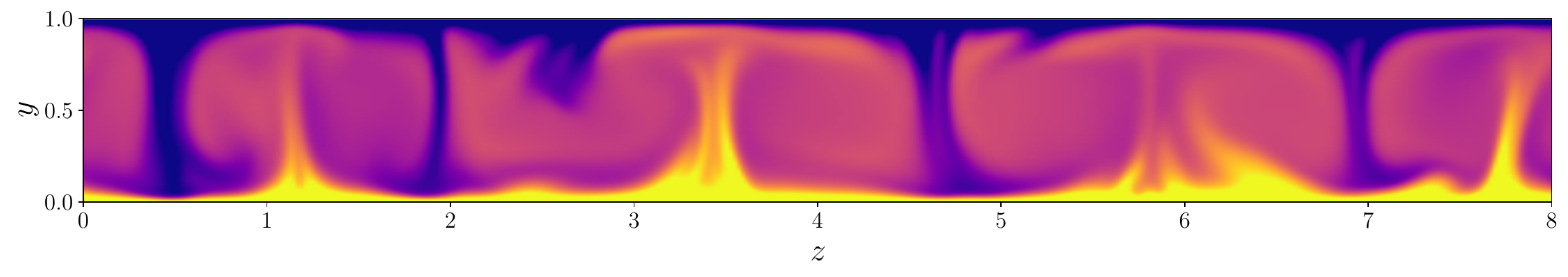}  
  \includegraphics[width=\linewidth]{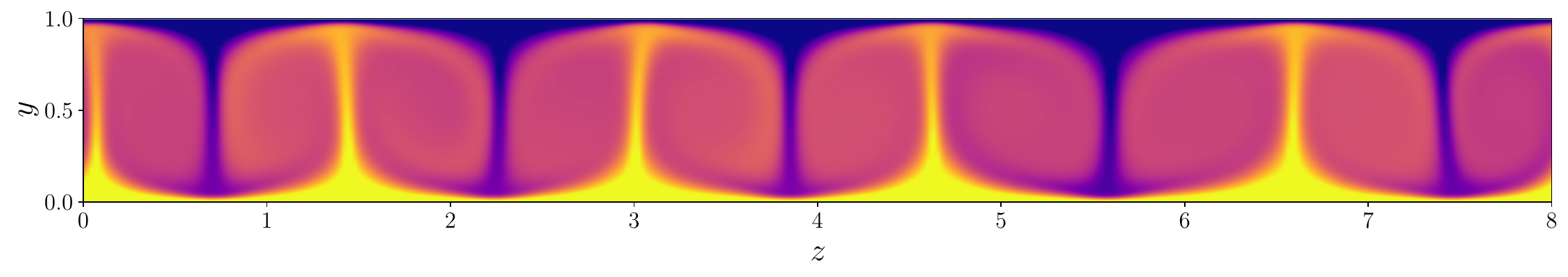}
  \includegraphics[width=0.6\linewidth]{figures/temperature_colorbar.pdf}
  
  \caption{Instantaneous temperature plots in the $y-z$ plane with a wall-parallel magnetic field. (Top to bottom) cases S, D, E and F respectively. $\textbf{g} \downarrow \textbf{B} \otimes$.}
  \label{fig:yz_temperature_plots_hax}
\end{figure}
\begin{figure}[htbp]
  \centering
  \includegraphics[width=\linewidth]{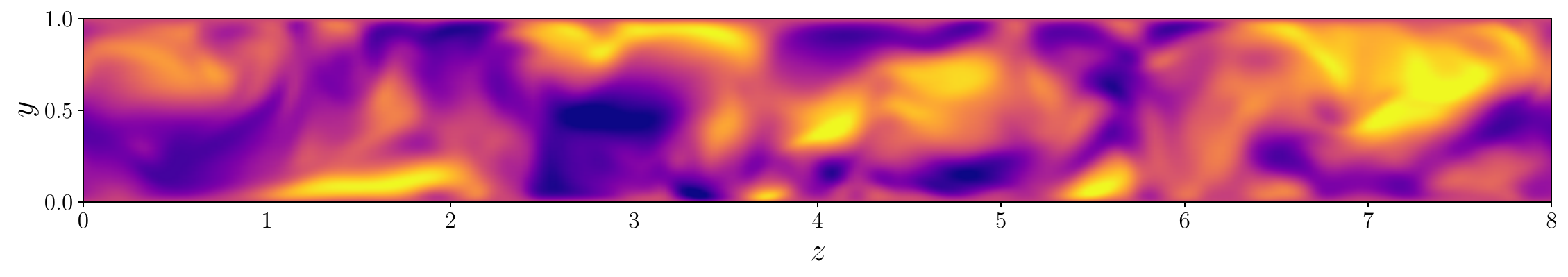}
  \includegraphics[width=\linewidth]{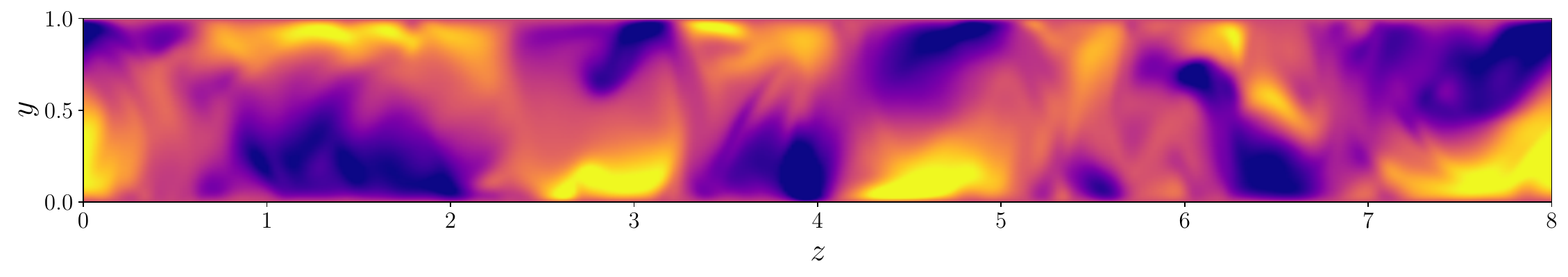}
  \includegraphics[width=\linewidth]{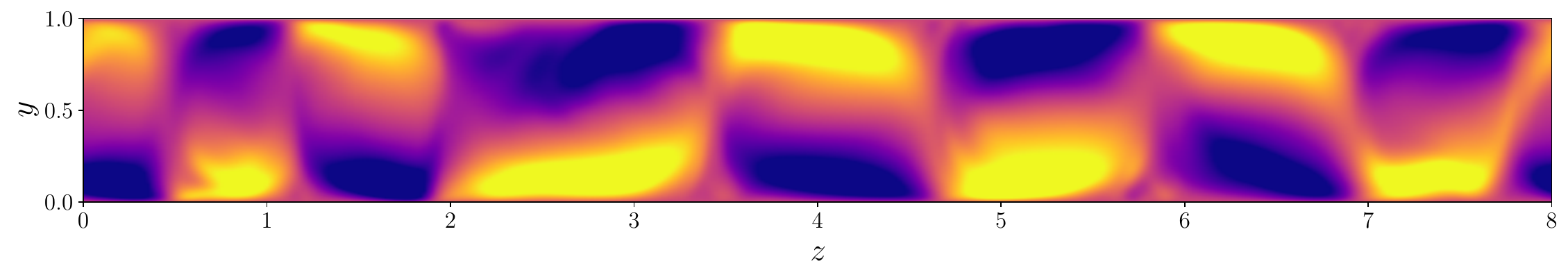}  \includegraphics[width=\linewidth]{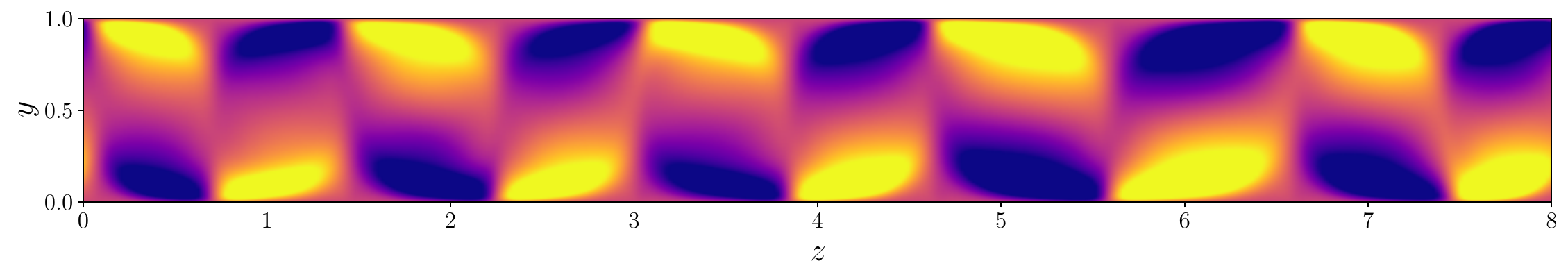}
  \includegraphics[width=0.6\linewidth]{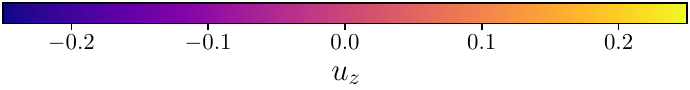}
  \caption{Instantaneous $w$ plots in the $y-z$ plane with a wall-parallel magnetic field. (Top to bottom) cases S, A, B and C respectively. $\textbf{g} \downarrow \textbf{B} \otimes$.}
  \label{fig:yz_uz_plots_hax}
\end{figure}
\begin{figure}[htbp]
  \centering
  \includegraphics[width=\linewidth]{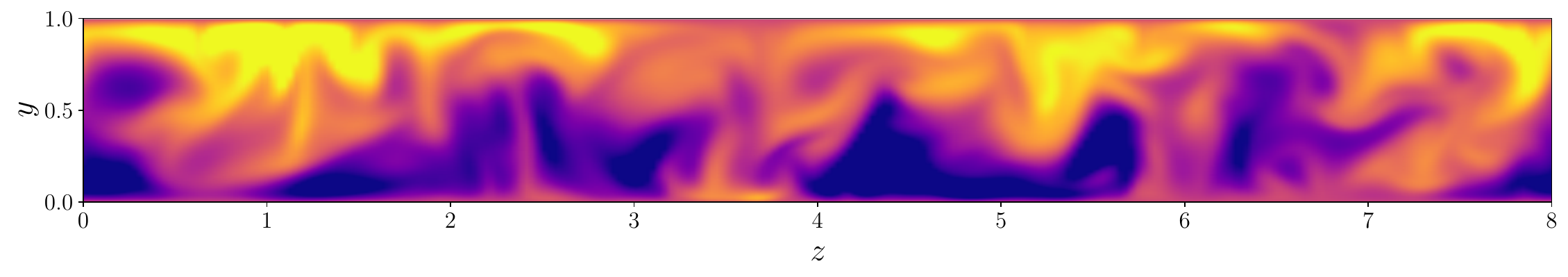}
  \includegraphics[width=\linewidth]{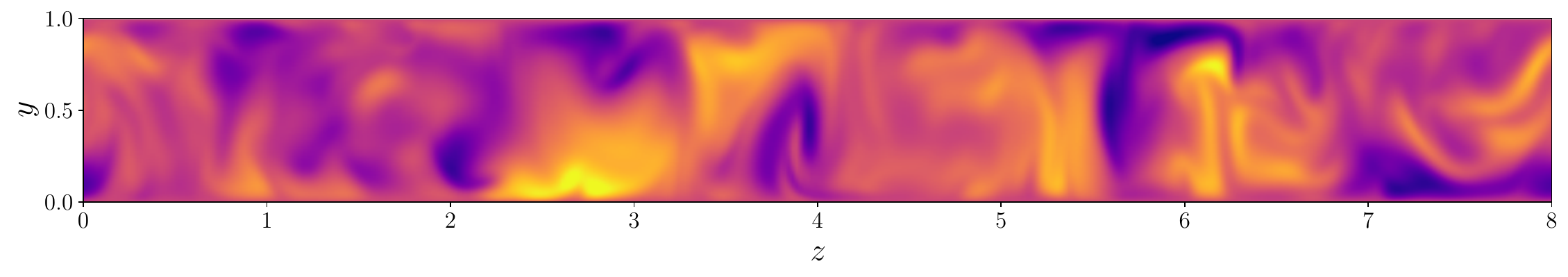}
  \includegraphics[width=\linewidth]{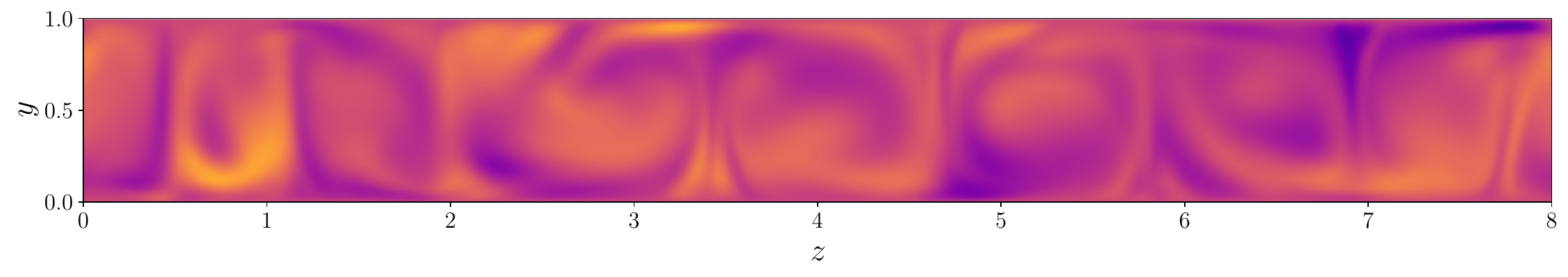}  \includegraphics[width=\linewidth]{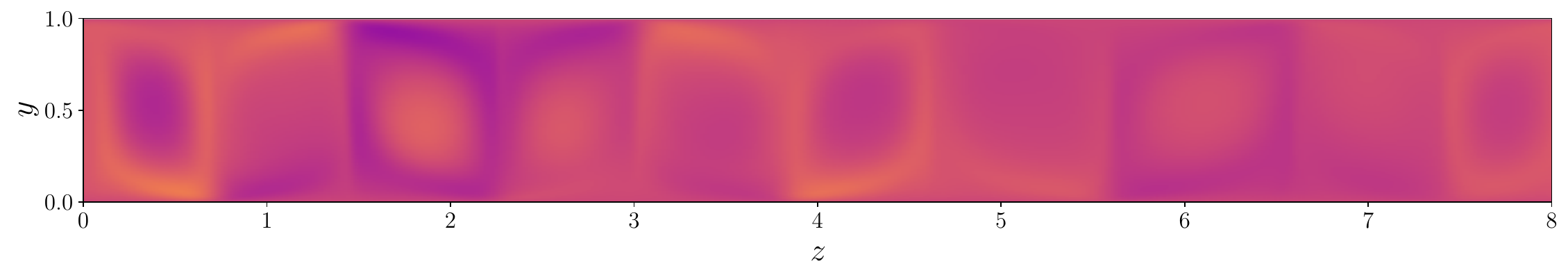}
  \includegraphics[width=0.6\linewidth]{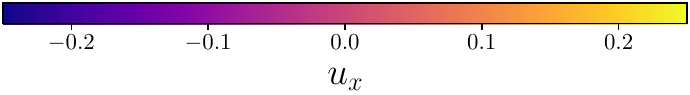}
  \caption{Instantaneous $u$ plots in the $y-z$ plane with a wall-parallel magnetic field. (Top to bottom) cases S, A, B and C respectively. $\textbf{g} \downarrow \textbf{B} \otimes$.}
  \label{fig:yz_ux_plots_hax}
\end{figure}
\begin{figure}[htbp]
  \centering
  \begin{minipage}[b]{0.48\linewidth}
    \includegraphics[width=\linewidth]{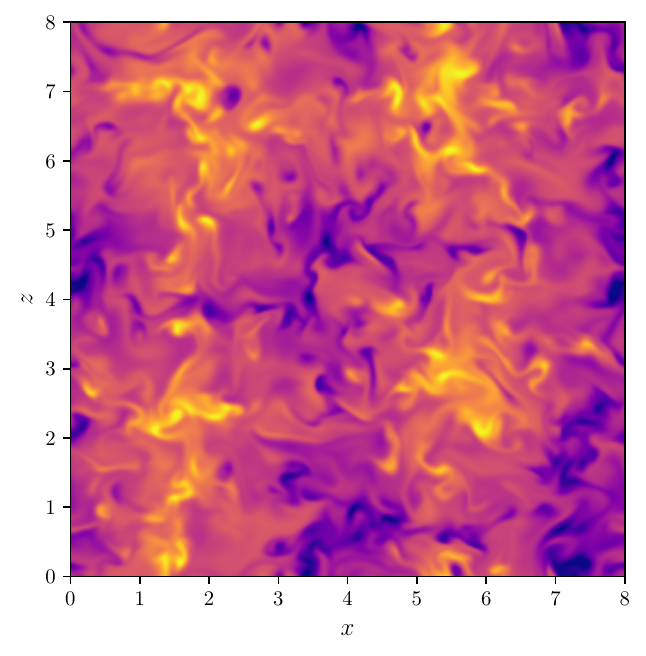}
  \end{minipage}%
  \hspace{0.03\linewidth}
  \begin{minipage}[b]{0.48\linewidth}
    \includegraphics[width=\linewidth]{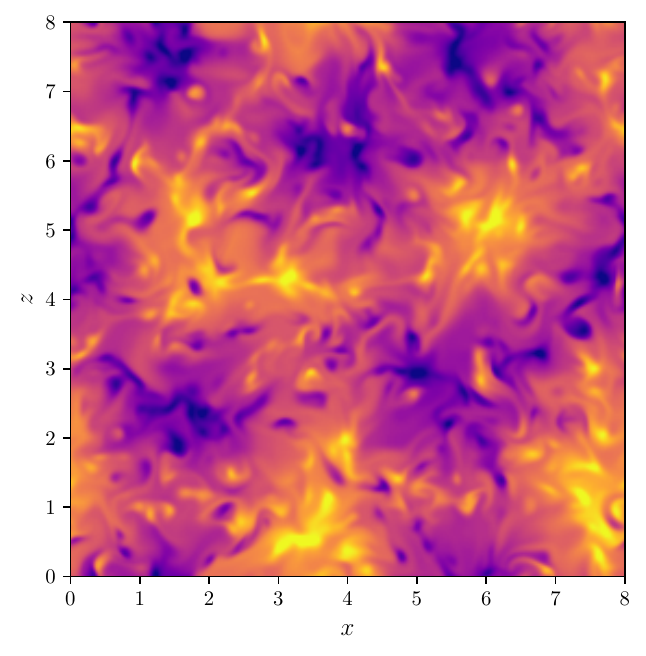}
  \end{minipage}

  \vspace{1ex}

  \begin{minipage}[b]{0.48\linewidth}
    \includegraphics[width=\linewidth]{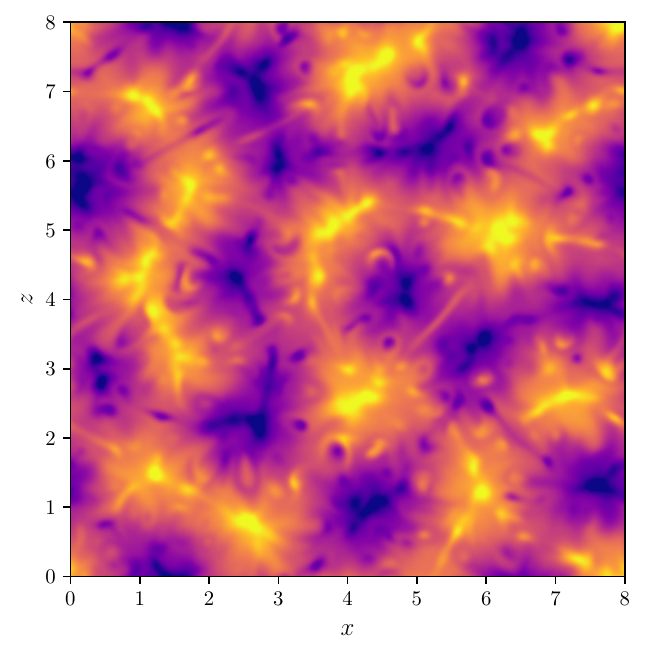}
  \end{minipage}%
  \hspace{0.03\linewidth}
  \begin{minipage}[b]{0.48\linewidth}
    \includegraphics[width=\linewidth]{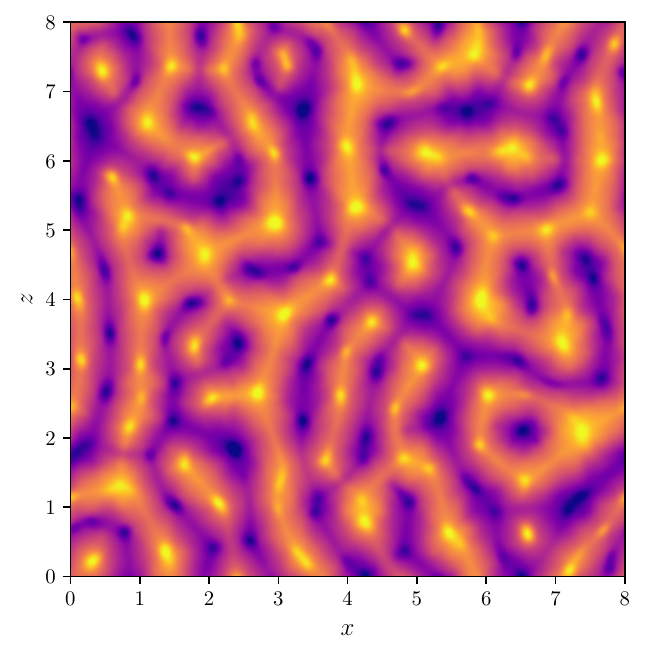}
  \end{minipage}

  \includegraphics[width=0.6\linewidth]{figures/temperature_colorbar.pdf}

  \caption{Instantaneous temperature in the $x\text{-}z$ plane for cases with a field in the $y$-direction: cases S (top-left), D (top-right), E (bottom-left), and F (bottom-right). $\textbf{g} \otimes \textbf{B} \odot$.}
  \label{fig:xz_temperature_plots_hay}
\end{figure}
\begin{figure}[htbp]
  \centering
  \begin{minipage}[b]{0.48\linewidth}
    \includegraphics[width=\linewidth]{figures/temperature_xz_ha0.pdf}
  \end{minipage}%
  \hspace{0.03\linewidth}
  \begin{minipage}[b]{0.48\linewidth}
    \includegraphics[width=\linewidth]{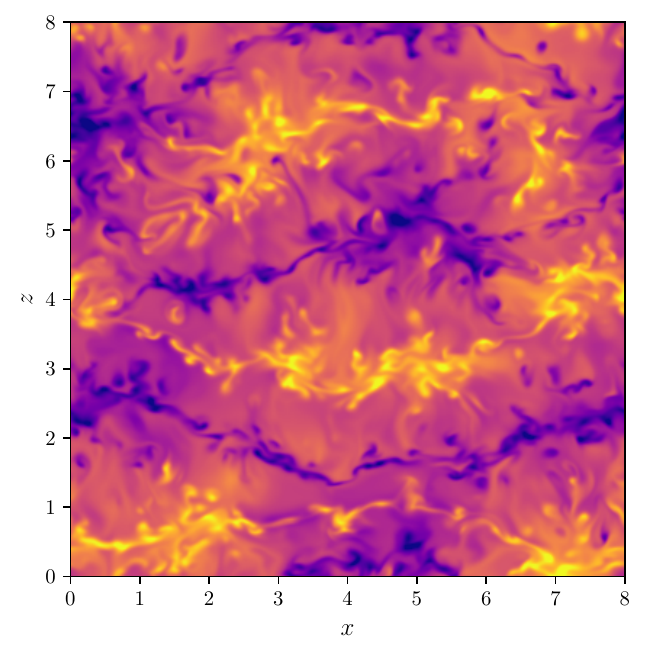}
  \end{minipage}

  \vspace{1ex}

  \begin{minipage}[b]{0.48\linewidth}
    \includegraphics[width=\linewidth]{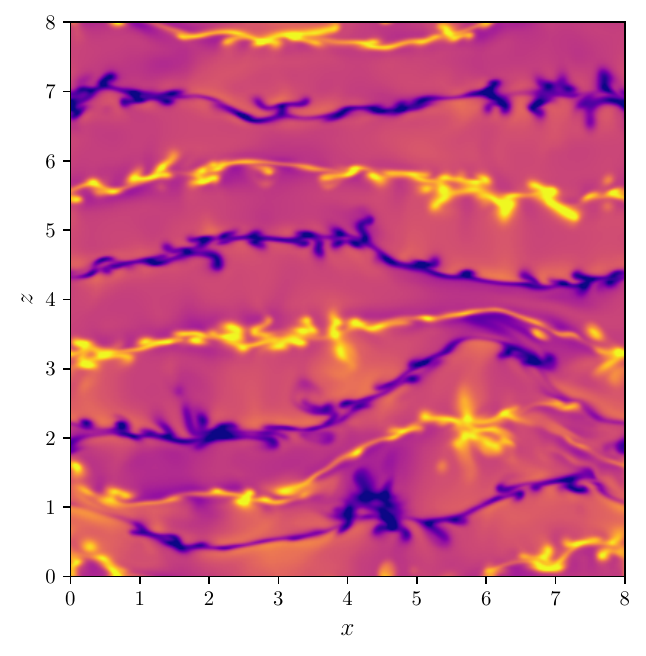}
  \end{minipage}%
  \hspace{0.03\linewidth}
  \begin{minipage}[b]{0.48\linewidth}
    \includegraphics[width=\linewidth]{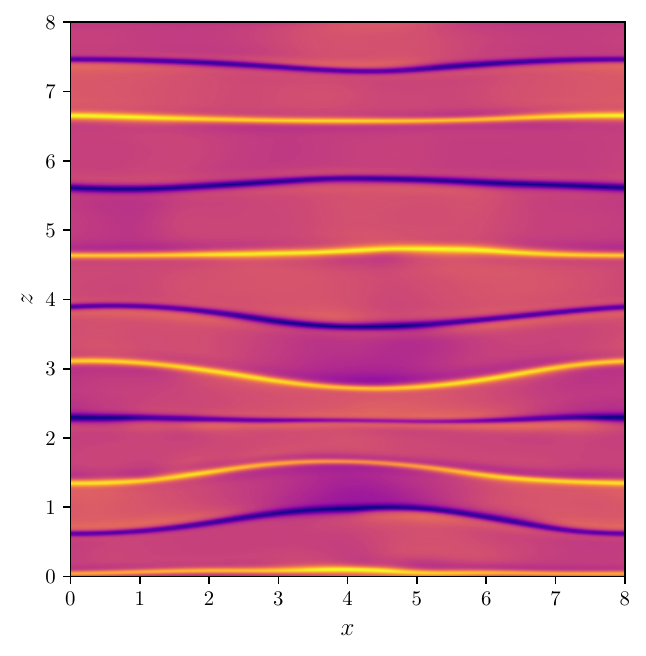}
  \end{minipage}

  \includegraphics[width=0.6\linewidth]{figures/temperature_colorbar.pdf}

  \caption{Instantaneous temperature in the $x\text{-}z$ plane for cases with a field in the $y$-direction: cases S (top left), A (top right), B (bottom left) and C (bottom right). $\textbf{g} \otimes \textbf{B} \rightarrow$.}
  \label{fig:xz_temperature_plots_hax}
\end{figure}

The influence of the magnetic field upon the thermal plumes and the velocity fields can be seen in figures \ref{fig:yz_temperature_plots_hay}-\ref{fig:xz_temperature_plots_hax}, showing drastic rearrangement depending upon both the field strength and orientation. Starting with the cases with a wall-parallel magnetic field (A,B,C), Q2D turbulence occurs where the flow variables become highly correlated along the field direction due to the Joule dissipation mechanism (figure \ref{fig:xz_temperature_plots_hax}). From the $y-z$ plane perspective (figure \ref{fig:yz_temperature_plots_hax}) the plumes closely resemble those of non-MHD convection, but with a notably wider spatial extent due to the presence of organised wall-parallel jets that form in close proximity to the wall (figure \ref{fig:yz_uz_plots_hax}). As $Ha$ increases, the $u$ velocity component is damped (figure \ref{fig:yz_ux_plots_hax}) whilst the $w$ velocity component remains similar in magnitude but organises into laminar jets. A likely cause of this is the interaction between the Lorentz force and the pressure-diffusion mechanism; as mentioned in \citet{togni2015physical} the role of the pressure diffusion mechanism is to redistribute the vertical velocity components to horizontal velocity components as the plumes approach the wall. However, in this case the velocity gradients in the field direction are hindered due to the Joule dissipation mechanism and it therefore becomes energetically favourable for the kinetic energy to be redistributed to the $z$ component. As a result, jets in $w$ form near the wall with significantly greater velocity than in the non-MHD case. It is also noted that the plumes feel the Lorentz force as they are accelerated through the bulk through buoyancy, but any excess kinetic energy that still remains as the plumes approach is favourably converted to the $w$ velocity component resulting in larger values of $w$ near the boundary, relative to $u$.

Moving onto the cases with wall-normal magnetic fields, full 3D flow structures are observed but are significantly warped by the presence of the magnetic field (figures \ref{fig:yz_temperature_plots_hay}, \ref{fig:xz_temperature_plots_hay}). In this case, the flow moves parallel to the field as the buoyancy accelerates it through the bulk meaning the Lorentz force does not act within the bulk region. However, as the plumes begin to impinge upon the wall, the flow crosses the magnetic field lines and the Lorentz force begins to act. As a result, the plumes are notably thinner due to the horizontal action of the Lorentz force which acts to inhibit the pressure diffusion mechanism. Since in this case the field is parallel to gravity, the Joule dissipation mechanism acts in direct competition with the buoyancy force. At sufficiently high $Ha$ one would expect no convective motion and purely conductive heat transport. This can be thought of as the critical Rayleigh number increasing with $Ha$ and is supported by the observed decline in $Nu$ with increasing $Ha$ (table \ref{tab:nu_values}).

\begin{figure}[htbp]
  \centering
  \begin{minipage}[b]{0.48\linewidth}
    \includegraphics[width=\linewidth]{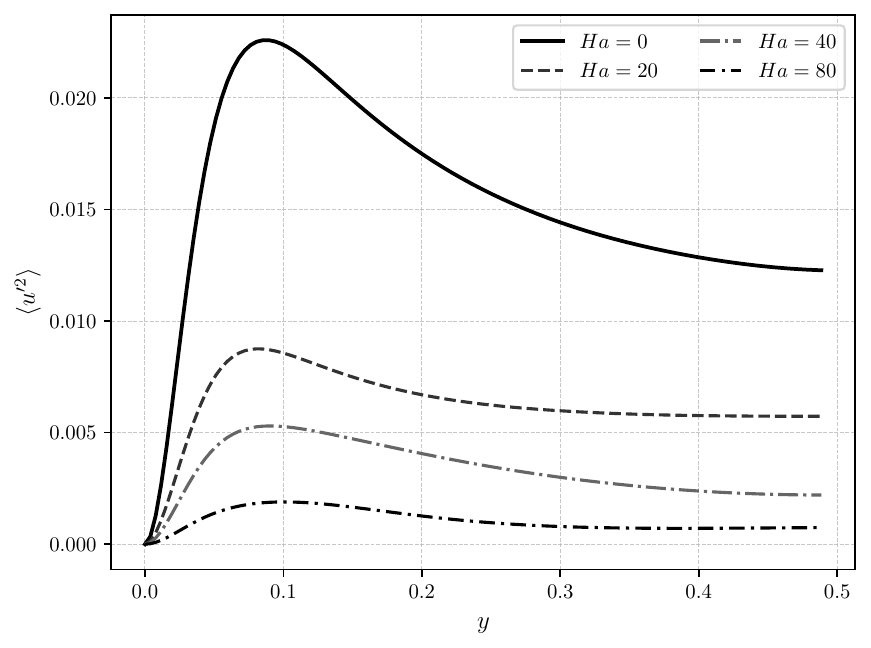}
  \end{minipage}%
  \hspace{0.03\linewidth}
  \begin{minipage}[b]{0.48\linewidth}
    \includegraphics[width=\linewidth]{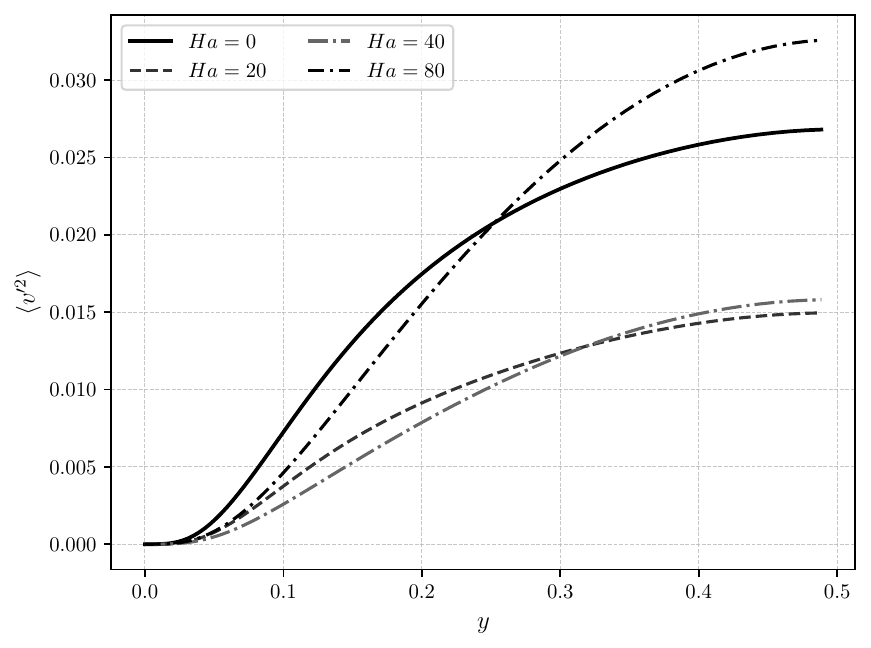}
  \end{minipage}

  \vspace{1ex}

  \begin{minipage}[b]{0.48\linewidth}
    \includegraphics[width=\linewidth]{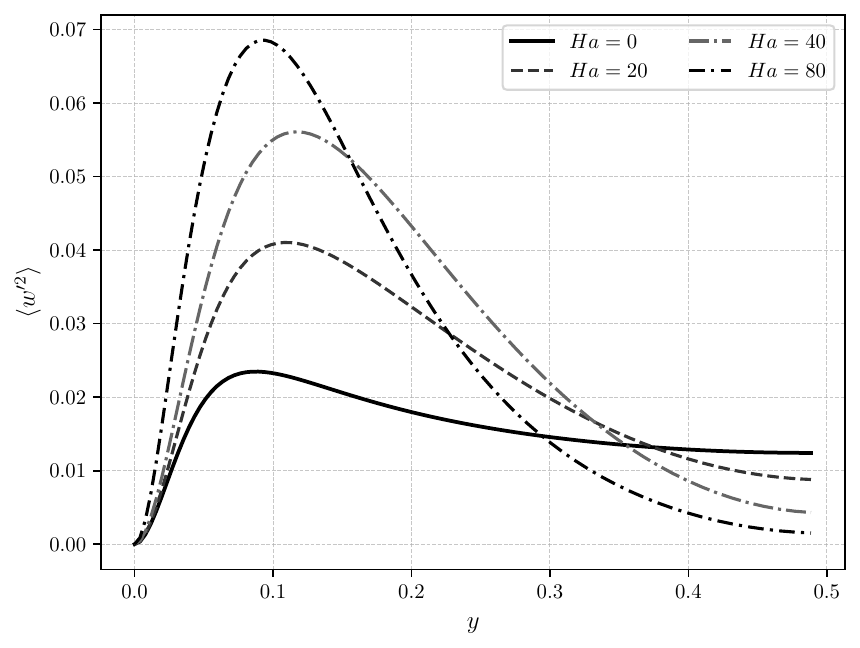}
  \end{minipage}%
  \hspace{0.03\linewidth}
  \begin{minipage}[b]{0.48\linewidth}
    \includegraphics[width=\linewidth]{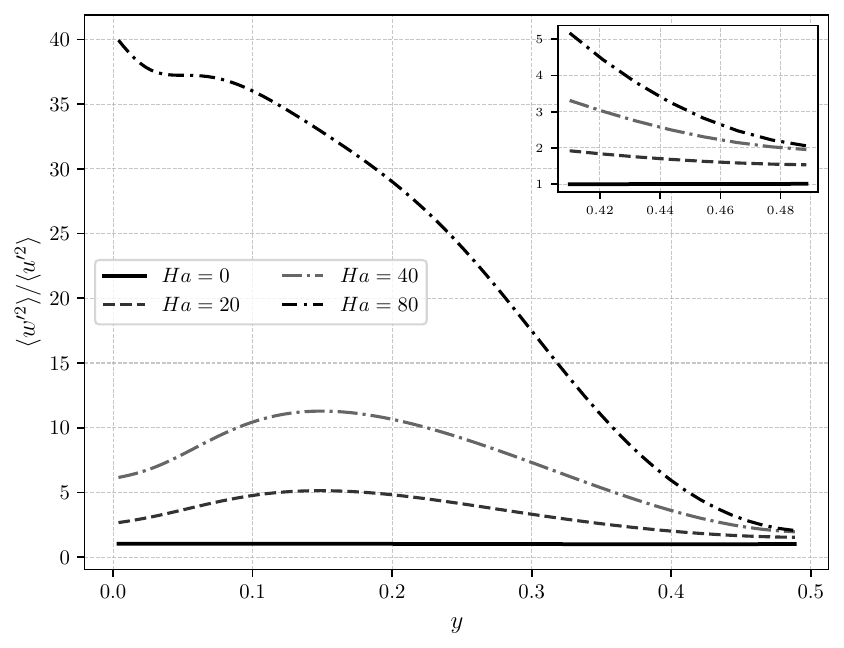}
  \end{minipage}

  \caption{$Ha$ dependence of second order velocity moments for cases A,B and C. (Top left) $\langle u'^2 \rangle$ (top right) $\langle v'^2 \rangle$, (bottom left) $\langle w'^2 \rangle$. Bottom right plots shows the ratio $\langle w'^2 \rangle / \langle u'^2\rangle$ describing the anisotropy of the velocity components where the divergence at $y=0$ is omitted.}
  \label{fig:velocity_variance_plots_hax}
\end{figure}

\begin{figure}[htbp]
  \centering
  \begin{minipage}[b]{0.48\linewidth}
    \includegraphics[width=\linewidth]{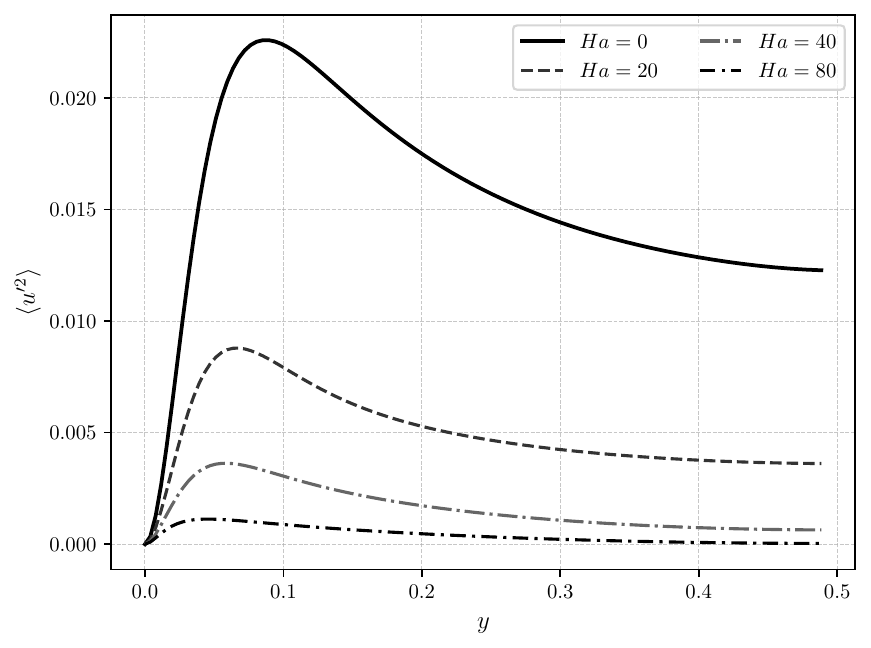}
  \end{minipage}%
  \hspace{0.03\linewidth}
  \begin{minipage}[b]{0.48\linewidth}
    \includegraphics[width=\linewidth]{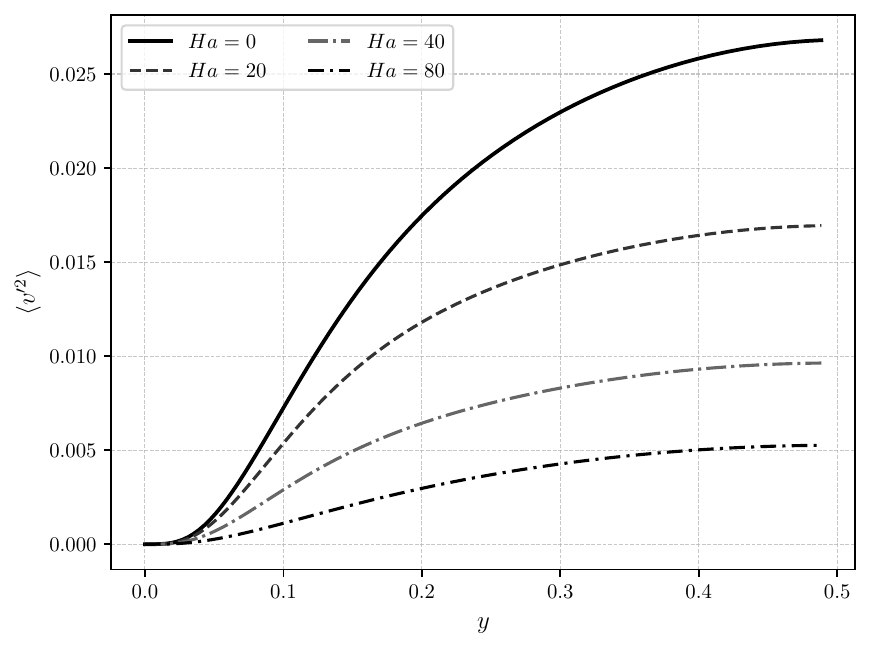}
  \end{minipage}

  \caption{$Ha$ dependence of second order velocity moments for cases D,E and F. (Left) $\langle u'^2 \rangle$ (right) $\langle v'^2 \rangle$.}
  \label{fig:velocity_variance_plots_hay}
\end{figure}

From the perspective of second-order velocity statistics, the long-term influence of the magnetic field on the velocity can be seen in figures \ref{fig:velocity_variance_plots_hax}-\ref{fig:velocity_variance_plots_hay}. The wall-normal field cases (figure \ref{fig:velocity_variance_plots_hay}), cases E,D,F show all three velocity variances declining in magnitude relative to case S, as $Ha$ increases. From the perspective of velocity, the Lorentz is purely dissipative in this case and for sufficiently large $Ha$ one would expect the velocity fluctuations to be completely damped out. 

The cases with wall-parallel fields (figure \ref{fig:velocity_variance_plots_hax}), cases A,B,C show a similar trend where $\langle u'^2 \rangle$ declines with the magnetic field. This can be related to the Joule dissipation mechanism damping the field parallel velocity component. However, the other two velocity components behave differently. The wall-normal velocity variance, $\langle v'^2 \rangle$ shows non-monotonic behaviour with increasing $Ha$. Initially the magnetic field damps $\langle v'^2 \rangle$ but for case F with $Ha=80$, the $\langle v'^2 \rangle$ increases to a magnitude compatible to case S. This change is likely a consequence of a regime change in the range $Ha \in (40,80]$, where the flow laminarises or becomes two-dimensional. This change can be seen qualitatively in figure \ref{fig:xz_temperature_plots_hax}. The variance of $w$, $\langle w'^2 \rangle$ also shows interesting behaviour as $Ha$ is increased. Relative to case S, with increasing $Ha$, $\langle w'^2\rangle$ becomes greater in magnitude near to the wall and lesser in magnitude at the centre of the domain. This is clearly a result of the organisation of near-wall jets that can be seen in figure  \ref{fig:yz_uz_plots_hax}. The likely cause of this is the favourable redistribution of $\langle v'^2\rangle$ into $\langle w'^2 \rangle$ resulting in lesser values of $\langle u'^2\rangle$. The result of this, the anisotropy between $\langle w'^2\rangle$ and $\langle u'^2\rangle$ can be seen in figure \ref{fig:velocity_variance_plots_hax} where the ratio $\langle w'^2\rangle/\langle u'^2\rangle$ is plotted. Observing this plot, it is clear that as $Ha$ increases the anisotropy increases in magnitude. For cases A,B, the anisotropy peaks in the near-wall region and decreases moving towards the bulk and the wall. For case C, the anisotropy greatly increases and now peaks in the boundary layer, having no local maxima between the wall and the bulk. Again, this change in behaviour is likely related to a regime change occurring in the range $Ha \in (40,80]$. 

\textbf{}

\section{Distribution of TKE and temperature variance in physical space}
\label{sec:physbudgets}

\begin{figure}[htbp]
  \centering
  \includegraphics[width=0.48\linewidth]{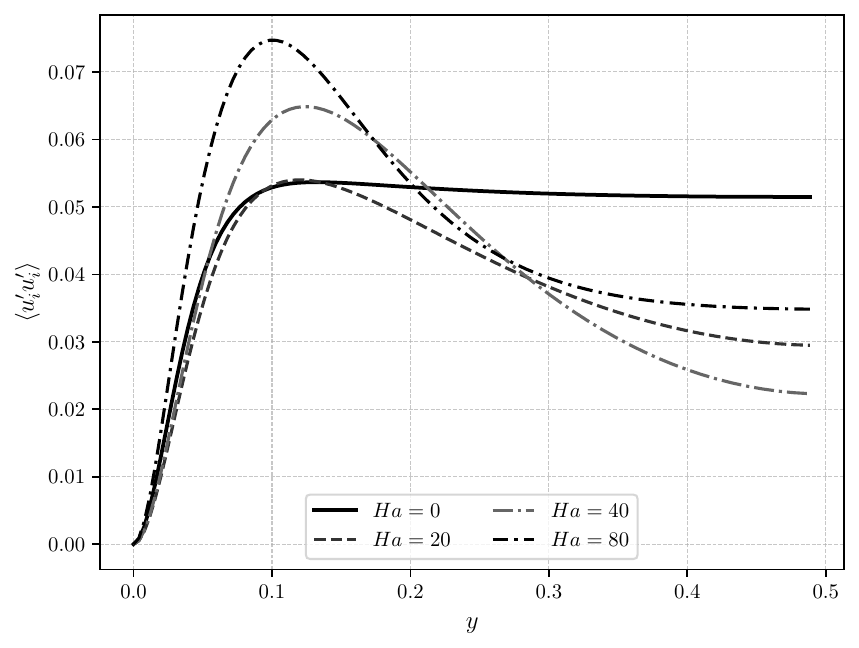}%
  \hfill
  \includegraphics[width=0.48\linewidth]{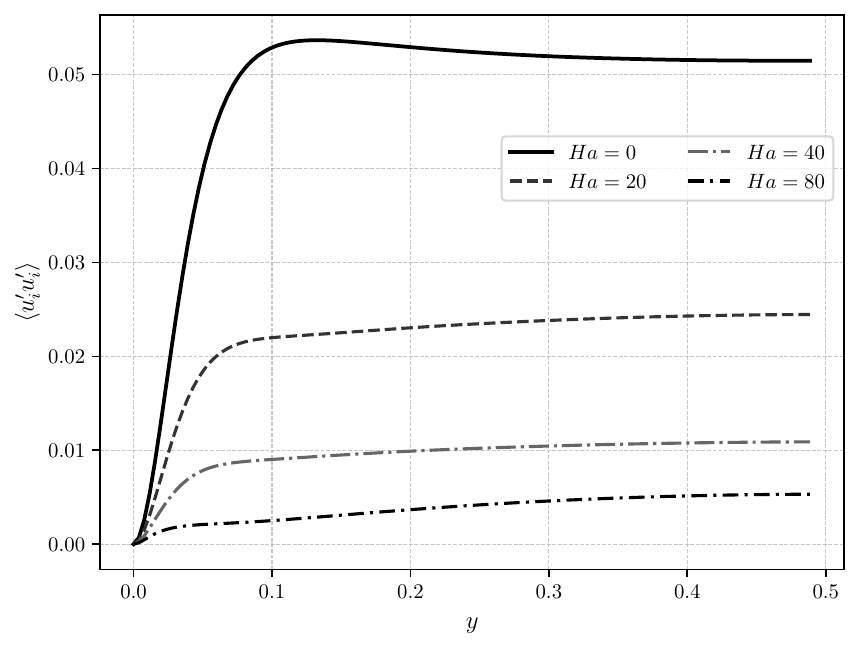}
  \caption{$Ha$ dependence of the turbulence kinetic energy, $k=\langle u_i' u_i'\rangle$. (left) cases A, B and C (right) cases D, E and F.}
  \label{fig:tke_plots}
\end{figure}

The focus of this section is to provide a quantitative description of the flow by studying transport equations describing the production, transport and dissipation of turbulent kinetic energy (TKE) and temperature variance, defined as $k = u'_iu'_i/2$ and $\theta'^2$ respectively. The TKE is plotted in figure \ref{fig:tke_plots}, showing similar trends to those discussed in section \ref{sec:topology}. That is, the wall-normal field damps all three velocity variances and hence the total TKE too. The wall-parallel field leads to an increasing anisotropy ratio $\langle w'^2\rangle/\langle u'^2 \rangle$ with increasing $Ha$.

\subsection{Turbulent Kinetic Energy budget}

\begin{figure}[htbp]
  \centering
  \includegraphics[width=0.48\linewidth]{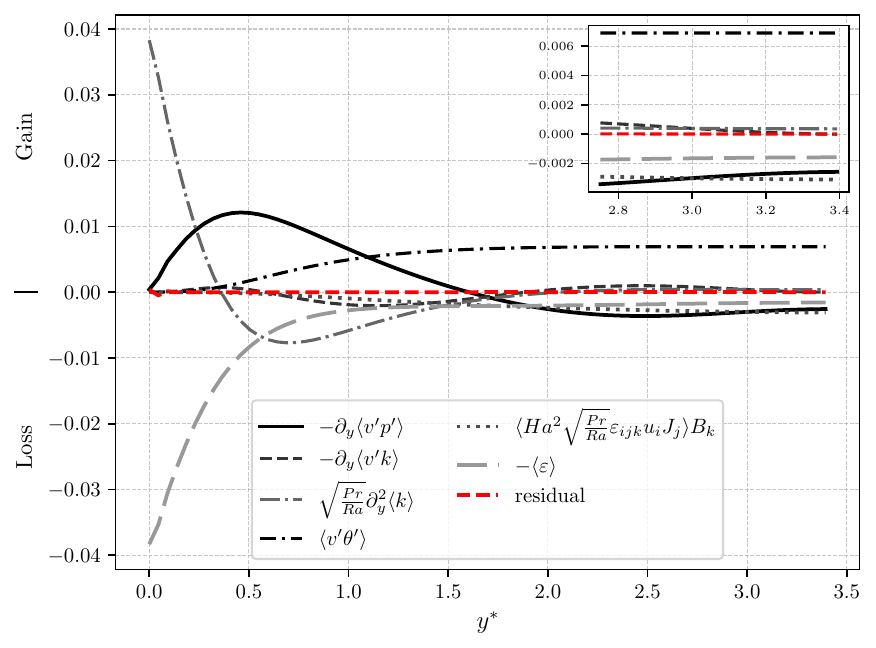}%
  \hfill
  \includegraphics[width=0.48\linewidth]{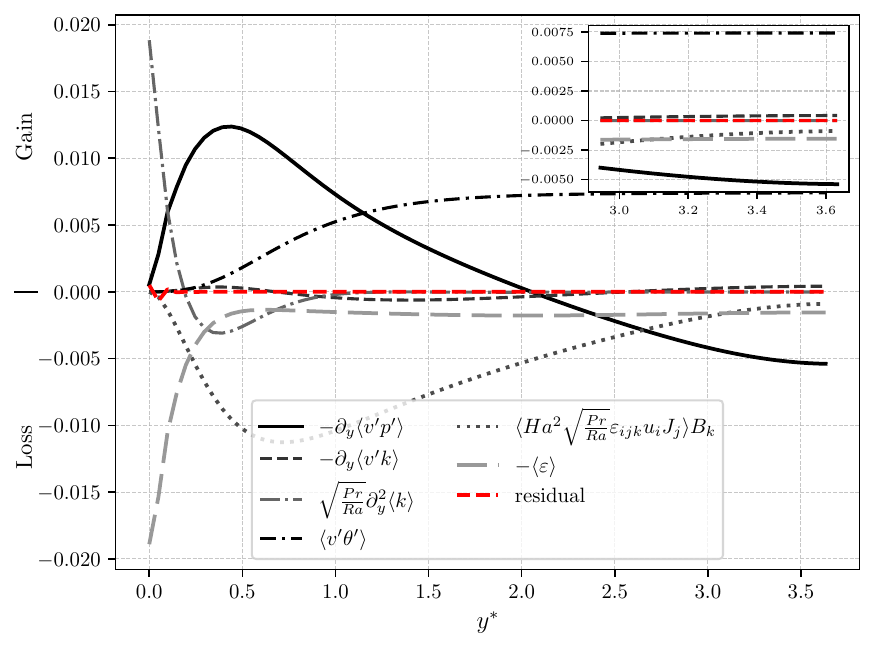}
  \caption{TKE budgets for cases B (left) and E (right), wall-parallel and wall-normal magnetic field cases respectively, with $Ha=40$. Here, $y^*$ is defined to be the distance from the wall in units of twice the thermal boundary layer thickness, $y^*= yNu$.}
  \label{fig:tke_budgets}
\end{figure}

We start with the analysis of the TKE budget,

\begin{equation}
\label{eq:tke}
    -\frac{d \langle kv' \rangle}{dy} 
    - \frac{d \langle p'v' \rangle}{dy}
    + \sqrt{\frac{Pr}{Ra}} \left( \frac{d^2 \langle k \rangle}{dy^2}
    + Ha^2\varepsilon_{ijk}\langle u'_i J'_j\rangle B_k \right)
    + \langle v' \theta'\rangle
    - \langle \epsilon \rangle
    = 0
\end{equation}

\noindent where the equation has been derived by applying the Reynolds decomposition, $u_i = \langle u_i \rangle + u'_i$ to equations (\ref{eq:momentum}-\ref{eq:energy}), $\langle u_i\rangle$ being the mean component and $u'_i$ the fluctuating component. The averages are all taken temporally and then spatially over the $x$ and $z$ directions, meaning the statistics only depend upon the wall-normal coordinate $y$. It is noted that simplifications have been made due to statistical stationarity as well as spatial homogeneity in the wall-parallel planes. The exact form of the work done by the Lorentz force, $\mathcal{L}=\varepsilon_{ijk}u_i J_jB_k$, depends upon the field direction. For the wall-parallel case with $B_i =\delta_{i1}$, $\mathcal{L} = v'J_z' - w'J_y'$ whilst for the wall-normal case with $B_i=\delta_{i2}$ it is $\mathcal{L}=u'J_z' - w'J_x'$.

The terms in equation \ref{eq:tke} are plotted in figure \ref{fig:tke_budgets}. As in \citet{togni2015physical}, three distinct regions can be defined within the flow: the bulk region where the buoyant production crosses the pressure transport, $\langle v'\theta'\rangle = - d\langle p'v' \rangle / dy$; the transitional region where the viscous transport intersects the pressure transport, $- d\langle p'v' \rangle / dy = \sqrt{Pr /Ra} \space d^2\langle k \rangle /dy^2$ and the boundary layer enclosed between the wall and the start of the transitional layer. 

This is now an opportune moment to discuss the influence of the Lorentz force upon the distribution and transport of TKE within the flow. Starting with the wall-parallel field case, the Lorentz force is mainly acting within the bulk of the flow, and tends towards zero close to the wall. This highlights the first key role of the Lorentz force, which is to drain from the TKE which is produced within the bulk via the buoyant production term, $\langle v'\theta'\rangle$. However, the buoyant production does exceed the combined value of the Lorentz force and the dissipation, creating an excess of TKE within the bulk, $\langle v'\theta' \rangle - \langle \varepsilon \rangle - \langle \mathcal{L}\rangle >0$. Similarly to the non-MHD case, this energy excess is then transported towards the transitional layer, mostly via inertial and pressure mechanisms, whilst the viscous transport remains negligible within the bulk. In the wall-normal cases, the Lorentz work is less significant in the bulk but peaks within the transitional layer. This results in a larger buoyant production within the bulk and therefore a larger energy excess is transported towards the transitional layer in comparison with the wall-parallel case. Within the transitional layer, the Lorentz work and the dissipation are greater than the buoyant production resulting in an energy deficit, which is compensated for by the positive pressure transport term. The viscous transport term becomes significant within the transitional layer and is negative meaning it acts to transport TKE towards the wall. The magnitude of the viscous transport is notably lower in the wall-normal field case, due to the Lorentz force peaking within the transitional layer, meaning the transport of TKE towards the wall is inhibited by the Lorentz force in this case. Finally, in both cases, within the boundary layer TKE is transported viscously and dissipated close to the wall. We note that the dissipation has a different functional form in the wall-parallel case whilst the wall-normal case has a form closely resembling the non-MHD case. 

The distribution of TKE can be studied in further detail by considering budgets for the variances of the each velocity component, also highlighting further the influence of the Lorentz force upon the flow,

\begin{subequations}
\label{eq:1D_tke_budgets}
\begin{equation}
    \label{eq:uu}
    -\frac{1}{2}\frac{d \langle v' u'^2\rangle}{dy}
    + \langle p' \frac{du'}{dx}\rangle
    + \frac{1}{2} \sqrt{\frac{Pr}{Ra}}\frac{d^2 \langle u'^2 \rangle}{dy^2}
    - \langle \varepsilon_x \rangle + \langle \mathcal{L}_x \rangle = 0,
\end{equation}

\begin{equation}
    \label{eq:vv}
    -\frac{1}{2}\frac{d \langle v'^3\rangle}
    {dy}
    + \langle p' \frac{dv'}{dy}\rangle
    - \frac{d \langle p'v' \rangle}{dy}
    + \frac{1}{2} \sqrt{\frac{Pr}{Ra}}\frac{d^2 \langle v'^2 \rangle}{dy^2}
    + \langle v' \theta' \rangle
    - \langle \varepsilon_y \rangle 
    + \langle \mathcal{L}_y \rangle= 0,
\end{equation}

\begin{equation}
    \label{eq:ww}
    -\frac{1}{2} \frac{d \langle v'w'^2 \rangle}{dy}
    + \langle p' \frac{dw'}{dz} \rangle
    + \frac{1}{2} \sqrt{\frac{Pr}{Ra}} \frac{d\langle w'^2\rangle}{dy^2} 
    - \langle \varepsilon_z \rangle
    + \langle \mathcal{L}_z \rangle = 0,
\end{equation}
\end{subequations}

\noindent where $\mathcal{L}_x, \mathcal{L}_y, \mathcal{L}_z$ are defined as,

\begin{equation}
\begin{aligned}
\mathcal{L}_x &= 
\begin{cases}
   0 & \text{if } B_i = \delta_{i1}, \\
   u'J'_z & \text{if } B_i = \delta_{i2},
\end{cases}
\\[1ex]
\mathcal{L}_y &= 
\begin{cases}
   v'J'_z & \text{if } B_i = \delta_{i1}, \\
   0 & \text{if } B_i = \delta_{i2},
\end{cases}
\\[1ex]
\mathcal{L}_z &= 
\begin{cases}
   -w'J'_y & \text{if } B_i = \delta_{i1}, \\
   -w'J'_x & \text{if } B_i = \delta_{i2},
\end{cases}
\end{aligned}
\end{equation}

\noindent and $\varepsilon_x, \varepsilon_y, \varepsilon_z$ are defined as,

\refstepcounter{equation}
$$
  \varepsilon_x =  \frac{\partial u'}{\partial x_j} \frac{\partial u'}{\partial x_j}, \quad
  \varepsilon_y = \frac{\partial v'}{\partial x_j} \frac{\partial v'}{\partial x_j}, \quad
  \varepsilon_z = \frac{\partial w'}{\partial x_j} \frac{\partial w'}{\partial x_j}.
  \eqno{(\theequation{\mathit{a},\mathit{b}},\mathit{c})}
$$

\begin{figure}[htbp]
  \centering
  \begin{minipage}{0.48\linewidth}
    \centering
    \includegraphics[width=\linewidth]{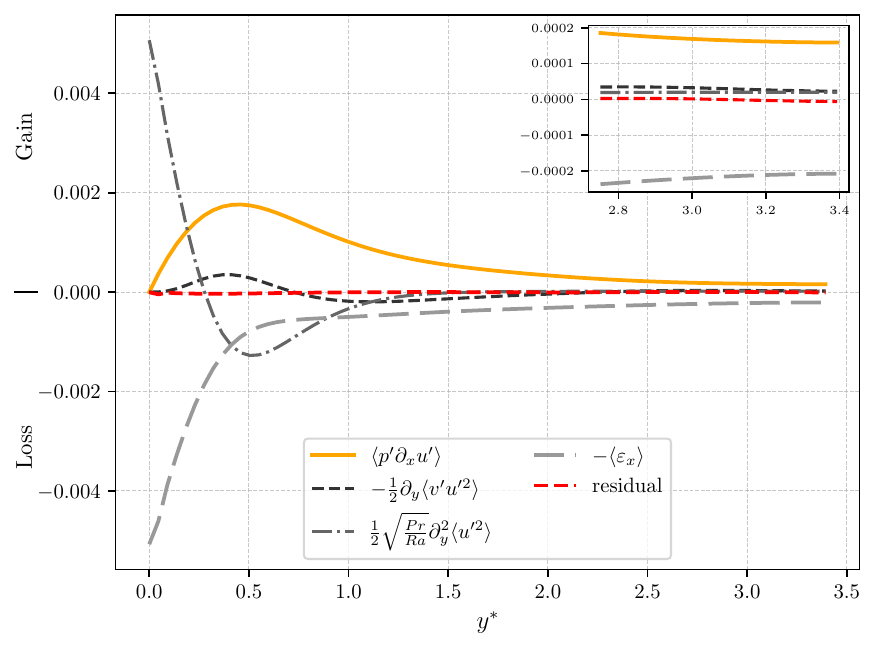}
    \subcaption{}
    \label{fig:uu_budget}
  \end{minipage}\hfill
  \begin{minipage}{0.48\linewidth}
    \centering
    \includegraphics[width=\linewidth]{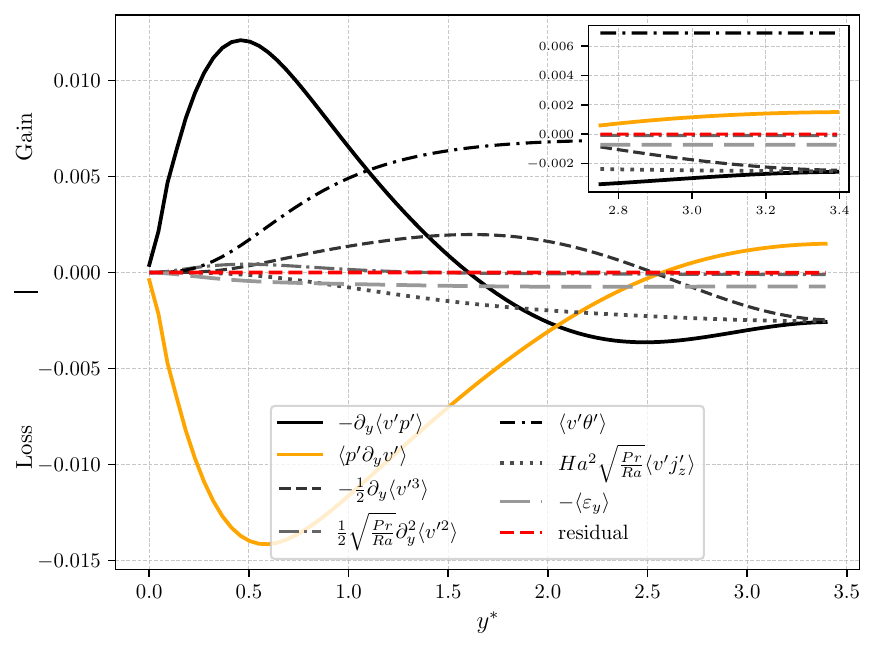}
    \subcaption{}
    \label{fig:vv_budget}
  \end{minipage}

  \vspace{0.5em} 
  \begin{minipage}{0.48\linewidth}
    \centering
    \includegraphics[width=\linewidth]{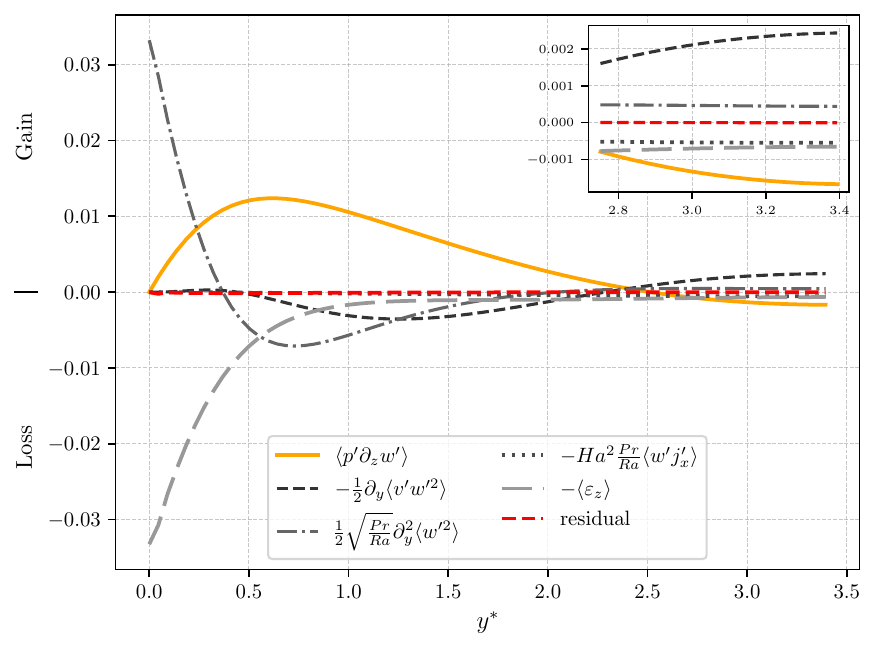}
    \subcaption{}
    \label{fig:ww_budget}
  \end{minipage}

  \caption{Budgets of (a) $\langle u'^2 \rangle$, (b) $\langle v'^2 \rangle$, and (c) $\langle w'^2 \rangle$ for case B, wall-parallel field case with $Ha=40$.}
  \label{fig:1D_tke_budget_ha40x}
\end{figure}

\begin{figure}[htbp]
  \centering
  \begin{minipage}{0.48\linewidth}
    \centering
    \includegraphics[width=\linewidth]{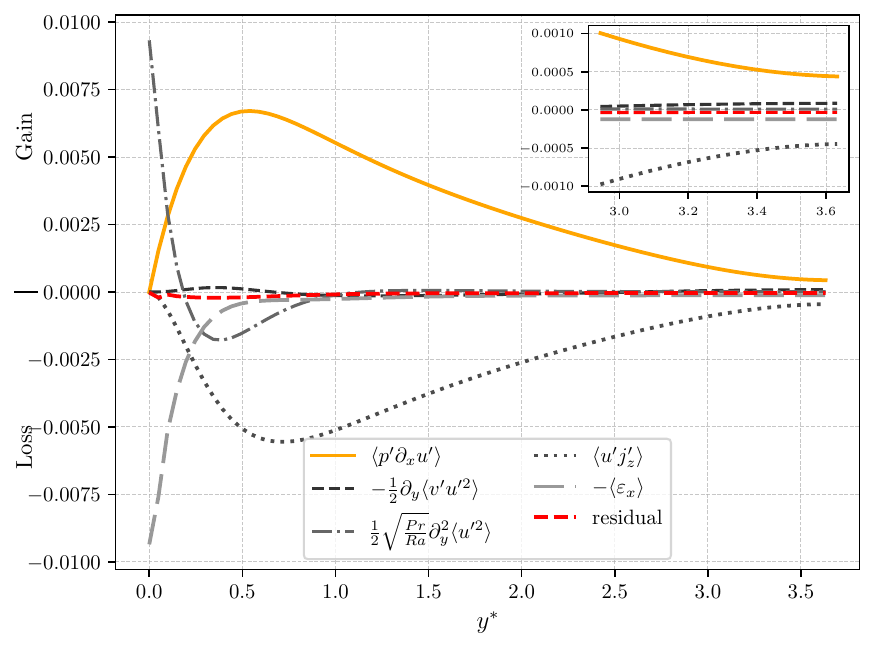}
    \subcaption{}
    \label{fig:uu_budget}
  \end{minipage}\hfill
  \begin{minipage}{0.48\linewidth}
    \centering
    \includegraphics[width=\linewidth]{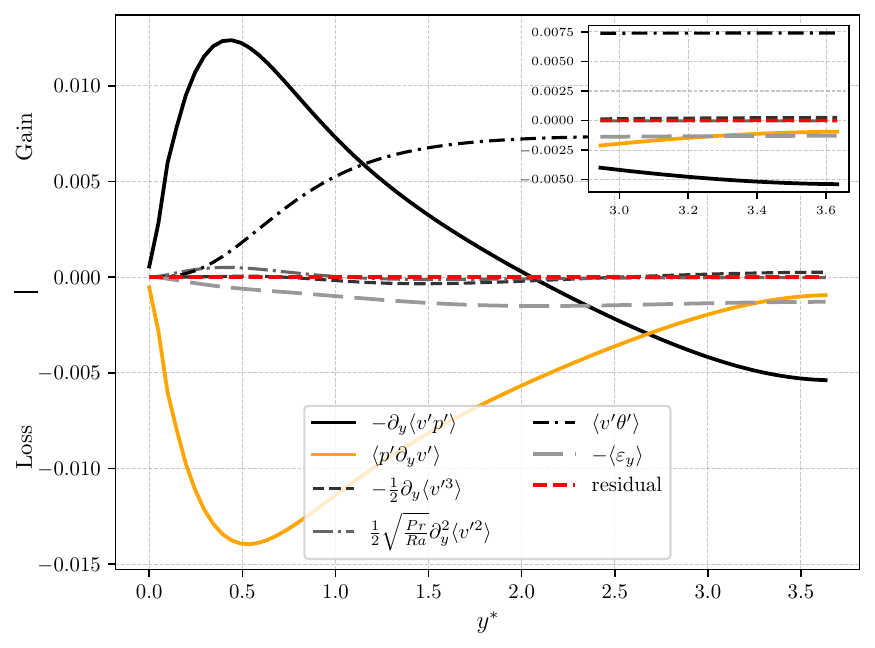}
    \subcaption{}
    \label{fig:vv_budget}
  \end{minipage}
  \caption{Same as figure \ref{fig:1D_tke_budget_ha40x} but for case E, wall-normal field case with $Ha=40$. The $\langle w'^2\rangle$ budget is not plotted, as it is identical to the $\langle u'^2\rangle$ budget due to statistical isotropy in the $x-z$ plane.}
  \label{fig:1D_tke_budget_ha20y}
\end{figure}

\noindent The terms in equations \ref{eq:uu}-\ref{eq:ww} are plotted in figures \ref{fig:1D_tke_budget_ha40x} and \ref{fig:1D_tke_budget_ha20y}, showing the budgets for the wall-parallel and wall-normal cases respectively. Both of these plots show that buoyant production in the bulk predominantly provides energy to the $v$ velocity component, which is then transported towards the wall. The pressure-strain mechanism then acts to convert $v$ velocity into the $u$ and $w$ components, where the TKE is then dissipated in the wall-parallel directions. The process discussed so far also highlights further the influence of the Lorentz force, and some key differences between the wall-parallel and wall-normal cases. In the wall-parallel case (figure \ref{fig:1D_tke_budget_ha40x}), the Lorentz force primarily acts upon the $v$ component, supporting the idea that the Lorentz force drains the from the TKE generated within the bulk of the flow. The influence of the Lorentz force upon the $w$ component is minor. In the wall-normal field case (figure \ref{fig:1D_tke_budget_ha20y}), instead the Lorentz force equally inhibits both the $u$ and $w$ components, acting within the transitional layer where the pressure-strain mechanism is converting $v$ TKE to $u$ and $w$ TKE. On the topic of the pressure-strain mechanism, and moving back to the wall-parallel case, the pressure diffusion mechanism favourably converts the $v$ kinetic to $w$ kinetic, and little energy is redistributed into the $u$ direction. This can be thought of as an interaction between the Joule dissipation mechanism and pressure, where it becomes energetically favourable for kinetic energy to be redistributed from $v$ to $w$. Given the typically low value of the Lorentz force term in the wall-parallel field case, most of the TKE produced by buoyancy makes it to the wall, which is then redistributed mostly to the $w$ velocity component. This explains the observed higher $\langle w'^2\rangle$ and lower $\langle u'^2 \rangle$.

\subsection{Temperature variance budget}

\begin{figure}[htbp]
  \centering
  \includegraphics[width=0.48\linewidth]{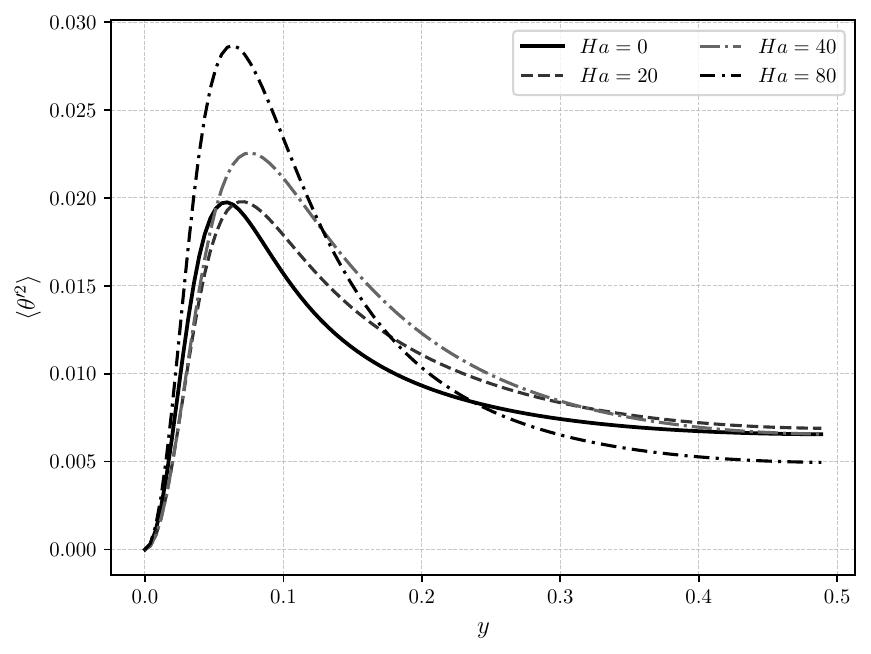}%
  \hfill
  \includegraphics[width=0.48\linewidth]{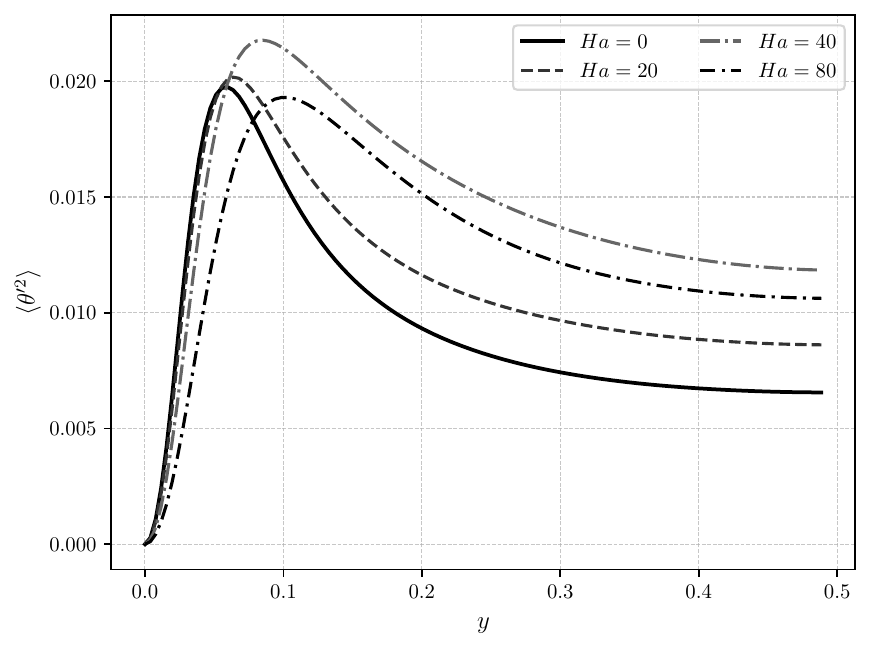}
  \caption{$Ha$ dependence of the temperature variance, $\theta'^2$. (left) cases A, B and C (right) cases D, E and F.}
  \label{fig:temperature_variance_plots}
\end{figure}

Given that buoyancy is the sole driving force of the flow, analysing the behaviour of the temperature variance is necessary to provide a more complete physical description. The temperature variance is plotted in figure \ref{fig:temperature_variance_plots}. The wall-parallel field cause the maximum of the temperature variance to increase as $Ha$ increases. In the wall-normal field cases, the temperature variance initially rises with $Ha$ and then drops for case F with $Ha=80$. The temperature variance budget equation is written as, 

\begin{equation}
    \label{eq:tempvar}
    -\frac{d \langle v' \theta'^2 \rangle}{dy}
    + \frac{1}{\sqrt{PrRa}} \frac{d^2 \langle \theta'^2 \rangle}{dy^2}
    - 2\langle v' \theta' \rangle \frac{d \langle \theta \rangle}{dy} 
    - 2 \langle \chi \rangle
    =0
\end{equation}

\noindent where $\chi = \frac{1}{\sqrt{PrRa}}\partial \theta' /\partial x_j  \partial \theta' / \partial x_j$ is the dissipation of temperature variance. The terms from left to right are the turbulent transport, viscous transport, production and dissipation of temperature variance. In the absence of Joule heating, this equation does not explicitly depend upon any MHD phenomena but the magnitude of the terms changes as $Ha$ is increased. 

\begin{figure}[htbp]
  \centering
  \includegraphics[width=0.48\linewidth]{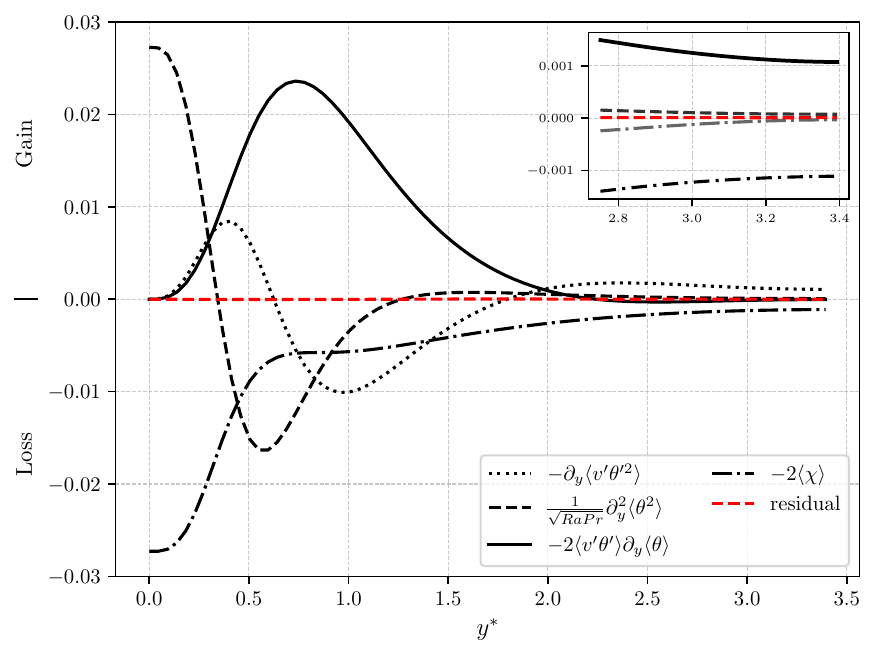}%
  \hfill
  \includegraphics[width=0.48\linewidth]{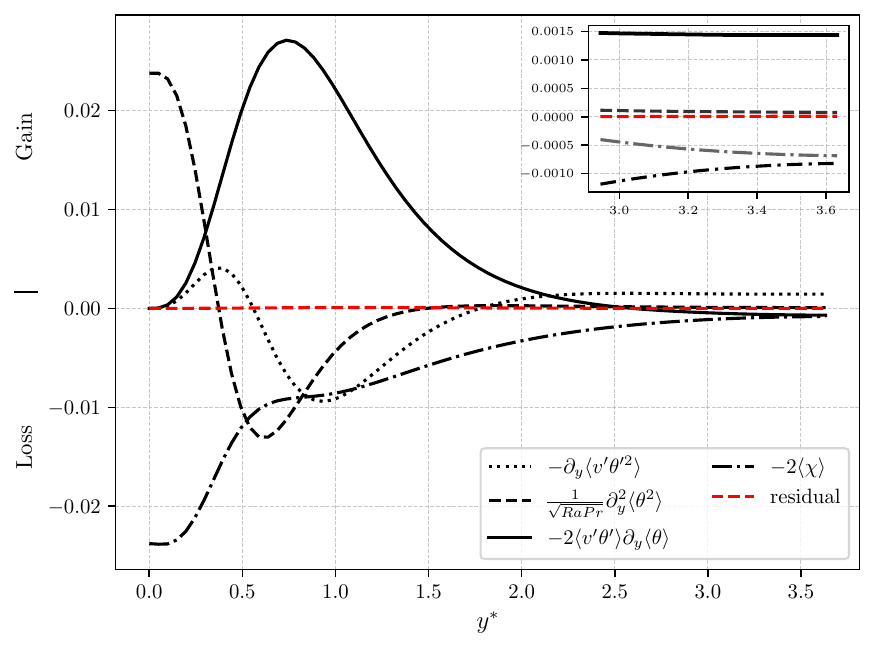}
  \caption{Temperature variance budgets for cases B (left) and E (right), wall-parallel and wall-normal magnetic field cases respectively, with $Ha=40$.}
  \label{fig:temperature_variance_budgets}
\end{figure}

The budget of temperature variance in both cases (figure \ref{fig:temperature_variance_budgets}) is found to be qualitatively similar to the non-MHD case, where the the only difference between the MHD and non-MHD cases is the magnitudes of the terms. This is reflective of the fact that the Lorentz force only acts directly upon the TKE produced through buoyancy, and how it affects the temperature field indirectly through its interaction with the velocity components. 

For clarity, the results plotted in figure \ref{fig:temperature_variance_budgets} are summarised here. Again, three regions can be defined; the thermal boundary layer, thermal transitional region and the thermal bulk. The transitional layer is defined as the region where the production dominates, the boundary layer is the region close to the wall where the viscous transport exceeds the production and the bulk is defined as the central region where the inertial transport exceeds the production. Temperature variance is produced within the transitional layer where it is transported towards the wall by viscous transport, and towards the bulk by the inertial transport. Temperature variance transported towards the wall is then dissipated whilst temperature variance carried towards the bulk is enriched by buoyancy and forms new plumes, creating the recurring cycle of RBC. The key takeaway here, is that the behaviour and energetics of MC works in a similar manner to that of non-MHD RBC. The magnetic field mainly influences the TKE budget, which then indirectly leads to subtle differences in the temperature variance budget.

\section{Multiscale analysis of TKE and temperature variance}
\label{sec:hill}

 This approach is extended to the space of scales by studying exact relationships between second- and third-order structure functions \citep{hill2002exact}. The $p$-th order increment of a variable $\psi$ is defined as $\delta \psi^p = (\psi(x_i)-\psi(x_i-r_i))^p$, which is the fluctuating increment of the variable between positions $x_i$ and $x_i + r_i$. Structure functions are then defined as averages of the increments, $\langle \delta \psi^p \rangle$.  In general, for a statistically stationary flow the structure function depends on the midpoint coordinate $X_i = (X,Y,Z) = x_i + r_i/2$ and the separation distance $r_i = (r_x, r_y, r_z)$. However, given that the cases considered possess spatially homogeneity in the $x$ and $z$ directions, after spatial averaging is applied structure functions are only dependent on the wall-normal midpoint coordinate $Y$ and the separation distance $r_i$, yielding a function of four variables $\delta \psi = \delta \psi(Y,r_i)$. Also, in this study only the case with $r_y=0$ is considered. The structure functions in the homogenous $x-z$ plane can be computed efficiently using Fast Fourier Transforms (FFTs) whilst the $r_y$ component requires an explicit calculation. The need for the explicit calculation stems from the inhomogeneity  present in the $y$-direction; FFT based approaches imply an average over the $y$ coordinate which would lose information regarding the inhomogeneous effects. To obtain a function of $(Y,r_y)$ an explicit calculation would be required yielding a time complexity of $\sim \mathcal{O}([N_x \log Nx][N_z \log Nz] N_y^2)$ as opposed to $\sim \mathcal{O}([N_x \log Nx][N_z \log Nz][N_y\log N_y])$ when using FFTs. The full explicit calculation is considered computationally infeasible in this study and is left as future work. The wall-normal field cases (D,E and F) also possess isotropy in the $x-z$ plane allowing the parameters $r_x$ and $r_z$ to be averaged over circles $C(r)$ of radius $r=\sqrt{r_x^2+r_z^2}$ through the operation,

\begin{equation}
\langle \delta \psi (Y, r)\rangle_r = \frac{1}{\pi r^2} \int_{C(r)} \langle \delta \psi (Y, r_x, r_z)\rangle dr_x dr_z.
\label{eq:r_average}
\end{equation}

\noindent However, the cases with wall-parallel fields (A,B and C) are anisotropic in the $x-z$ plane, meaning the structure functions retain unique dependence on both $r_x$ and $r_z$, implying $\delta u_i=\delta u_i(Y, r_x, r_z)$. To analyse the anisotropic cases in more detail, we consider the directional structure functions,

\begin{subequations}
\begin{equation}
\delta \psi_{\parallel}(Y,l) = \delta \psi(Y, r_x=l, r_z=0)
\label{eq:directional1}
\end{equation}
\begin{equation}
\delta \psi_{\perp}(Y,l) = \delta \psi
(Y, r_x=0, r_z=l).
\label{eq:directional2}
\end{equation}
\end{subequations} 

\subsection{Scale by scale budgets of $\delta u^2$ and $\delta \theta^2$}

The first quantity of interest is the scale-energy, defined as $\langle \delta u^2 \rangle = \langle \delta u_i \delta u_i\rangle$ which can be thought of as a multi-scale analogue of the single point TKE, $k$. The quantity $\langle \delta u^2(Y,r_i)\rangle$ can be thought of as the TKE at wall-normal position $Y$ and scale $r_i$. The budget equation for the scale TKE is,

\begin{align}
    \label{eq:KH_budget}
    & -\frac{\partial \langle v^* \delta u^2 \rangle}{\partial Y}
    - 2 \frac{\partial \langle \delta v \delta p \rangle}{\partial Y}
    + \frac{1}{2} \sqrt{\frac{Pr}{Ra}} \frac{\partial^2 \langle \delta u^2 \rangle}{\partial Y^2}
    \nonumber \\
    & + 2 \sqrt{\frac{Pr}{Ra}} \frac{\partial^2 \langle \delta u^2 \rangle}{\partial r_i \partial r_i}    
    + 2Ha^2 \sqrt{\frac{Pr}{Ra}} \varepsilon_{ijk} \langle \delta u_i \delta J_j \rangle B_k 
    - \frac{\partial \langle \delta u_i \delta u^2 \rangle}{\partial r_i}
    + 2\langle \delta v \delta \theta \rangle
    - 4 \langle \epsilon^* \rangle = 0
\end{align}

\noindent where $\psi^* = (\psi(x_i) + \psi(x_i+r_i) /2$ is the midpoint average of $\psi$. This equation is similar to \ref{eq:tke} but with the addition of terms appearing under $\partial / \partial r_i$. These terms are interpreted to represent the transfer of scale TKE between scales of motion $r_i$ whilst the terms appearing under $\partial / \partial Y$ correspond to transport of scale TKE between physical positions within the flow, $Y$. Terms appearing under $\partial / \partial r_y$ are considered separately and are written as $R(r_i, Y)$,

\begin{equation}
R(r_i,Y) = -\frac{\partial \langle \delta v \delta u^2\rangle}{\partial r_y}
+
2\sqrt{\frac{Pr}{Ra}} \frac{\partial ^2 \langle \delta u^2 \rangle}{\partial r_y^2}
\end{equation}

\noindent This redefines equation \ref{eq:KH_budget} as,

\begin{align}
\label{eq:KH_budget_complete}
R(r_i,Y) =\;&
- \frac{\partial \langle v^* \delta u^2 \rangle}{\partial Y}
- 2 \frac{\partial \langle \delta v \delta p \rangle}{\partial Y}
+ \frac{1}{2} \sqrt{\frac{Pr}{Ra}}
  \frac{\partial^2 \langle \delta u^2 \rangle}{\partial Y^2}
\nonumber\\[4pt]
&+ 2 \sqrt{\frac{Pr}{Ra}}
  \left(
    \frac{\partial^2 \langle \delta u^2 \rangle}{\partial r_x^2}
  + \frac{\partial^2 \langle \delta u^2 \rangle}{\partial r_z^2}
  \right)
+ 2Ha^2 \sqrt{\frac{Pr}{Ra}}
  \varepsilon_{ijk} \langle \delta u_i \delta J_j \rangle B_k
\nonumber\\[4pt]
&- \left(
    \frac{\partial \langle \delta u \, \delta u^2 \rangle}{\partial r_x}
  + \frac{\partial \langle \delta u \, \delta u^2 \rangle}{\partial r_z}
  \right)
+ 2\langle \delta v \delta \theta \rangle
- 4 \langle \epsilon^* \rangle .
\end{align}

\noindent For simplicities, equation \ref{eq:KH_budget_complete} is rewritten as, 

\begin{equation}
    R(r_i,Y) = I_c(r_i, Y) +  P(r_i, Y) + D_c(r_i,Y) 
    + D_r(r_i,Y) + L(r_i,Y) + I_r(r_i,Y) 
    + \Pi(r_i,Y)
    +E(Y) ,
\end{equation}

\noindent where each of the terms right-hand-side corresponds to its respective term in \ref{eq:KH_budget_complete}. That is inertial transport, pressure transport, viscous transport, inertial transfer, viscous transfer, buoyant production, Lorentz dissipation and viscous dissipation of scale TKE. The term $R(r_i,Y)$ is the overall transfer of scale energy between wall-normal scales. Then the overall transport and transfer terms can be defined as,

\refstepcounter{equation}
$$
  T_Y(r_i,Y) = I_Y(r_i,Y) + P(r_i,Y) + D_Y(r_i,Y), \quad 
  T_r(r_i,Y) = I_r(r_i,Y)+D_r(r_i,Y)-R(r_i,Y), \quad
  \eqno{(\theequation{\mathit{a},\mathit{b}})}
$$

\noindent respectively, resulting in a final simplified form for \ref{eq:KH_budget},

\begin{equation}
    T_Y(r_i,Y) +T_r(r_i,Y)
    + \Pi(r_i,Y)
    +L(r_i,Y) +E(Y) = 0.
\end{equation}

\noindent Similarly to the single point analysis, a complete description of the flow requires studying the behaviour of the temperature in scale-space. The quantity of choice here is the scale variance, $\delta \theta^2$, which also obeys a transport equation,  

\begin{align}
    & -\frac{\partial \langle v^* \delta \theta^2 \rangle}{\partial Y}
    + \frac{1}{2} \frac{1}{\sqrt{PrRa}} \frac{\partial^2 \langle \delta \theta^2 \rangle}{\partial Y^2}
    - \frac{\partial \langle \delta \theta^2 \delta u_i \rangle}{\partial r_i} \nonumber \\
    & + 2 \frac{1}{\sqrt{PrRa}} \frac{\partial^2 \langle \delta \theta^2 \rangle}{\partial r_i \partial r_i}
    - 2 \langle \delta v \delta \theta \rangle \frac{d \langle \theta^* \rangle}{d Y} 
    - 4 \langle \chi^* \rangle = 0.
    \label{eq:KY_budget}
\end{align}

\noindent The wall-normal flux of scale energy, $R^\theta(r_i, Y)$, is then separated from equation \ref{eq:KY_budget}, resulting in,

\begin{align}
    R^\theta(r_i,Y) =
    & -\frac{\partial \langle v^* \delta \theta^2 \rangle}{\partial Y}
    + \frac{1}{2} \frac{1}{\sqrt{PrRa}} \frac{\partial^2 \langle \delta \theta^2 \rangle}{\partial Y^2}
    - \left(\frac{\partial \langle \delta \theta^2 \delta u_i \rangle}{\partial r_x} 
    + \frac{\partial \langle \delta \theta^2 \delta u_i \rangle}{\partial r_z} \right)
    \nonumber \\
    & + 2 \frac{1}{\sqrt{PrRa}} \left(\frac{\partial^2 \langle \delta \theta^2 \rangle}{\partial r_x^2}
     +  \frac{\partial^2 \langle \delta \theta^2 \rangle}{\partial r_z^2} \right)
    - 2 \langle \delta v \delta \theta \rangle \frac{d \langle \theta^* \rangle}{d Y} 
    - 4 \langle \chi^* \rangle 
    \label{eq:KY_budget_complete}
\end{align}

\noindent where, 

\begin{equation}
    R^\theta(r_i, Y) = -\frac{\partial \langle \delta \theta^2 \delta u_i \rangle}{\partial r_y} 
    +
    2 \frac{1}{\sqrt{PrRa}} \frac{\partial^2 \langle \delta \theta^2 \rangle}{\partial r_y^2}
\end{equation}

\noindent Again equation \ref{eq:KY_budget_complete} can be written in a simpler form as follows, 

\begin{equation}
    R^\theta(r_i, Y) = I_c^{\theta}(r_i,Y) + D_c^{\theta}(r_i,Y)
    + I_r^{\theta}(r_i,Y) + D_r^{\theta}(r_i,Y)
    + \Pi^{\theta}(r_i,Y) + E^{\theta}(Y)
    \label{eq:reduced_KH_budget}
\end{equation}

\noindent where each term on the right-hand-side corresponds to its respective term in \ref{eq:KY_budget}. That is inertial transport, viscous transport, inertial transfer and viscous transfer of scale variance. $R^\theta(r_i,Y)$ is the overall transfer of scale variance between wall-normal scales. The overall transport and transfer terms can then be written as,

\refstepcounter{equation}
$$
  T^\theta_c(r_i,Y) = I^\theta_c(r_i,Y) + D^\theta_c(r_i,Y), \quad 
  T^\theta_r(r_i,Y) = I^\theta_r(r_i,Y)+D^\theta_r(r_i,Y) -R^\theta(r_i,Y), \quad
  \eqno{(\theequation{\mathit{a},\mathit{b}})}
$$

\noindent respectively, resulting in the final form for \ref{eq:KY_budget},
 
\begin{equation}
    T^\theta_c(r_i,Y)+T^\theta_r(r_i,Y) + \Pi^\theta(r_i,Y) + E^\theta(Y) = 0.
    \label{eq:reduced_KY_budget}
\end{equation}

\noindent A detailed derivation of scale by scale transport equations for velocity structure functions can be found in  appendix A, and similar methods can be used to derive transport equations for scalar structure functions. After averaging, considering the forcing and accounting for the symmetries relevant to this study (homogeneity and stationarity), equations \ref{eq:KH_budget} and \ref{eq:KY_budget} can be derived. These equations represent generalisations of the previously defined TKE and temperature variance budgets, equations \ref{eq:tke} and \ref{eq:tempvar}, where the dependence on the correlation vector $r_i$ is now included. This can be seen explicitly, by considering the large $r_i$ limit, where the structure functions de-correlate and tend towards their single-point expectations, leading to the recovery of the midpoint average of equations \ref{eq:tke} and \ref{eq:tempvar}, within a constant factor. The choice to separate the terms $R(r_i,Y)$ and $R^\theta(r_i,Y)$ was based upon the choice of $r_y=0$. This equation is exact as written but the $R$ terms are calculated as the converged residual, which is the sum of all the remaining terms. As far as the authors are aware, other methods would involve approximations of derivatives in $r_y$, that require computing the structure functions for $r_y>0$ whilst retaining dependence on the wall-normal position $Y$. This limits the analysis from the perspective of wall-normal transfer mechanisms, given that this approach leaves wall-normal viscous and inertial transfer indistinguishable but beneficially saves time on convergence and computation.

\subsection{Isotropic scale by scale analysis of  wall-normal field cases}

The scale by scale analysis of MC begins here, by first considering the cases with wall-normal magnetic fields (Cases D, E and F). In comparison to the wall-parallel cases, the analysis here is simplified as the cases remain isotropic in the $x-z$ plane. As a result, the r-averaging operation defined in equation \ref{eq:r_average} can be applied to each term within the budgets individually, which is done for all budgets and results shown within this section. 

\begin{figure}[htbp]
  \centering
  \begin{minipage}{0.48\linewidth}
    \centering
    \includegraphics[width=\linewidth]{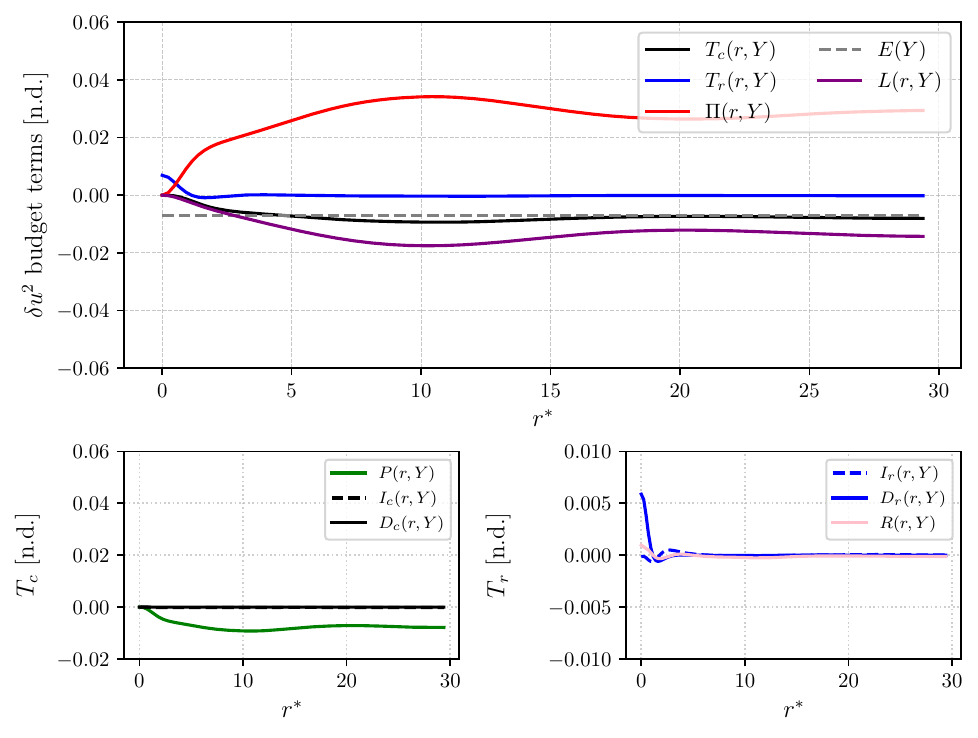}
    \subcaption{}
    \label{fig:KH_budget_ha40y50}
  \end{minipage}\hfill
  \begin{minipage}{0.48\linewidth}
    \centering
    \includegraphics[width=\linewidth]{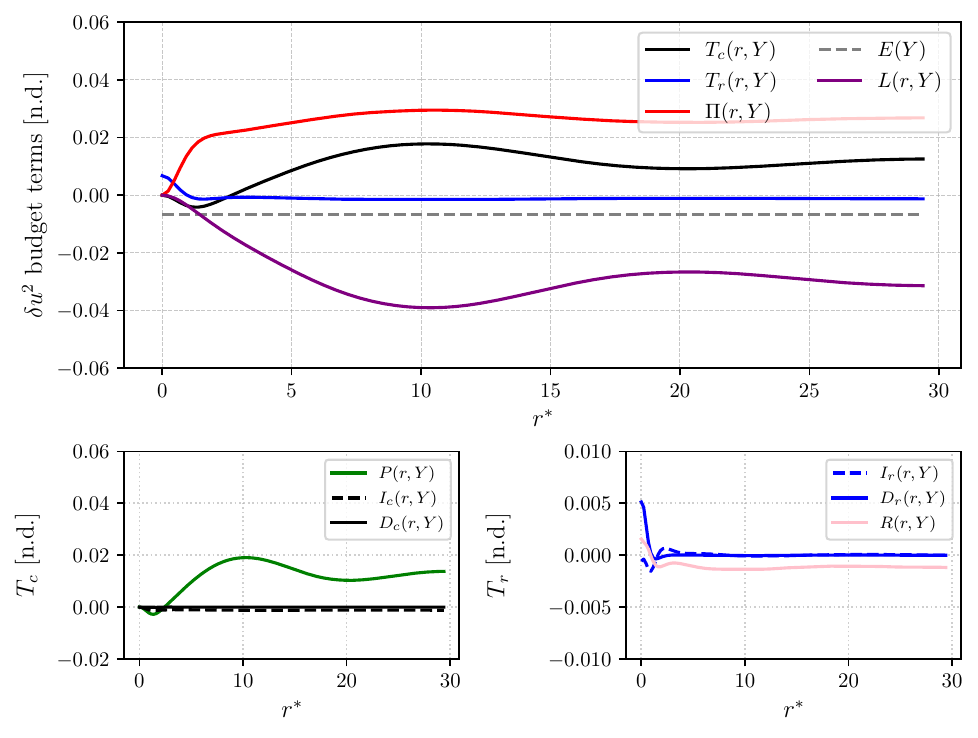}
    \subcaption{}
    \label{fig:KH_budget_ha40y30}
  \end{minipage}

  \vspace{0.5em}
  \begin{minipage}{0.48\linewidth}
    \centering
    \includegraphics[width=\linewidth]{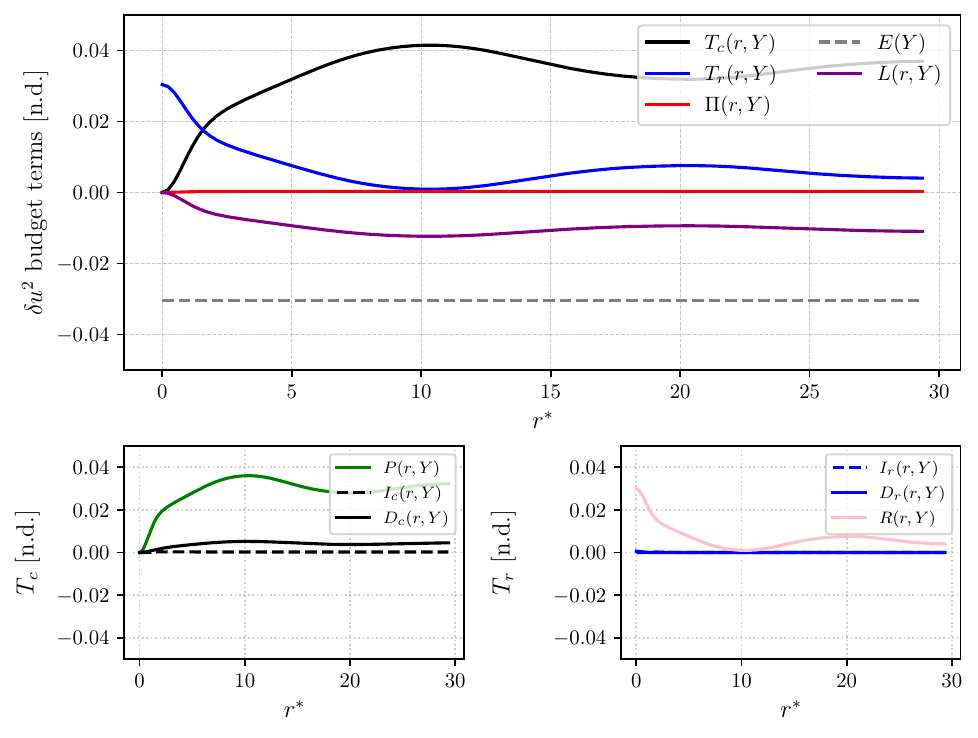}
    \subcaption{}
    \label{fig:KH_budget_ha40y1}
  \end{minipage}

  \caption{Budgets of $\delta u^2$ plotted in (a) the bulk, (b) the transitional layer and (c) the boundary layer. The left-hand subplots contain the contribution terms of $T_c$ whilst right-hand subplots contain the contributing terms of $T_r$. Plotted for case E at $Ha=40$ with a wall-normal magnetic field.}
  \label{fig:KH_budgets_ha40y}
\end{figure}

\begin{figure}[htbp]
  \centering
  \begin{minipage}{0.48\linewidth}
    \centering
    \includegraphics[width=\linewidth]{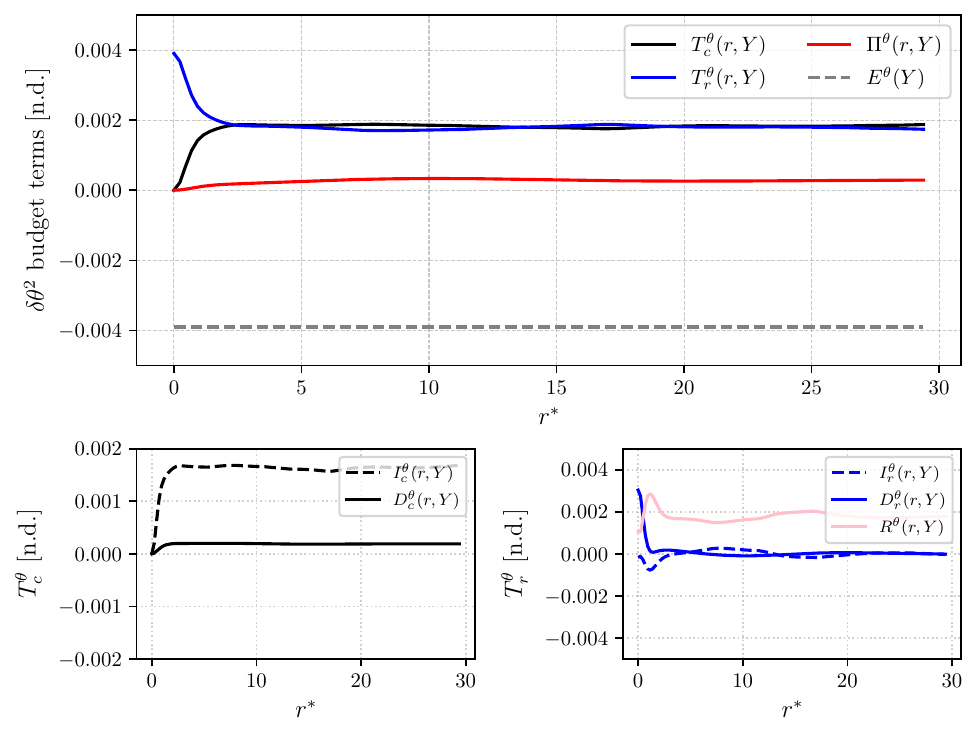}
    \subcaption{}
    \label{fig:KY_budget_ha40y50}
  \end{minipage}\hfill
  \begin{minipage}{0.48\linewidth}
    \centering
    \includegraphics[width=\linewidth]{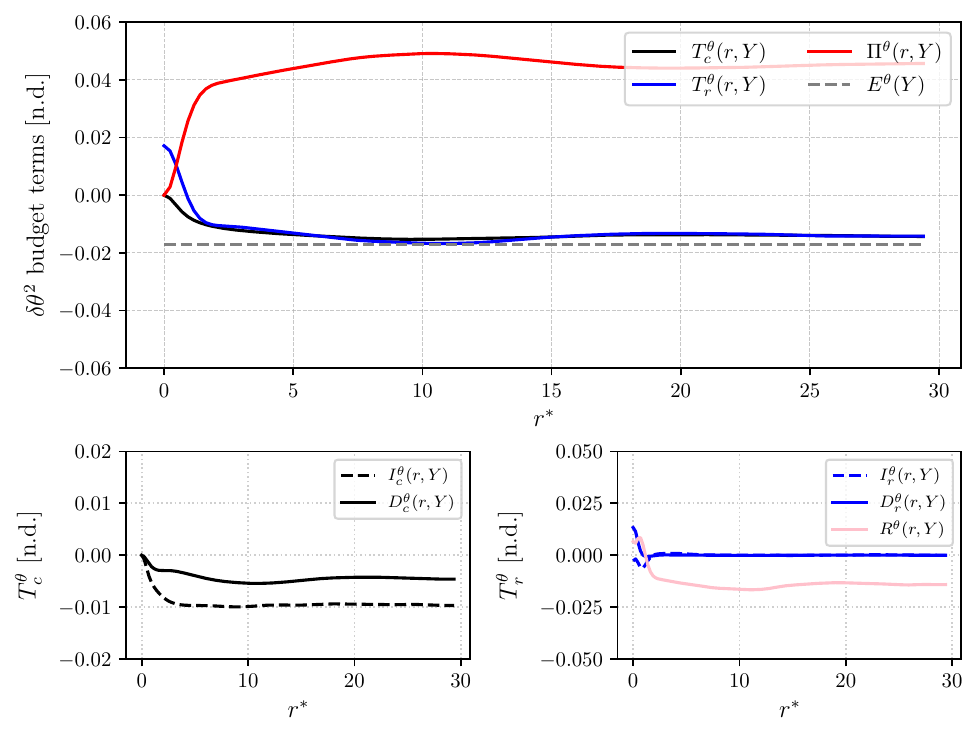}
    \subcaption{}
    \label{fig:KY_budget_ha40y20}
  \end{minipage}

  \vspace{0.5em}
  \begin{minipage}{0.48\linewidth}
    \centering
    \includegraphics[width=\linewidth]{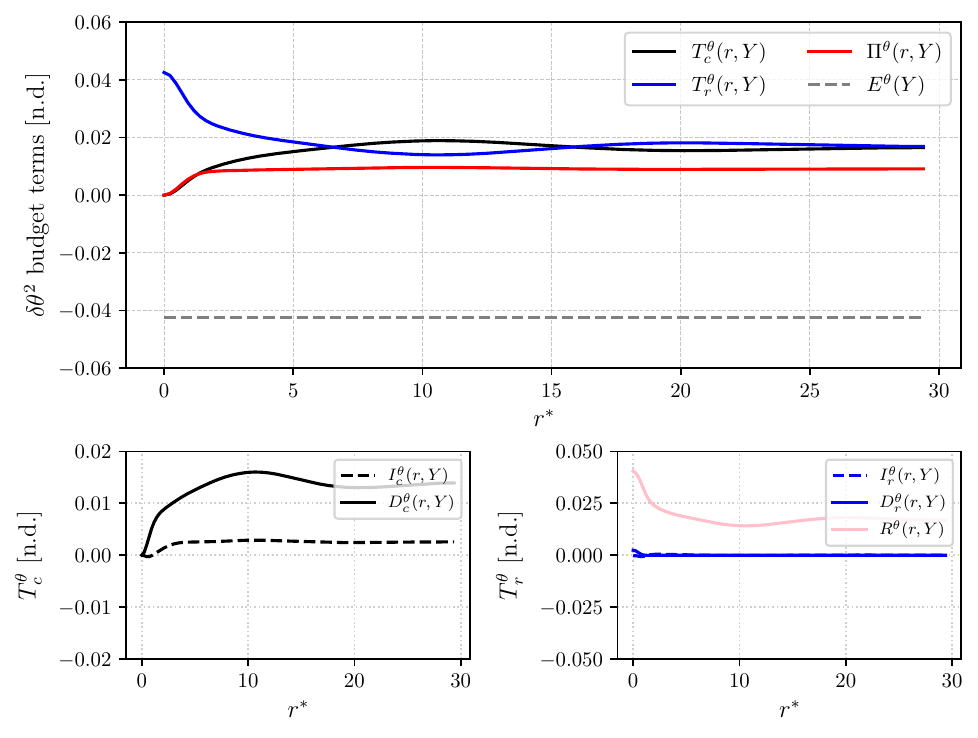}
    \subcaption{}
    \label{fig:KY_budget_ha40y5}
  \end{minipage}

  \caption{Budgets of $\delta \theta^2$ plotted in (a) the bulk, (b) the transitional layer and (c) the boundary layer. The left-hand subplots contain the contribution terms of $T_c^\theta$ whilst right-hand subplots contain the contributing terms of $T_r^\theta$. Plotted for case E at $Ha=40$ with a wall-normal magnetic field.}
  \label{fig:KY_budgets_ha40y}
\end{figure}

The scale energy balance described in equation \ref{eq:reduced_KH_budget} is plotted against $r$ for various $Y$ positions in figure \ref{fig:KH_budgets_ha40y}, for case E. Similar the scale variance balance (equation \ref{eq:reduced_KY_budget}) is plotted in figure \ref{fig:KY_budgets_ha40y}. Again, the focus here is on the Lorentz forcing term, $L$, and how this affects the overall balance of the budget, this time in scale-space. Then the second-hand influence of velocity upon the scale-variance budget is discussed.

\begin{figure}[htbp]
  \centering
\includegraphics[width=\linewidth]{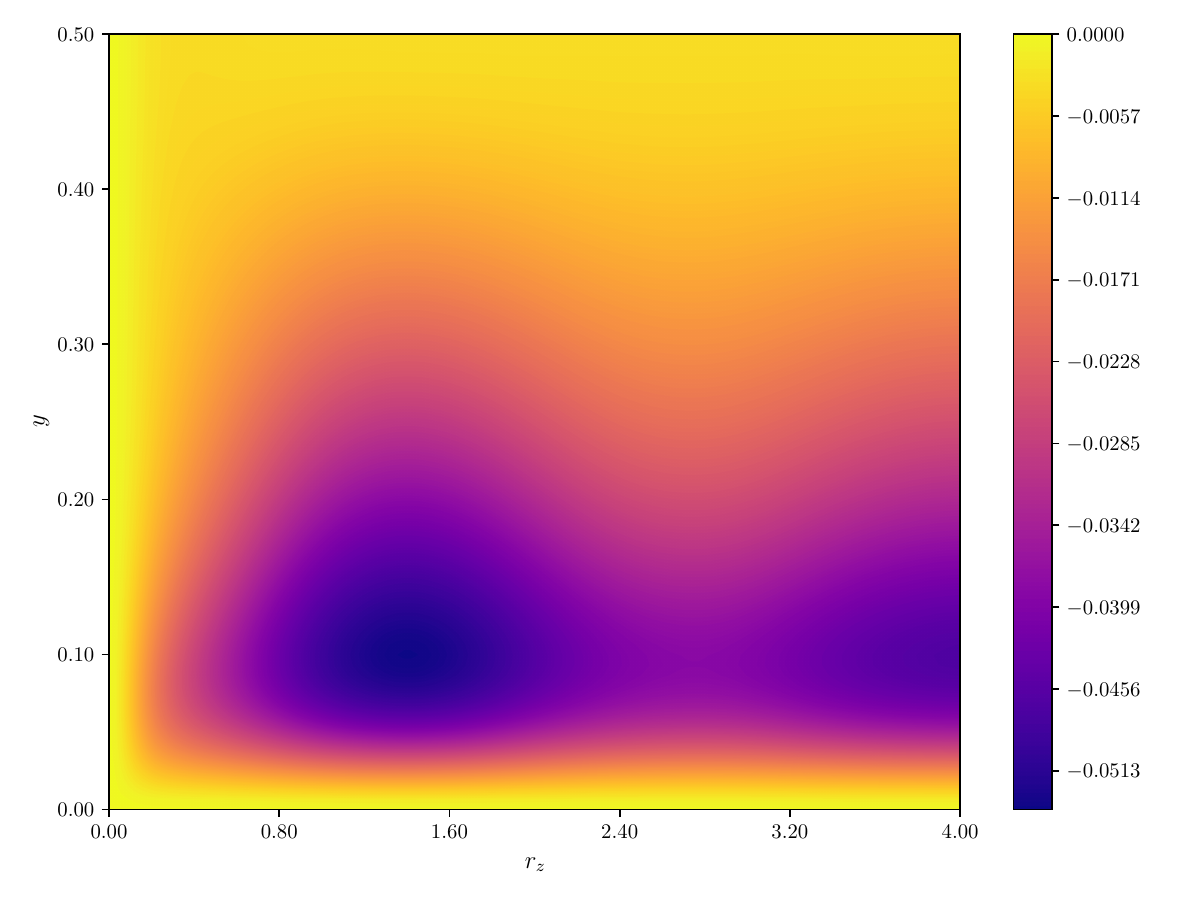}
  \caption{The mean work done by the Lorentz force, $L(r,Y)=Ha^2\sqrt{\frac{Pr}{Ra}} \varepsilon_{ijk} \langle \delta u_i \delta J_j \rangle$ plotted in the $Y$-$r_z$ plane at $r_x=0$ for case E.}
  \label{fig:lorentz_struct_ha40y}
\end{figure}

Firstly we focus on the bulk region. We see that for most scales, the most significant terms in the budget are the production $\Pi$ and the Lorentz work $L$. The Lorentz term, $L$, is plotted in figure \ref{fig:lorentz_struct_ha40y} and represents the work done by the Lorentz force at position $Y$ and scale $r$. The transport term $T_c$ and the dissipation $E$ are also significant whilst the transfer term $T_r$ is small for most of the budget. At most scales, there is a positive energy excess $\Pi-E-L > 0$, which is redistributed to different scales and physical positions through the transfer and transport terms, $T_r$ and $T_c$ respectively. The strictly negative Lorentz force arises in the budget and is large in magnitude relative to $E$, meaning most of the produced energy is dissipated through Lorentz mechanisms $L$. In the wall-normal MHD cases, $E$ is observed to be much smaller. From the perspective of transport, the only non-negligible term contributing to $T_c$ is the pressure transport, $P$, which is negative and acts to redistribute energy towards near wall regions. Again this differs to non-MHD RBC, where $I_c$ and $D_c$ contribute to transporting scale energy through the bulk. Looking at the transfer term, $T_r$, it is clear that transfer between scales is damped by the Lorentz force. All of $I_r$, $D_r$ and $R$ are mostly close to zero implying energy produced through buoyancy at large and intermediate scales remains at these scales, and is then redistributed to near wall regions through $P$, or dissipated prior to this through $L$. That is, with the exception of the smallest scales, where $D_r$ becomes positive implying an energy cascade towards smaller scales of motion, where the energy is then dissipated through viscous mechanisms, $E$.

Moving onto the transitional layer, $L$ is now greater in magnitude than $\Pi$, meaning more energy is dissipated than produced. However, a positive energy excess still exists here due to the positive value of $T_c$ within the transitional layer, which is mostly due to the largely positive $P$ corresponding to scale energy being redistributed from the bulk to the transitional layer through pressure mechanisms. $I_c$ and $D_c$ remain negligible in the transitional layer. Additionally, for most scales $R<0$; this implies an inverse energy cascade within the transitional layer. $I_r$ remains negligible. This reverse cascade can be associated to the plumes being stretched out in the $y$ direction. Conversely, $D_r$ is zero for most scales meaning little energy is being transferred to wall-parallel scales. Relative to the non-MHD case, this term is damped explaining the thinner plumes observed as $Ha$ increases (figure \ref{fig:xz_temperature_plots_hay}). $I_r$ is damped out almost completely by the Lorentz force. Usually in non-MHD convection, this term is positive corresponding to the cascade of energy towards smaller scales. The lack of small scale velocity features in wall-normal MC can be explained by the Lorentz force damping $I_r$ meaning energy is not passed to smaller scales as effectively. 

The picture in the boundary layer is relatively simple. There is no significant buoyant production here, and the Lorentz force has no long-term energetic influence. The boundary layer is sustained purely by positive $D_c$ and $D_r$ terms which are compensated by a largely negative $E$. The energy provided by $P$ and $D_c$ is transported towards smaller scales, completely through $D_r$ where it is eventually dissipated. Again, inertial contributions are insignificant here. This picture is close to that of the non-MHD MC, as one would expect given that $L$ is insignificant within the boundary layer. In the non-MHD case inertial terms are small in magnitude and in the MHD case they are damped out almost completely.

To give a complete picture of MC the scale-variance balance described in equation \ref{eq:reduced_KY_budget} is plotted in figure \ref{fig:KY_budgets_ha40y}. This balance is similar in many ways to what is described in the non-MHD case, so to avoid repetition, focus is explicitly given to differing behaviour in the MHD cases. A lot of this behaviour is a consequence of the previous scale-energy analysis given that the Lorentz force does not act directly upon the temperature field. In the thermal bulk region, scale-variance is sustained through the transport $T^\theta_c$ and transfer $T_r^\theta$ terms whilst production $\Pi^\theta$ is negligible. The thermal dissipation $E^\theta$ acts to balance the budget. Looking closer, the bulk receives scale-variance mostly through the inertial transport $I^\theta_c$ which transports variance from the transitional layer into the bulk region. With regards to transfer, $T_r^\theta$ is positive and it is mostly $R^\theta$ that is acting to transport scale variance to smaller wall-normal scales. In the transitional layer, $\Pi$ is the dominant term which contributes scale-variance to the transitional layer. The variance is then redistributed to different spatial positions through $T_c<0$ and also contributed to larger scales of motion $T_r$. Both $I_c$ and $D_c$ contribute to $T_c$ in this case. However, with respect to $T_r$, the contribution is mainly from $R^\theta$ suggesting the energy is transferred to larger wall-normal scales. This is likely associated with the formation of new thermal plumes; energy is produced through buoyancy, stretched in the wall-normal direction ($R^\theta>0$) and carried away towards the bulk ($T_c<0$). This differs to \citet{togni2015physical} who associated the negative $T_r$ with the plumes spreading out in wall-parallel directions as they impinge near the wall. But the present analysis based upon $R^\theta$ being the predominant transfer term suggests this is related to the stretching of thermal structures in the wall-normal direction and hence is associated with plume emission instead. Finally, within the boundary layer the behaviour is similar to the bulk. Variance is received from the transitional layer, this time through $D_c$ and is instead transferred to smaller wall-normal scales instead through $R^\theta$ where it is finally dissipated through $E^\theta$.

From the mechanistic perspective the balance of scale-energy in MC behaves similarly to the non-MHD case. However, the magnitudes of terms differ significantly which must be an indirect result of scale-energy balance, given that $L$ only directly acts on scale-energy. The first important observation is the overall lower mean temperature gradient and the larger observed temperature variance. The key to this is the damping of the inertial terms in scale space. This results in lower magnitude velocity fluctuations, slower plume dynamics and correlations like $\langle v' \theta' \rangle$ being smaller in magnitude. As a result, convective heat transfer is less efficient and a lower mean temperature gradient is observed. Conversely, the lack of inertial transport in the velocity field leads to a lack of a classic forward turbulence cascade in the velocity meaning less small scale turbulence structure, less mixing of the large scale coherent structures and a longer lifetime of coherent structures. However, this coherence leads to locally large variations in the temperature variance resulting in a larger observed $\langle \theta'^2\rangle$ and $\delta \theta^2$. As a result, terms like $T_c$ and $T_r$, which depend upon the local value of $\delta \theta$ are observed to be larger in MC. 

Summarising the overall scale energy behaviour of wall-normal MC, it is clear that the behaviour in scale-space described by \citep{togni2015physical} is robust and a lot of the mechanisms extend well, at least phenomenologically, to MC. Nevertheless, it is still essential to account for the extra dissipation due to the presence of the Lorentz force which comes at the expense of the magnitudes of the transport and transfer terms within the budget. Specifically, inertial mechanisms $I_c$ and $I_r$ are damped to the point where their contribution to the budget is negligible. This suggests that over the long-term the Lorentz force heavily suppresses non-linearity explaining the slower evolution of coherent structures and the larger time-scales typically associated with wall-normal MC. Additionally, outside of the boundary layer, $D_c$ and $D_r$ are notably suppressed relative to the non-MHD case, with $R$ only contributing a small inverse scale transfer within the transitional layer giving explanation to the thinner plumes observed in figure \ref{fig:xz_temperature_plots_hay}. The redistribution of scale energy in wall-normal MC is mostly due to $P$, where the behaviour is mechanistically similar to non-MHD RBC. Finally, the boundary layer behaves similarly to the non-MHD case, with the exception of $I_r$ being negligible, and a less significant $E$, which is explained given that a significant amount of energy is dissipated through $L$ before reaching the boundary layer. This behaviour is then responsible for magnitude differences in the mechanistically invariant scale-variance budget (in contrast to non-MHD RBC). The damping of non-linear inertial terms leads to a lack of turbulent mixing in the velocity. As a result, convective heat transfer is less effective resulting in a lesser global temperature gradient but large scale coherence means a locally greater temperature variance. This behaviour is reflected in the differing magnitudes of the terms in the MHD scale-variance budgets. 

\begin{figure}[htbp]
  \centering
  \begin{minipage}[b]{0.48\linewidth}
    \centering
    \includegraphics[width=\linewidth]{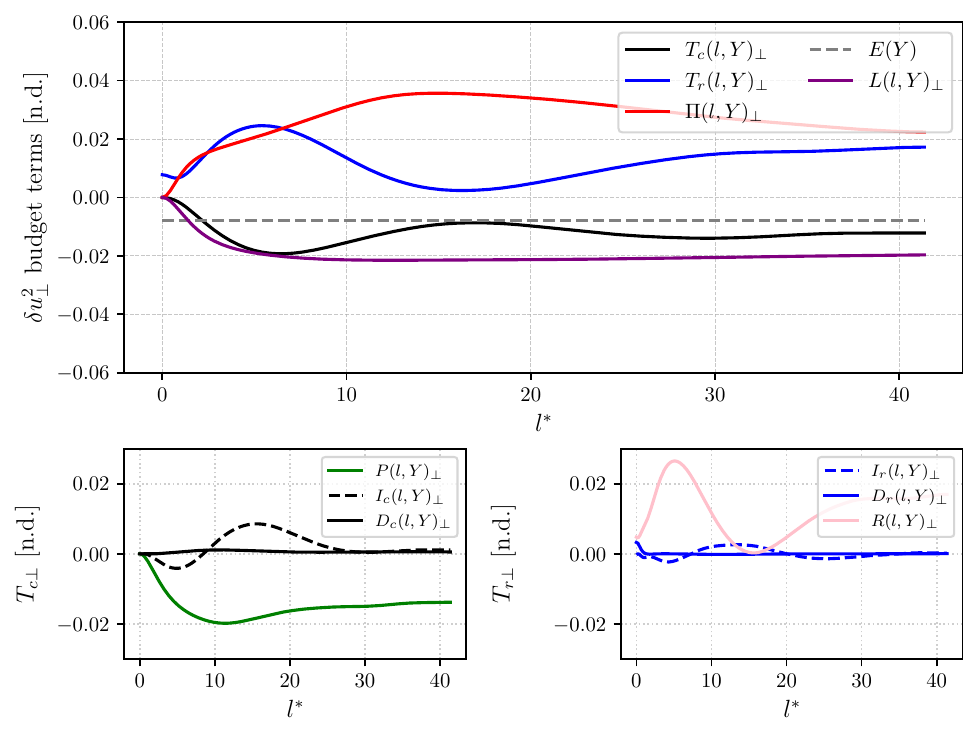}
    \subcaption{}
    \label{fig:KH_budget_perp_ha40x50}
  \end{minipage}%
  \hspace{0.03\linewidth}
  \begin{minipage}[b]{0.48\linewidth}
    \centering
    \includegraphics[width=\linewidth]{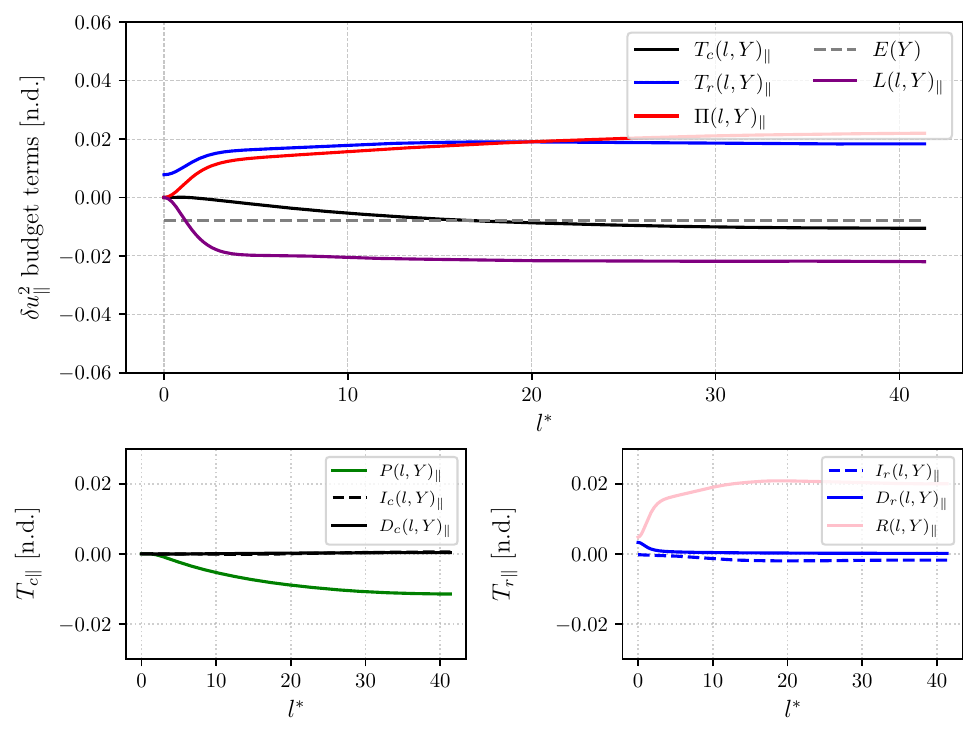}
    \subcaption{}
    \label{fig:KH_budget_parallel_ha40x50}
  \end{minipage}

  \vspace{1ex}

  \begin{minipage}[b]{0.48\linewidth}
    \centering
    \includegraphics[width=\linewidth]{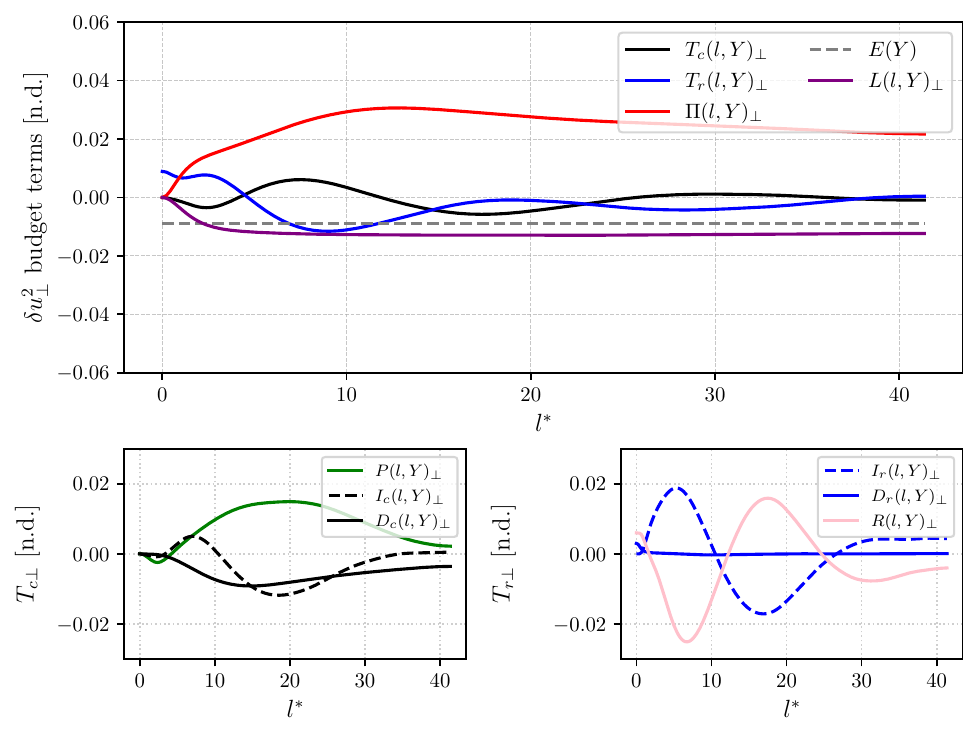}
    \subcaption{}
    \label{fig:KH_budget_perp_ha40x30}
  \end{minipage}%
  \hspace{0.03\linewidth}
  \begin{minipage}[b]{0.48\linewidth}
    \centering
    \includegraphics[width=\linewidth]{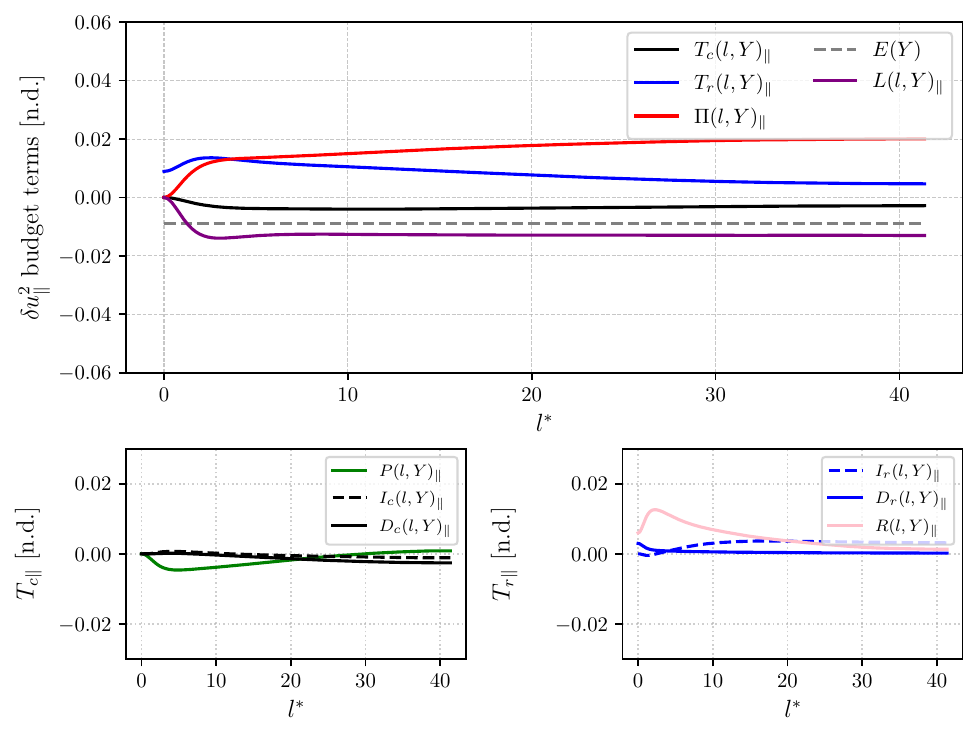}
    \subcaption{}
    \label{fig:KH_budget_parallel_ha40x30}
  \end{minipage}

  \vspace{1ex}

  \begin{minipage}[b]{0.48\linewidth}
    \centering
    \includegraphics[width=\linewidth]{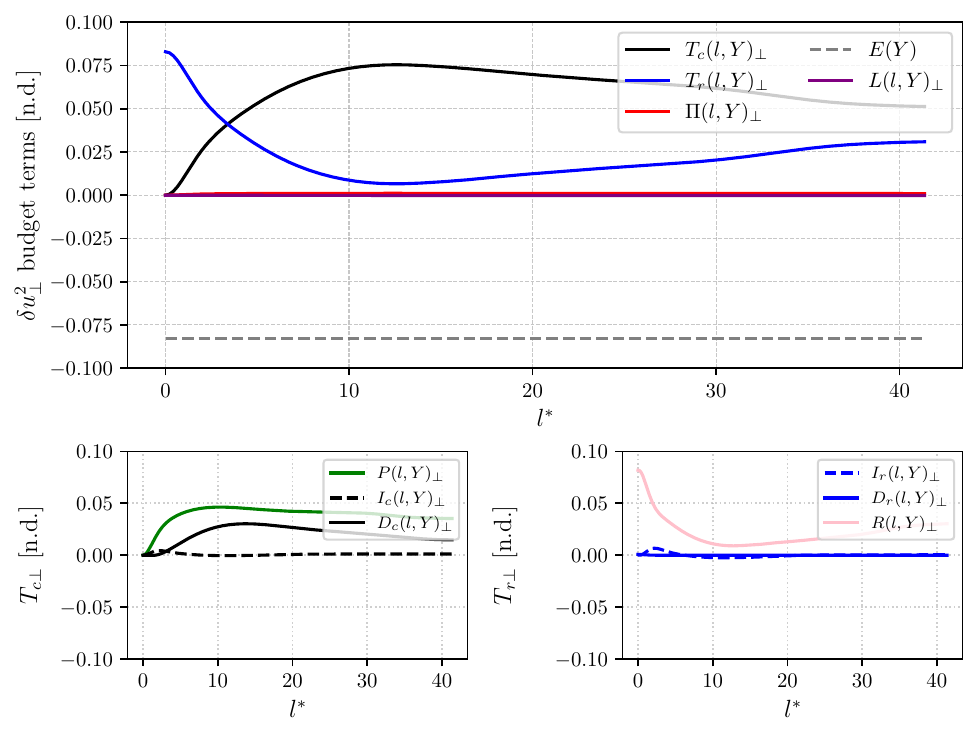}
    \subcaption{}
    \label{fig:KH_budget_perp_ha40x10}
  \end{minipage}%
  \hspace{0.03\linewidth}
  \begin{minipage}[b]{0.48\linewidth}
    \centering
    \includegraphics[width=\linewidth]{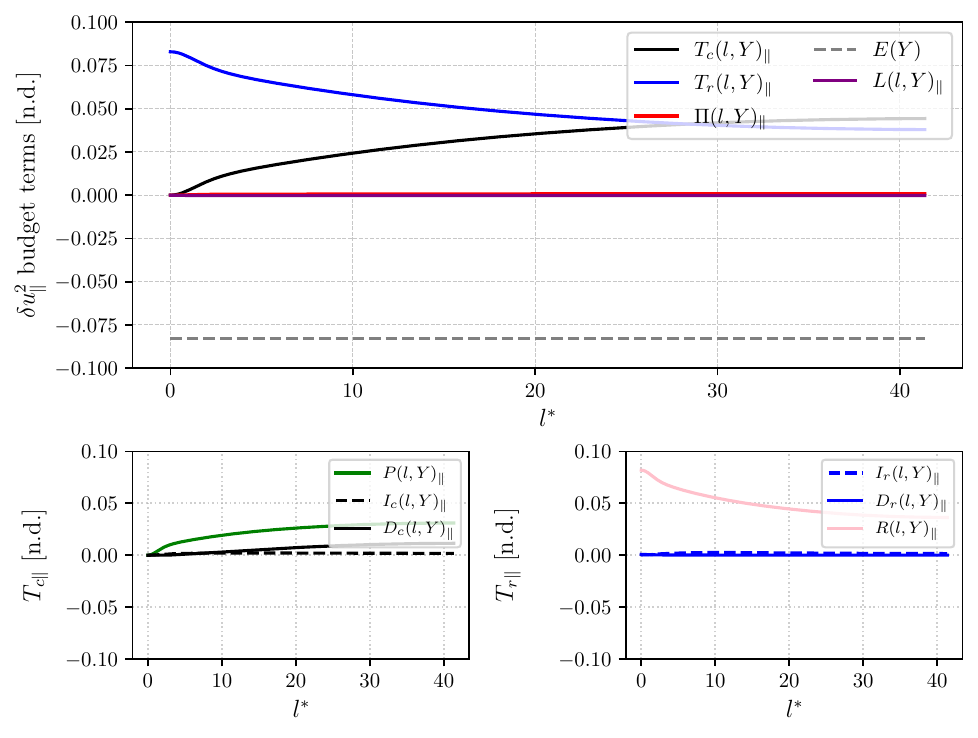}
    \subcaption{}
    \label{fig:KH_budget_parallel_ha40x10}
  \end{minipage}

  \caption{Budgets of $\delta u^2$ plotted in (a)(b) the bulk, (c)(d) the transitional layer and (e)(f) the boundary layer. The left-hand subplots contain the contribution terms of $T_c$ whilst right-hand subplots contain the contributing terms of $T_r$. Perpendicular components are plotted on the left (a)(c)(e) and parallel components on the right (b)(d)(f). Plotted for case B at $Ha=40$ with a wall-parallel magnetic field.}
  \label{fig:KH_budgets_ha40x}
\end{figure}

\subsection{Anisotropic scale by scale analysis of wall-parallel field cases}

The focus on this section will be extending the previous scale by scale analysis to cases with wall-parallel magnetic fields (cases A, B and C). As is clear from previous results in this study (figures \ref{fig:xz_temperature_plots_hax} and \ref{fig:1D_tke_budget_ha40x}) these cases introduce strong anisotropy into the system, where the symmetry of $u$ and $w$ is broken by the presence of the magnetic field along the $x$ direction. One would also expect that the symmetry in $r_x$ and $r_z$ of previously computed structure functions is also broken. To extend this analysis we consider directional structure functions, defined in equations \ref{eq:directional1}-\ref{eq:directional2}, to  describe the anisotropy. This will allow distinguishing between scales of motion in the $x$ and $z$ directions. The approach will be the same as the previous section, studying the production, transport, transfer and dissipation of scale-energy and scale-variance, but this time considering the slices $(r_x=0, Y,r_z)$ and $(r_x,Y,r_z=0)$ separately. This is to describe the directional influence of the magnetic field upon the overall energy balance. Trivially, the full budgets written in equations \ref{eq:reduced_KH_budget} and \ref{eq:reduced_KY_budget} hold in these planes and the terms plotted are subsets of the full $(r_x, Y,r_z)$ space. Use of the $r$-averaging operator is avoided in this section (equation \ref{eq:r_average}).

Again, to avoid repetition, the analysis will focus on a single case, case B this time, which has a wall-parallel magnetic field and a Hartmann number of $Ha=40$. The scale-energy budget for this case is shown in figure \ref{fig:KH_budgets_ha40x}, plotted directionally for various wall-normal positions. The anisotropy between the $\delta u^2_\parallel$ and $\delta u^2_\perp$ budgets can be observed immediately. The field-parallel budget terms show monotonic behaviour whilst the field-perpendicular terms have minima and maxima throughout the budget. This field-perpendicular behaviour can be associated with the Q2D coherent structures that are present within the flow. This means structure functions are dominated by the two-point-correlation as opposed to the contribution due to single point expectations. As a result of this, the terms in the budget have preferred length scales which can be associated with the large scale coherent structures that are aligned perpendicular to the field. By contrast, the terms in the field-parallel budgets are monotonic suggesting parallel scale energy is not concentrated at specific scales. Naturally, this is expected in the field-parallel budgets given that there is little variation in the flow variables along this direction (figure \ref{fig:xz_temperature_plots_hax}). Given the magnitude of the terms in the parallel budgets, it is clear that production, transport and transfer are still active in these directions. 

\begin{figure}[htbp]
    \centering
    
    \begin{subfigure}{0.8\textwidth}
        \centering
        \includegraphics[width=\linewidth]{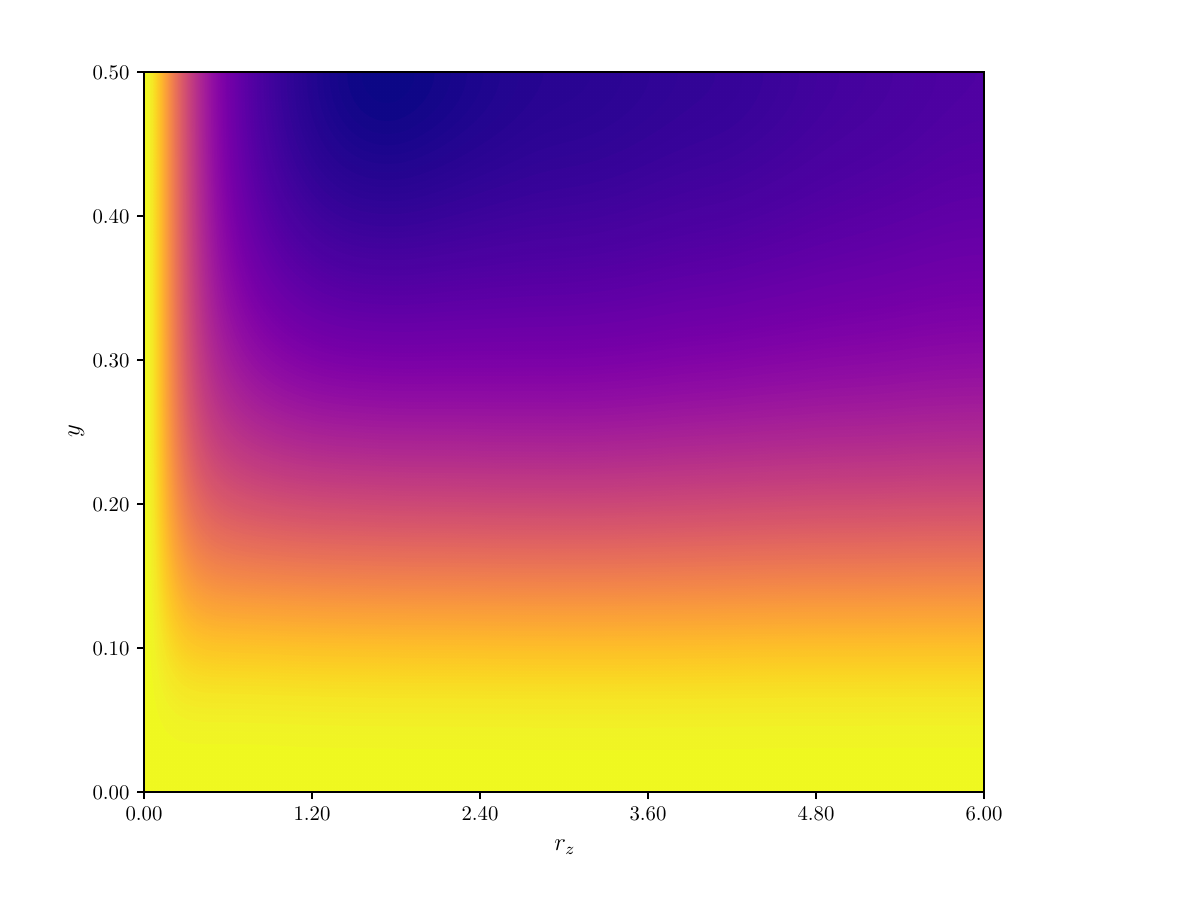}
    \end{subfigure}
    
    \begin{subfigure}{0.8\textwidth}
        \centering
        \includegraphics[width=\linewidth]{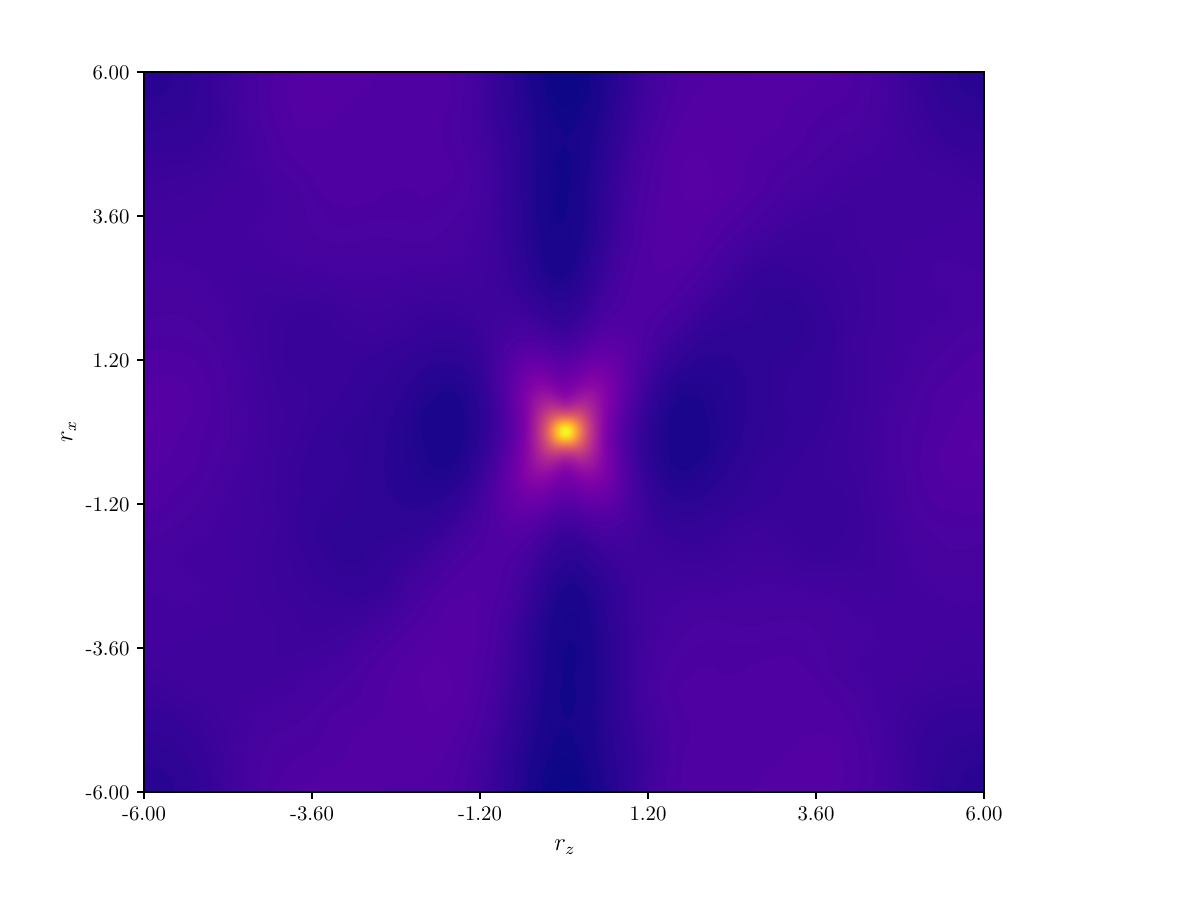}
    \end{subfigure}
    
    \begin{subfigure}{0.8\textwidth}
        \centering
        \includegraphics[width=\linewidth]{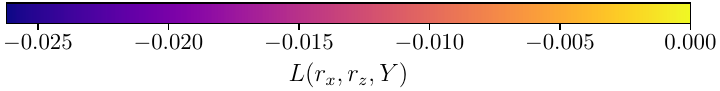}
    \end{subfigure}
    \caption{The mean work done by the Lorentz force, $L(r_x,r_z,Y)=Ha^2\sqrt{\frac{Pr}{Ra}} \varepsilon_{ijk} \langle \delta u_i \delta J_j \rangle$ plotted for case E. (Top) $Y-r_z$ plane at $r_x=0$ (bottom) $r_x$-$r_z$ plane plotted at $Y=0.5$.}
    \label{fig:lorentz_struct_ha40x}
\end{figure}

To study the influence of the anisotropy, differing budget terms are discussed. The roles of the terms remain mostly similar to what is discussed in the previous section, so repetition of this is avoided. First of all, the only approximately isotropic term is the Lorentz term $L$ (figure \ref{fig:lorentz_struct_ha40x}), where $L_\parallel \approx L_\perp$ for all $Y$ positions shown. $L$ acts equally on all intermediate and large scales of motion and is predominantly active outside of the boundary layer. The smooth negative contribution of the Lorentz force across these scales and directions indicates it acts approximately as a scale-independent isotropic dissipative sink in scale space. Therefore, the anisotropy in the budgets must result from the reorganisation of production, transport and transfer terms instead. The first feature of the plots is that $\Pi_\perp \ge \Pi_\parallel$ at all $Y$ positions meaning production is providing more energy to field-perpendicular scales. With respect to the redistribution of energy, $T_{c\perp}$ and $T_{r\perp}$ can be seen to be anti-correlated in all of the perpendicular budgets. This is related to the the large scale coherence giving the system preferred length scales. Some length scales are more sensitive to transport whilst others are more sensitive to transfer. Looking closer at the transport mechanisms, it can be seen that in the parallel budgets most energy is redistributed through $P_\parallel$ whilst $D_{c\parallel}$ and $I_{c\parallel}$ are suppressed. The perpendicular transport receives contributions from $P_\perp$, $D_{c\perp}$ and $I_{c\perp}$. Looking at transfer in the bulk, both the parallel and perpendicular budgets mainly transfer energy through $D_r$. The difference between the two plots is that $D_{r\parallel}$ term is roughly scale independent ($\partial_l T_{r\parallel} \approx 0$) whilst $D_{r\perp}$ shows the oscillatory behaviour discussed before. This trend is shown in the transitional layer too, except $D_{r\parallel}$ now has a gradient. In the boundary layer, the transfer behaviour is similar between the parallel and perpendicular directions, differing in magnitude at intermediate scales, where $D_{r\perp}$ is lower in magnitude.

\begin{figure}[htbp]
  \centering

  \begin{subfigure}{0.48\linewidth}
    \includegraphics[width=\linewidth]{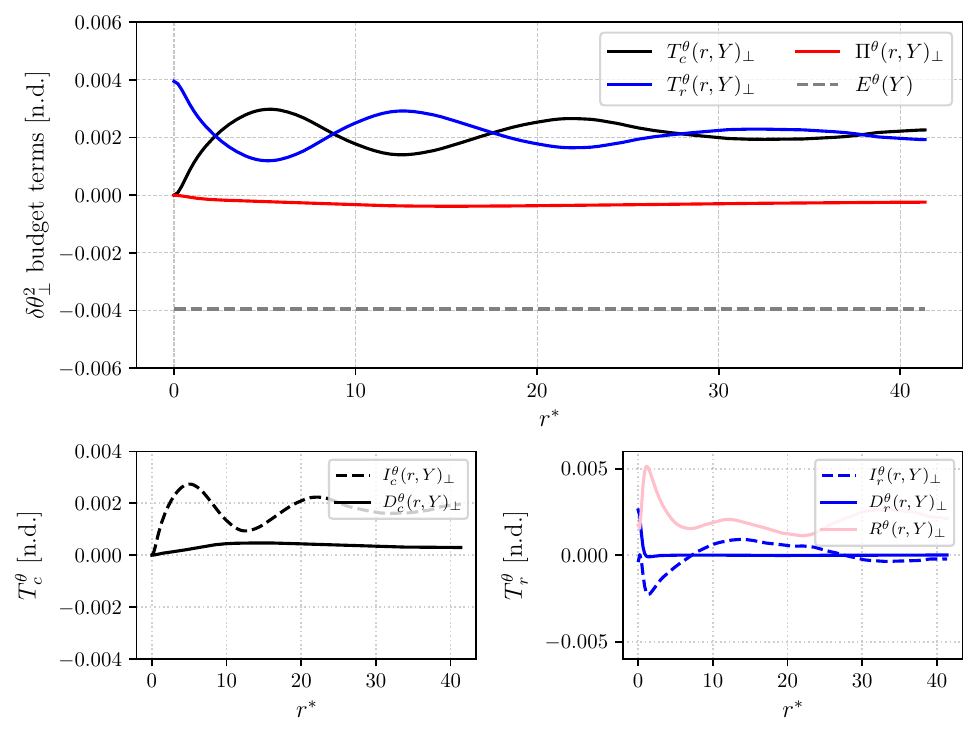}
    \caption{}
  \end{subfigure}\hfill
  \begin{subfigure}{0.48\linewidth}
    \includegraphics[width=\linewidth]{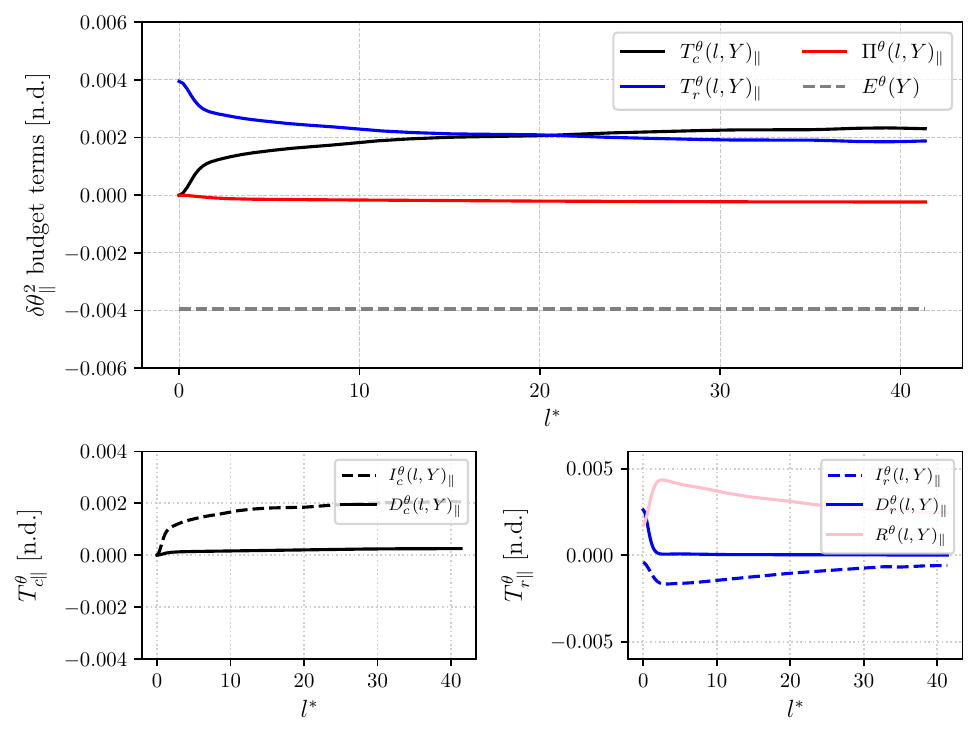}
    \caption{}
  \end{subfigure}

  \vspace{1ex}

  \begin{subfigure}{0.48\linewidth}
    \includegraphics[width=\linewidth]{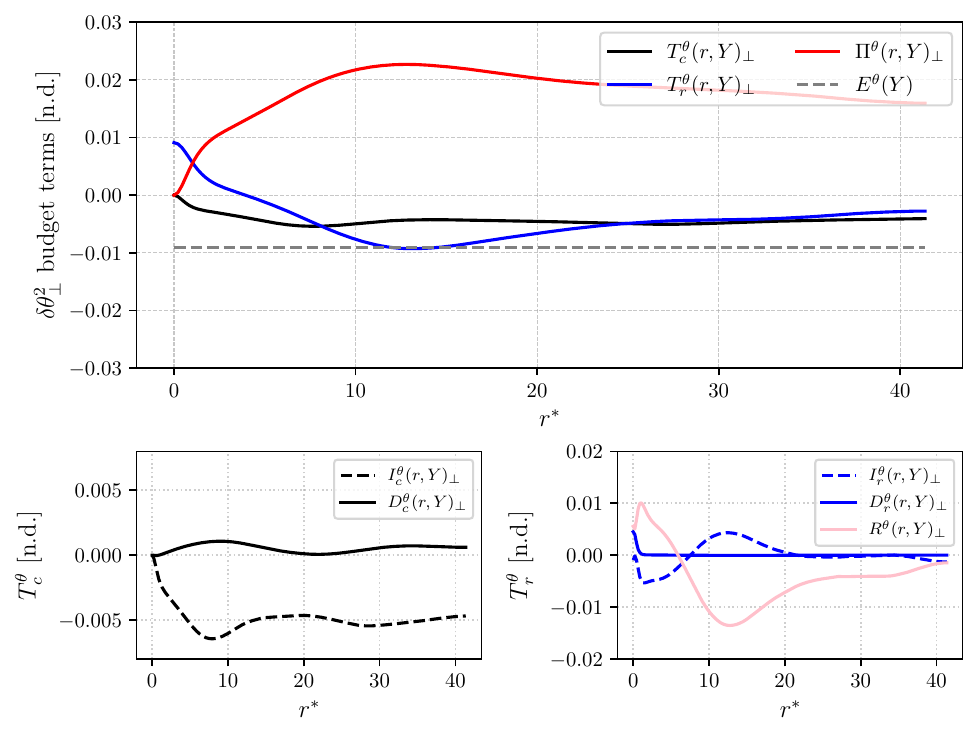}
    \caption{}
  \end{subfigure}\hfill
  \begin{subfigure}{0.48\linewidth}
    \includegraphics[width=\linewidth]{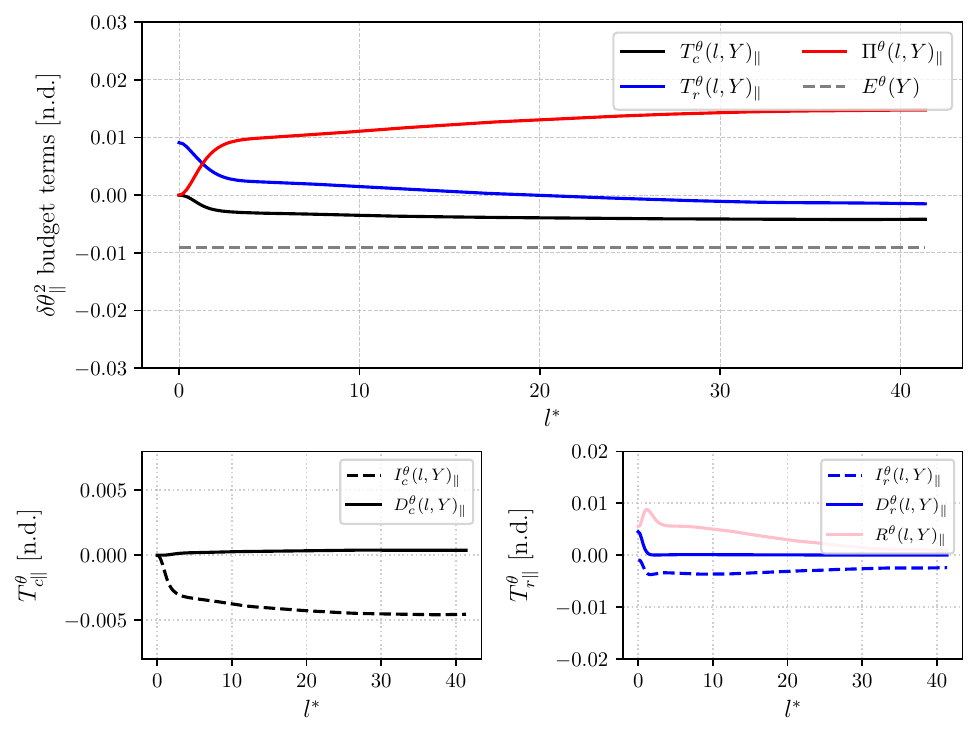}
    \caption{}
  \end{subfigure}

  \vspace{1ex}

  \begin{subfigure}{0.48\linewidth}
    \includegraphics[width=\linewidth]{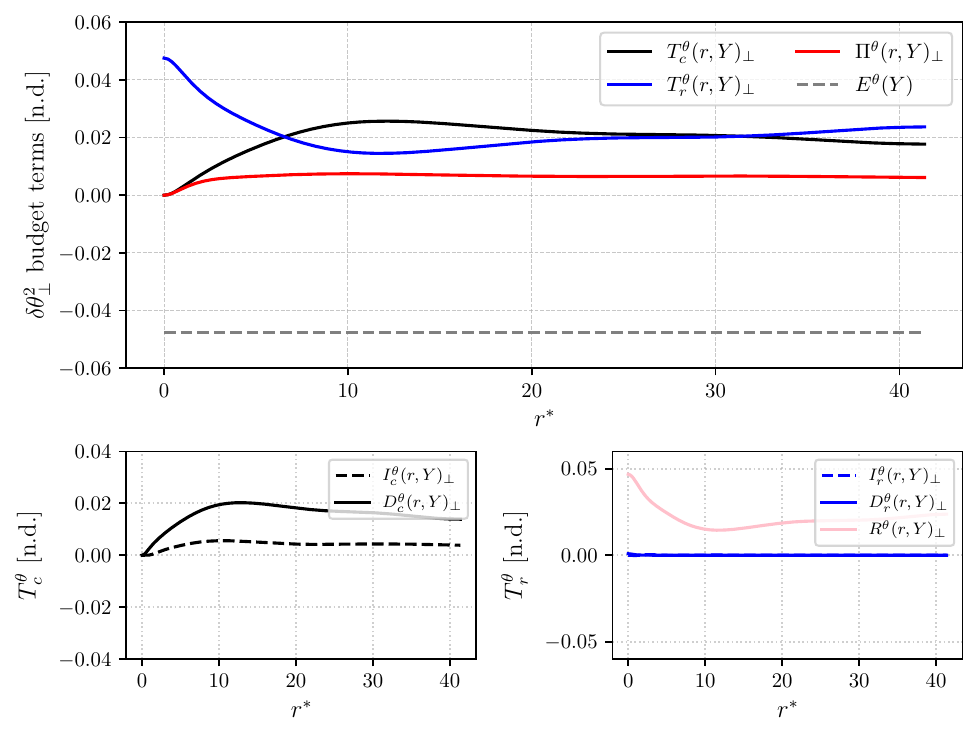}
    \caption{}
  \end{subfigure}\hfill
  \begin{subfigure}{0.48\linewidth}
    \includegraphics[width=\linewidth]{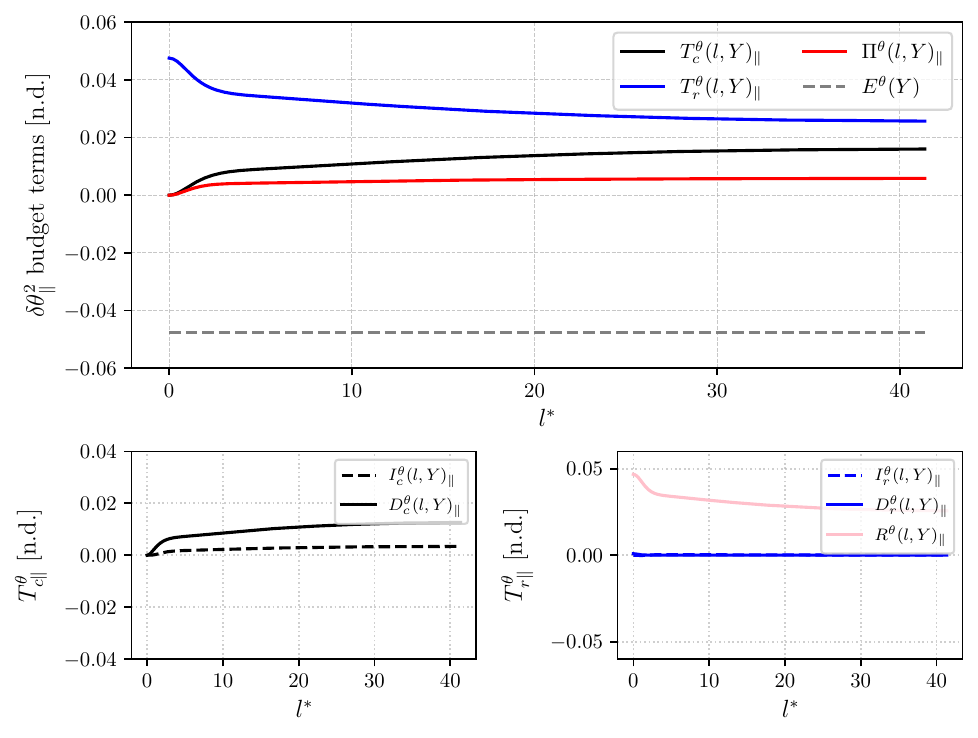}
    \caption{}
  \end{subfigure}
  \caption{Budgets of $\delta \theta^2$ plotted in (a)(b) the bulk, (c)(d) the transitional layer and (e)(f) the boundary layer. The left-hand subplots contain the contribution terms of $T_c^\theta$ whilst right-hand subplots contain the contributing terms of $T_r^\theta$. Perpendicular components are plotted on the left (a)(c)(e) and parallel components on the right (b)(d)(f). Plotted for case B at $Ha=40$ with a wall-parallel magnetic field.}
  \label{fig:KY_budgets_ha40x}
\end{figure}

Moving onto the scale-variance analysis, equation \ref{eq:reduced_KY_budget} is plotted directionally in figure \ref{fig:KY_budgets_ha40x}.These budgets exhibit behaviour that closely resembles the scale-energy budgets. That is, the parallel budget terms show monotonic behaviour whilst the perpendicular budgets have minima and maxima and can show oscillatory behaviour. Clearly this anisotropy is an indirect consequence of the Lorentz force. The anisotropy is initially imposed upon the velocity field and this carries into the temperature budget through inertial mechanisms. To avoid repetition, the roles of terms in the scale-variance budget are not discussed given they are similar to results in previous sections and the effect of anisotropy is similar to what has been discussed in this section.

\section{Conclusion}
\label{sec:conclusion}

This study has considered magnetoconvection in the classical canonical Rayleigh-Bénard configuration under the influence of both wall-normal and wall-parallel magnetic fields. The magnetic field strongly modifies the structure of convective turbulence through anisotropic Joule dissipation leading to different flow dynamics depending upon the field orientation.

For the cases with wall-parallel magnetic fields, the flow develops quasi-two-dimensional (Q2D) turbulence characterised by flow structures that align perpendicularly with the magnetic field and show weak variation along the field direction. This anisotropy arises from the field preferentially damping field parallel velocity components which imposes a preferred direction upon the flow. A notable feature of this regime is the formation of laminar wall jets in the field-perpendicular, wall-parallel velocity component. This reorganisation of velocity is indirectly transferred to the temperature field resulting in Q2D thermal structures too. In contrast, the wall-normal magnetic field preserves symmetry in the horizontal directions and maintains three-dimensional flow structures, although these are strongly damped by the Lorentz force. Increasing the Hartmann number leads to thinner plumes that lack in small-scale turbulent features, accompanied by a reduction in global heat transfer. In this configuration, Joule dissipation acts in the same direction as buoyancy and sufficiently strong magnetic fields would be expected to completely suppress convective motion.

Energy and temperature variance budgets reveal that the magnetic field alters the redistribution of kinetic energy through its interaction with pressure-strain mechanisms. In the wall-parallel field cases, the Lorentz dissipation primarily affects the bulk flow and leads to anisotropic redistribution that favourably redistributes kinetic energy to the wall-parallel, field-normal velocity component. For the wall-normal case, the Lorentz dissipation peaks in the transitional layer and damps both wall-parallel velocity components reducing the efficiency of pressure-driven redistribution mechanisms as the thermal plumes impinge on the wall. Analysis in scale-space further shows that the Lorentz force acts as an additional isotropic dissipation sink, and that it damps intermediate and large scales of motion, significantly weakening transfer processes between scales. As a result, energy produced by buoyancy generally remains near its production scale and is mostly dissipated by Lorentz dissipation. This explains the dominance of large scale structures and the suppression of small scale turbulence in magnetoconvective flows.

Overall, these results establish a direct connection between observed flow structures and underlying energy-transport and -transfer mechanisms in magnetoconvection. In particular, the interaction between the Lorentz force and pressure-mediated redistribution mechanisms plays a central role, as well as the appearance of the Lorentz dissipation as an isotropic sink in scale-space. The fundamental processes described here provide significant physical insight and should be accounted for in future modelling and theoretical attempts to describe magnetoconvective turbulence.

\section{Acknowledgements}
The authors would like to thank the EPSRC for the computational time made available on the UK national supercomputing facility ARCHER2 via the UK Turbulence Consortium (EP/X035484/1).

\appendix
\appendix
\renewcommand{\thesection}{Appendix \Alph{section}.}
\section{Derivation of scale-energy equation under buoyancy and Lorentz forces}\label{appA}

In this appendix a full derivation of equation \ref{eq:KH_budget} is shown, following the procedure defined in \citet{hill2002exact}. The goal is to derive a transport equation for the second-order velocity increment between two points $x_i$ and $x_i'$, $\delta u_i^2 = (u_i(x_i')-u_i(x_i))^2$. We start by defining the short-hand notation $u_i = u_i(x_i)$ and $u'_i=u_i(x_i')$ (not to be confused with the fluctuation) and the new coordinates as $X_i=(x_i+r_i)/2$ and $r_i=x_i-x_i'$. Equation \ref{eq:momentum} is satisfied independently at both points $x_i$ and $x_i'$,

\begin{equation}
    \label{eq:momentum_not_dash}
    \frac{\partial u_i}{\partial t}
    + u_j \frac{\partial u_i}{\partial x_j}
    = -\frac{\partial p}{\partial x_i}
    + \sqrt{\frac{Pr}{Ra}} \frac{\partial^2u_i}{\partial x_j \partial x_j}
    +f_i'
\end{equation}

\begin{equation}
    \label{eq:momentum_dash}
    \frac{\partial u'_i}{\partial t}
    + u'_j \frac{\partial u'_i}{\partial x'_j}
    = -\frac{\partial p'}{\partial x'_i}
    + \sqrt{\frac{Pr}{Ra}} \frac{\partial^2u'_i}{\partial x'_j \partial x'_j}
    + f_i'
\end{equation}

\noindent where the Lorentz and buoyancy forces are contained within the body force term, $f_i =\theta g_i +Ha^2\sqrt{Pr/Ra}J_jB_k$. The derivatives with respect to $x_i$, $x_i'$, $r_i$ and $X_i$ can be written as follows

\begin{equation}
\label{eq:deriv1}
\frac{\partial }{\partial x_i} = \frac{\partial}{\partial r_i} + \frac{1}{2}\frac{\partial}{\partial X_i}
\end{equation}

\begin{equation}
\label{eq:deriv2}
\frac{\partial }{\partial x'_i} = -\frac{\partial}{\partial r_i} + \frac{1}{2}\frac{\partial}{\partial X_i}.
\end{equation}

\begin{equation}
    \label{eq:deriv3}
    \frac{\partial}{\partial X_i} = \frac{\partial}{\partial x_i} + \frac{\partial}{\partial x_i'}
\end{equation}

\begin{equation}
\label{eq:deriv4}
\frac{\partial}{\partial r_i} = \frac{1}{2}\left(\frac{\partial}{\partial x_i} - \frac{\partial}{\partial x_i'}\right)
\end{equation}

\noindent Adding together equations \ref{eq:momentum_dash} and \ref{eq:momentum_not_dash} leads to

\begin{equation}
    \frac{\partial \delta u_i}{\partial t} 
    +u'_j\frac{\partial u_i}{\partial x_j} + u_j'\frac{\partial u_i'}{\partial x_j'}
    = 
    -\frac{\partial p}{\partial x_i} +\frac{\partial p'}{\partial x_i'}
    +\sqrt{\frac{Pr}{Ra}} \left( \frac{\partial^2 u_i}{\partial x_j \partial x_j} - \frac{\partial^2 u'_i}{\partial x'_j \partial x'_j} \right)
    + \delta f_i
\end{equation}, 

\noindent Applying the identities in equations \ref{eq:deriv1} and \ref{eq:deriv2} yields the following form for the viscous terms, 

\begin{equation}
\frac{\partial^2u_i}{\partial x_j \partial x_j} - \frac{\partial^2u'_i}{\partial x'_j \partial x'_j}
=
\frac{\partial^2\delta u_i}{\partial r_j \partial r_j} + \frac{1}{4} \frac{\partial^2\delta u_i}{\partial X_j\partial X_j}
\end{equation}

\noindent the convective terms,

\begin{equation}
u_j\frac{\partial u_i}{\partial x_j} - u'_j\frac{\partial u'_i}{\partial x'_j}
= u_j \frac{\partial \delta u_i}{\partial x_j} + u'_j \frac{\partial \delta u_i}{\partial x_j'} = \delta u_j \frac{\partial \delta u_i}{\partial r_j}
+ \frac{1}{2} (u_j + u'_j) \frac{\partial \delta u_i}{\partial X_j}
\end{equation}

\noindent and the pressure terms, 

\begin{equation}
    -\frac{\partial p}{\partial x_i} +\frac{\partial p'}{\partial x_i'}
    = -\frac{\partial (p+p')}{\partial r_i} - \frac{1}{2}\frac{\partial \delta p}{\partial X_i}
    = - \frac{\partial \delta p}{\partial X_i},
\end{equation}

\noindent where the identity, $\partial/\partial r_i[f(x) \pm g(x')] = \partial/\partial X_i[f(x) \mp g(x')]/2$ has been used. Therefore the resulting transport equation for the first-order increment can be written as

\begin{equation}
    \label{eq:1st_increment}
    \frac{\partial \delta u_i}{\partial t} +
    \delta u_j \frac{\partial \delta u_i}{\partial r_j}
    + \frac{1}{2} (u_j + u'_j) \frac{\partial \delta u_i}{\partial X_j}
    = -\frac{\partial \delta p}{\partial X_i}
    +\sqrt{\frac{Pr}{Ra}} \left( \frac{\partial^2\delta u_i}{\partial r_j \partial r_j} + \frac{1}{4} \frac{\partial^2\delta u_i}{\partial X_j\partial X_j} \right)
    + \delta f_i.
\end{equation}

\noindent The equation for the second-order increments, $\delta u_i \delta u_i $ is obtained by multiplying through by $\delta u_k$ and taking the trace of the resulting equation. Starting with the viscous terms, 

\begin{equation}
\delta u_i \left( \frac{\partial^2\delta u_i}{\partial r_j \partial r_j} + \frac{1}{4} \frac{\partial^2\delta u_i}{\partial X_j\partial X_j} \right)
= \frac{1}{2}\frac{\partial^2(\delta u_i\delta u_i)}{\partial X_j\partial X_j}
+ 2\frac{\partial^2(\delta u_i\delta u_i)}{\partial r_j \partial r_j}
- 2\left(\frac{\partial u_i}{\partial x_j}\frac{\partial u_i}{\partial x_j} + \frac{\partial u_i'}{\partial x_j'}\frac{\partial u_i'}{\partial x_j'}\right)
\end{equation}

\noindent Again, the identities in equations \ref{eq:deriv1}-\ref{eq:deriv4} have been used along with the relation $f_i \frac{\partial^2 f_i}{\partial x_j \partial x_j} = \frac{1}{2}\frac{\partial^2 (f_if_i)}{\partial x_j \partial x_j}
- \frac{\partial f_i}{\partial x_j}\frac{\partial f_i}{\partial x_j}$. The advective term can be written as follows, 

\begin{equation}
\delta u_i \delta u_j \frac{\partial \delta u_i}{\partial r_j}
+ \frac{1}{2} \delta u_i  (u_j + u'_j) \frac{\partial \delta u_i}{\partial X_j}
= \frac{1}{2} \frac{\partial(\delta u_j \delta u_i \delta u_i)}{\partial r_j}
+ \frac{1}{4} \frac{\partial [(u_j + u_j')\delta u_i \delta u_i]}{\partial X_j}
\end{equation}

\noindent using the product rule, and the incompressibility conditions for increments $\partial\delta u_i/\partial X_i=0$ and $\partial \delta u_i / \partial r_i = 0$. The remaining terms are relatively simple; $\delta u_i \partial p/ \partial X_i = \partial (\delta u_i \delta p)/ \partial X_i$ (incompressibility) and $\delta u_i\partial\delta u_i/\partial t = 1/2\partial (\delta u_i \delta u_i)/\partial t$ (product rule). So far, the forcing term has been left generally as $\delta u_i \delta f_i$. Trivially, for a buoyancy and Lorentz force this appears as $\delta u_i \delta f_i = \delta u_i \delta \theta g_i + Ha^2 \sqrt{Pr / Ra} \delta u_i \delta J_j B_k$, under the assumption that $B_k$ is constant in space. The final form of the instantaneous equation for the second order increment can now be written as,

\begin{align}
\frac{\partial (\delta u^2)}{\partial t}
+ \frac{\partial(\delta u_j \delta u^2)}{\partial r_j}
+ \frac{\partial (u_j^*\delta u^2)}{\partial X_j}
&= \nonumber \\ -\frac{\partial (\delta u_i \delta p)}{\partial X_i}
+  \sqrt{\frac{Pr}{Ra}}
\left(
\frac{1}{2}\frac{\partial^2(\delta u^2)}{\partial X_j\partial X_j}
+ 2\frac{\partial^2(\delta u^2)}{\partial r_j \partial r_j}
+ 2Ha^2 \delta u_i \delta J_j B_k
\right) 
& - 4(\varepsilon^*)
+ 2\delta u_i \delta \theta g_i
\end{align}

\noindent where we have written $\delta u^2 = \delta u_i \delta u_i$, $u_j^* = (u_j+u_j')/2$ and $\varepsilon=\sqrt{\frac{Pr}{Ra}}\frac{\partial u_i}{\partial x_j}\frac{\partial u_i}{\partial x_j}$. The final step is averaging the equations; here we assume stationarity and spatial homogeneity in the $x$ and $z$ directions. Here we denote the combined time and spatial average as $\langle \cdot\rangle$. Averaging the equations causes any terms of the form $\partial/\partial t \langle\cdot\rangle$, $\partial/\partial X\langle\cdot\rangle$ and $\partial / \partial Z\langle \cdot \rangle$ to vanish resulting in the transport equation for the second-order velocity structure function,

\begin{align}
\frac{\partial\langle\delta u_j \delta u^2\rangle}{\partial r_j}
+  \frac{\partial \langle v^*\delta u^2\rangle}{\partial Y}
&= -\frac{\partial \langle\delta v \delta p\rangle}{\partial Y}
+ \sqrt{\frac{Pr}{Ra}}
\left(
\frac{1}{2}\frac{\partial^2\langle\delta u^2\rangle}{\partial Y\partial Y}
+ 2\frac{\partial^2\langle\delta u^2\rangle}{\partial r_j \partial r_j}
+ 2Ha^2  \langle \delta u_i \delta J_j \rangle B_k
\right) \nonumber \\
&\quad - 4\langle \varepsilon^*\rangle
+ 2\langle \delta v \delta \theta \rangle g_i
\end{align}

\noindent This same procedure can be followed to derive a transport equation for $\langle \delta \theta^2 \rangle$ which we omit to avoid repetition. 

\clearpage
\bibliographystyle{plainnat} 
\bibliography{references}

\end{document}